\documentclass{aa}  

\usepackage{graphicx}
\graphicspath{{../fig/}} 
\usepackage{txfonts}
\usepackage{hyperref}
\newcommand{\eq}[1]{\begin{equation}  #1 \end{equation}}

\newcommand{\eqa}[1]{\begin{align}   #1 \end{align}}

\newcommand{\br}[1]{\left( #1 \right)}
\newcommand{\bc}[1]{\left\{ #1 \right\}}
\newcommand{\bb}[1]{\left[ #1 \right]}
\newcommand{\ba}[1]{\left\langle #1 \right\rangle}

\newcommand{\nn}{\nonumber}

\newcommand{\dd}{{\rm d}}

\newcommand{\expo}[1]{~{\rm e}^{ #1 }}

\newcommand{\ic}{{\rm i}}

\usepackage{xcolor}
\definecolor{darkred}{HTML}{660022}

\usepackage{amsmath}
\DeclareMathOperator*{\argmax}{\arg\!\max}

\begin{document} 

\title{KiDS-1000 Methodology: Modelling and inference for joint weak gravitational lensing and spectroscopic galaxy clustering analysis}
\titlerunning{KiDS-1000 Methodology}

\author{
  B. Joachimi\inst{1}
  \and
  C.-A. Lin\inst{2}
  \and
  M. Asgari\inst{2}
  \and
  T. Tr\"oster\inst{2}
  \and
  C. Heymans\inst{2,3}
  \and  
  H. Hildebrandt\inst{3}
  \and
  F. K{\"o}hlinger\inst{3}
  \and
  A. G. S{\'a}nchez\inst{4}
  \and
  A. H. Wright\inst{3}
  \and 
  M. Bilicki\inst{5}
  \and
  C. Blake\inst{6}
  \and
  J. L. van~den~Busch\inst{3}
  \and
  M. Crocce\inst{7,8}
  \and
  A. Dvornik\inst{3}
  \and
  T. Erben\inst{9}
  \and
  F. Getman\inst{10}
  \and
  B. Giblin\inst{2}
  \and
  H. Hoekstra\inst{11}
  \and
  A. Kannawadi\inst{12,11}
  \and
  K. Kuijken\inst{11}
  \and
  N. R. Napolitano\inst{13}
  \and
  P. Schneider\inst{9}
  \and
  R. Scoccimarro\inst{14}
  \and
  E. Sellentin\inst{11}
  \and
  H. Y. Shan\inst{15,16}
  \and
  M. von~Wietersheim-Kramsta\inst{1}
  \and
  J. Zuntz\inst{2}  
}
\authorrunning{
Joachimi, Lin, Asgari, Tr\"oster, Heymans et al. 
}

\institute{
  Department of Physics and Astronomy, University College London, Gower Street, London WC1E 6BT, UK\\
  \email{b.joachimi@ucl.ac.uk}
  \and
  Institute for Astronomy, University of Edinburgh, Royal Observatory, Blackford Hill, Edinburgh, EH9 3HJ, UK\\
  \email{calin@roe.ac.uk}
  \and
  Ruhr-Universit\"at Bochum, Astronomisches Institut, German Centre for Cosmological Lensing (GCCL), Universit\"atsstr. 150, 44801 Bochum, Germany
  \and
  Max-Planck-Institut f\"ur extraterrestrische Physik, Postfach 1312, Gie{\ss}enbachstr., 85741 Garching, Germany
  \and
  Center for Theoretical Physics, Polish Academy of Sciences, al. Lotnik{\'o}w 32/46, 02-668, Warsaw, Poland
  \and
 Centre for Astrophysics \& Supercomputing, Swinburne University of Technology, P.O. Box 218, Hawthorn, VIC 3122, Australia
  \and
  Institute  of  Space  Sciences  (ICE,  CSIC),  Campus  UAB, Carrer  de  Can  Magrans,  s/n,  08193 Barcelona,  Spain
  \and
  Institut  d\lq Estudis  Espacials  de  Catalunya  (IEEC),  08034  Barcelona,  Spain
  \and
   Argelander-Institut f{\"u}r Astronomie, Universit{\"a}t Bonn, Auf dem H{\"u}gel 71, 53121 Bonn, Germany
   \and
   INAF - Astronomical Observatory of Capodimonte, Via Moiariello 16, 80131 Napoli, Italy
   \and
  Leiden Observatory, Leiden University, P.O.Box 9513, 2300RA Leiden, The Netherlands 
  \and
  Department of Astrophysical Sciences, Princeton University, 4 Ivy Lane, Princeton, NJ 08544, USA 
  \and
  School of Physics and Astronomy, Sun Yat-sen University, Guangzhou 519082, Zhuhai Campus, P.R. China 
  \and
   Center  for  Cosmology  and  Particle  Physics,  Department  of  Physics, New  York  University,  NY  10003,  New  York,  USA
   \and
  Shanghai Astronomical Observatory (SHAO), Nandan Road 80, Shanghai 200030, China 
  \and
  University of Chinese Academy of Sciences, Beijing 100049, China 
}

\date{Received ; accepted }

\abstract{
We present the methodology for a joint cosmological analysis of weak gravitational lensing from the fourth data release of the ESO Kilo-Degree Survey (KiDS-1000) and galaxy clustering from the partially overlapping Baryon Oscillation Spectroscopic Survey (BOSS) and the 2-degree Field Lensing Survey (2dFLenS).
Cross-correlations between BOSS and 2dFLenS galaxy positions and source galaxy ellipticities have been incorporated into the analysis, necessitating the development of a hybrid model of non-linear scales that blends perturbative and non-perturbative approaches, and an assessment of signal contributions by astrophysical effects.
All weak lensing signals were measured consistently via Fourier-space statistics that are insensitive to the survey mask and display low levels of mode mixing.
The calibration of photometric redshift distributions and multiplicative gravitational shear bias has been updated, and a more complete tally of residual calibration uncertainties was propagated into the likelihood.
A dedicated suite of more than $20\,000$ mocks was used to assess the performance of covariance models and to quantify the impact of survey geometry and spatial variations of survey depth on signals and their errors.
The sampling distributions for the likelihood and the $\chi^2$ goodness-of-fit statistic have been validated, with proposed changes for calculating the effective number of degrees of freedom.
The prior volume was explicitly mapped, and a more conservative, wide top-hat prior on the key structure growth parameter $S_8=\sigma_8\,(\Omega_{\rm m}/0.3)^{1/2}$ was introduced.
The prevalent custom of reporting $S_8$ weak lensing constraints via point estimates derived from its marginal posterior is highlighted to be easily misinterpreted as yielding systematically low values of $S_8$, and an alternative estimator and associated credible interval are proposed.
Known systematic effects pertaining to weak lensing modelling and inference are shown to bias $S_8$ by no more than 0.1 standard deviations, with the caveat that no conclusive validation data exist for models of intrinsic galaxy alignments.
Compared to the previous KiDS analyses, $S_8$ constraints are expected to improve by $20\,\%$ for weak lensing alone and by $29\,\%$ for the joint analysis.
}  


\maketitle

\tableofcontents

\section{Introduction}

The $\Lambda$-cold dark matter ($\Lambda$CDM) concordance model of cosmology remains a resounding success, explaining a plethora of observations with little more than a handful of free parameters. Without alternative theories on the horizon that come anywhere close to rivalling $\Lambda$CDM in terms of completeness and predictiveness, the focus is turning increasingly towards nascent discrepancies between cosmological probes when interpreted within the standard model. This approach is fruitful: we are either witnessing the first hints towards new physics that will eventually lead to a revision or extension of $\Lambda$CDM, or we have encountered an important and persistent limitation in the fidelity of one or more key cosmological probes that is critical to push back for forthcoming cosmological experiments to succeed.

While tension in the value of the Hubble constant $H_0$ as inferred from local probes and early-Universe physics in combination with fluctuations in the cosmic microwave background (CMB) has received the most widespread attention to date \citep[e.g.][]{verde19}, a similarly puzzling systematic difference between CMB and low-redshift measurements of the amplitude of matter density fluctuations has persisted for at least the past seven years \citep[cf.][]{heymans13}, albeit at a lower significance. CMB constraints on the amplitude of the primordial matter power spectrum can be extrapolated to redshift zero assuming standard model structure growth. Alternatively, low-redshift probes of the cosmic large-scale structure provide a more direct measure of the growth amplitude, usually expressed in terms of the standard deviation of matter density fluctuations in spheres of radius $8\,h^{-1}\,{\rm Mpc}$ (where $H_0 = 100 h\, {\rm km\,s^{-1}\,Mpc^{-1}}$), denoted by $\sigma_8$.

Cosmic shear, the coherent distortions of faint galaxy images through weak gravitational lensing by the intervening large-scale matter distribution, is highly sensitive to the strength of density fluctuations projected along the line of sight, as well as ratios of angular diameter distances between observer, lenses, and sources. Consequently, it primarily constrains a degenerate combination of $\sigma_8$ and the density parameter of matter (both dark and baryonic matter), $\Omega_{\rm m}$, by convention usually expressed via the derived parameter $S_8=\sigma_8\,(\Omega_{\rm m}/0.3)^{1/2}$. The most recent, final analysis of the ESA \textit{Planck} mission produced a constraint of $S_8=0.83 \pm 0.02$ ($68\,\%$ credible region) from its primary CMB probes \citep{planck18_parameters,planck18_likelihood}.

By contrast, recent cosmic shear measurements resulted in $S_8=0.74 \pm 0.04$ for the first $450\,{\rm deg}^2$ of data from the ESO Kilo-Degree Survey (KiDS; \citealp{kv450}, KV450 hereafter) and $S_8 = 0.79^{+0.02}_{-0.03}$ for the first year of data from the Dark Energy Survey (DES; \citealp{troxel18}). Early data from the Hyper-SuprimeCam (HSC) Survey yielded $S_8 = 0.78 \pm 0.03$ \citep{hikage19}. At least part of the variation between the weak lensing-derived $S_8$ values can be attributed to the differences in the calibration of galaxy redshift distributions, with the central value of the DES analysis decreasing by 0.03 when switching to the approach taken by KiDS \citep{joudaki20}, and the KV450 $S_8$ best-fit decreasing by 0.02 when selecting galaxy samples that allow for robust calibration \citep{wright20b}. Independently of the calibration approach taken, however, the three concurrent surveys are in good agreement, with DES and KiDS in combination reaching $\sim \!\! 3\sigma$ tension with \textit{Planck} \citep{asgari20,joudaki20}. Intriguingly, fully independent measurements of the full redshift-space power spectrum of spectroscopic galaxy clustering with the Baryon Oscillation Spectroscopic Survey (BOSS) show the same tendency towards lower values of $S_8$ \citep{damico19,colas20,ivanov19,troester19}.

While cosmic shear is attractive as a direct probe of the dark and bright matter distribution with few assumptions about the nature or state of this matter, its statistical power derives mainly from non-linear scales that cannot be modelled analytically from first principles. The most recent cosmic shear analyses have already reduced the statistical errors to the same level as residual systematic uncertainty caused by both astrophysical and measurement biases that need to be carefully removed or modelled (see \citealp{mandelbaum18} for a recent review). The joint analysis of cosmic shear with galaxy clustering in overlapping sky areas, including the cross-correlations between galaxy positions and gravitational shear in the background commonly known as galaxy-galaxy lensing (GGL), offers additional cosmological constraining power and the removal of parameter degeneracies, as well as self-calibration of some of the systematics \citep{bernstein09,joachimi10}. Therefore, the combination of cosmic shear with foreground samples of galaxy \lq lenses\rq\ is being established as the default analysis approach to extracting gravitational lensing information from large-scale structure \citep{uitert18,joudaki18,desy1}.

This work details the approach to the joint probes modelling and likelihood analysis of Data Release 4 (DR4; \citealp{kuijken19}) of KiDS. The data set will be referred to as KiDS-1000, covering a raw survey area of $1006\,{\rm deg}^2$ out of the final KiDS area of $1350\,{\rm deg}^2$. The KiDS footprint has near-complete coverage by spectroscopic galaxy surveys -- BOSS \citep{eisenstein11} in the north and the 2-degree Field Lensing Survey (2dFLenS; \citealp{blake16}) in the south. The latter is designed to provide similar galaxy samples to BOSS albeit with sparser sampling. We opt to combine KiDS-1000 cosmic shear with GGL in the BOSS and 2dFLenS overlap and an existing galaxy clustering analysis over the full BOSS survey area by \citet[S17 hereafter]{sanchez17}. The approach is an extension of the earlier analyses by \citet{troester19} who combined S17 with cosmic shear only from KV450 and \citet{joudaki18} who combined the first KiDS cosmic shear results \citep{hildebrandt17} with clustering and GGL measured in 2dFLenS and BOSS. An analogous investigation for DES was carried out by \citet{des_y1_methods}, where both lens and source samples were derived from the same survey.

The paper is structured as follows: Section~\ref{sec:modelling} presents our choice of observables and the model to jointly predict the corresponding cosmological and astrophysical signals. Section~\ref{sec:data} summarises the characteristics and processing of our data sets and details the calibration and modelling of measurement systematics. An extensive suite of survey simulations is introduced in Sect.~\ref{sec:mocks}, and the impact of survey geometry and the spatial variation of survey characteristics on the cosmological signals is assessed. In Sect.~\ref{sec:covariances} a detailed discussion of our approach to modelling the joint covariance of the observables, and its accuracy, is provided. Section~\ref{sec:inference} covers the choice of model parameters and the form of the likelihood, as well as how we report goodness of fit and parameter constraints in the context of large, highly non-Gaussian parameter spaces. The analysis pipeline and associated modelling choices are validated in Sect.~\ref{sec:parameter_constraints} before concluding in Sect.~\ref{sec:conclusions}.

Companion papers present the construction and calibration of gravitational shear catalogues \citep{giblin20} and redshift distributions \citep{hildebrandt20}, the cosmological analysis of cosmic shear \citep{asgari20c} and of joint clustering and weak lensing \citep{heymans20}, and an investigation of extensions to the spatially flat $\Lambda$CDM model considered throughout here \citep{troester20}. Unless otherwise specified, calculations and plots use the fiducial values of our set of model parameters specified in Table$\,$\ref{tab:fiducialpars}.

\section{Signal modelling}
\label{sec:modelling}

Including non-linear scales is critical to unlock the full constraining power of large-scale structure cosmological probes. Traditionally, galaxy clustering analyses have opted for perturbative approaches to modelling signals into the mildly non-linear regime to wavenumbers of $k \sim \! 0.3\,h\,{\rm Mpc}^{-1}$, which enables closed-form and consistent expressions for non-linear structure growth, galaxy bias, and redshift-space effects (see \citealp{desjacques18} for a review). Cosmic shear, in contrast, routinely accesses scales at $k > 1\,h\,{\rm Mpc}^{-1}$, where its signal-to-noise is high, via empirical models based on the halo model paradigm that have been calibrated on N-body and/or full hydro-dynamical simulations (beginning with \citealp{hamilton91}; see also the seminal work by \citealp{smith03}). This makes the consistent modelling of a joint clustering and cosmic shear analysis, especially for the GGL cross-correlation, a major challenge for current and forthcoming analyses \citep[e.g.][]{bose19}. We choose a hybrid approach that blends models previously used successfully by the BOSS and KiDS teams, as detailed in Sects.~\ref{sec:matterps} and \ref{sec:pgm}.

The second major topic of this section is the choice of two-point statistics used in the analysis. As we demonstrate, the clustering and weak lensing (i.e. cosmic shear and GGL) probes can be treated independently, so that we can simply adopt the original statistics employed in S17, that is to say correlation functions as a function of three-dimensional separation and the angle subtended with the line of sight, binned into wedges. For GGL and cosmic shear we opt for band powers derived from angular correlation functions, closely following \citet{uitert18}. They are straightforward to compute and combine good scale localisation in both configuration and Fourier space, which helps to separate interpretable cosmological information from contamination by systematics and astrophysical signals (cf. \citealp{asgari20b,park20} for recent examples in GGL). Examples include additive shear bias on large configuration-space scales, blending of galaxy images on small configuration-space scales, baryonic effects on the matter power spectrum at high angular frequencies, and computationally expensive curved-sky expressions at low angular frequencies.
 
While the majority of weak lensing analyses still rely on correlation functions, which have sensitivity to a very broad range of Fourier modes of the matter distribution \citep[e.g.][]{asgari20}, some direct Fourier-space measures like pseudo-Cls \citep{hikage19} and quadratic maximum likelihood estimators \citep{koehlinger17} have seen recent applications. \citet{becker16} calculated both pseudo-Cl and band power spectra, but only used real-space measures in their main likelihood analysis, as did the most recent cosmic shear papers from KiDS and DES. Correlation function-derived band powers have two additional advantages over direct Fourier-space statistics in that they are not sensitive to the survey geometry, and in that they do not require an estimate of the noise power, which is trivially removed by excluding zero-lag correlation functions in the band power computation. A detailed comparison between correlation functions, band powers, and compressed statistics with exact separation of E- and B-modes (versus the approximate separation offered by band powers) is presented in \citet{asgari20c}.

\subsection{Matter power spectra}
\label{sec:matterps}

Linear matter power spectra $P_{\rm m, lin}(k,z)$ as a function of wavenumber $k$ and redshift $z$ are calculated with \textsc{Camb}\footnote{Code for Anisotropies in the Microwave Background; \texttt{https://camb.info}} \citep{lewis00,lewis02,howlett12}. Massive standard-model neutrinos are included assuming the normal hierarchy at the minimum sum of masses, $\sum m_\nu = 0.06\,{\rm eV}$. We keep this parameter fixed in our default setup, in line with \citet{planck18_parameters}, and consider constraints on neutrino mass in a companion paper \citep{troester20}.

We derive two non-linear matter power spectra from the linear one. First, we compute a non-perturbative model using the \citet{mead15} \textsc{HMCode} included in \textsc{Camb}, whose validity extends into the deeply non-linear regime, recovering N-body results at the few per-cent level to $k=10\,h\,{\rm Mpc}^{-1}$. The halo model approach in \textsc{HMCode} incorporates baryonic feedback (primarily through Active Galactic Nuclei; AGN) on the matter distribution via a halo bloating parameter $\eta_0$ and the amplitude of the halo mass-concentration relation, denoted by $A_{\rm bary}$. Following \citet{joudaki18}, we fix the relation between the two feedback parameters to $\eta_0 = 0.98 - 0.12 A_{\rm bary}$ and only use $A_{\rm bary}$ as a free parameter in our analysis. The resulting matter power spectrum is denoted by $P_{\rm m, nl}$.

The choice of a one-parameter feedback model is justified as, on relevant spatial scales, it captures the range of matter power spectrum models extracted from hydro-dynamic simulation which are consistent with observations (see \citealp{chisari19} for a review). KiDS analyses have only very weak sensitivity to the scale substantially affected by baryon feedback as evidenced by the unconstrained posterior on $A_{\rm bary}$ \citep{kv450}; this remains the case for KiDS-1000 \citep{asgari20c}. More conservative modelling choices, including a parametrisation of the possible ranges of matter power spectrum modifications due to feedback via principal component analysis \citep{eifler15}, were investigated by \citet{huang19}. For scales slightly less conservative than those used in KiDS-1000, they found minimal impact on $S_8$ and moderate shifts along the $\Omega_{\rm m} - \sigma_8$ degeneracy, which are captured well by $A_{\rm bary}$.

The second, perturbative non-linear power spectrum, $P_{\rm m, nl-pt}$, is calculated using renormalised perturbation theory (RPT; \citealp{crocce06}) to one-loop order. We adopt the updated version applied in S17 (see also \citealp{eggemeier20}), dubbed gRPT, that incorporates a re-summation of the mode coupling terms consistently with the propagator ones (i.e. those proportional to the linear power spectrum) to make the theory Galilean invariant (see also \citealp{taruya12}).  S17 demonstrated with N-body simulations that this model is accurate to $2\,\%$ or better out to scales of at least $k=0.25\,h\,{\rm Mpc}^{-1}$.

\begin{figure}
	\includegraphics[width=\columnwidth]{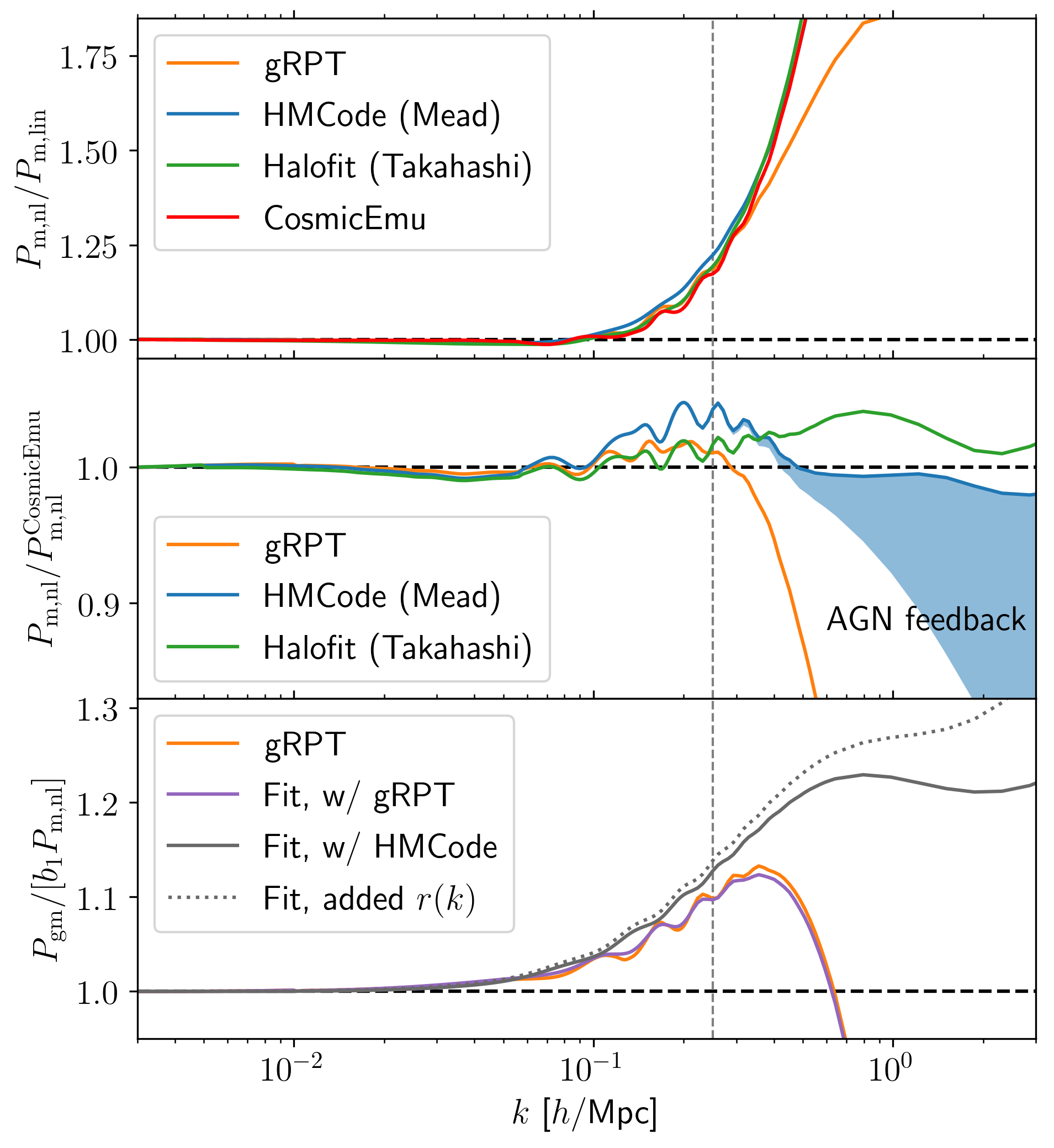}
	\caption{Comparison of 3D power spectra, computed at $z=0.38$ and the parameters listed in Table$\,$\ref{tab:fiducialpars}. \textit{Top}: Non-linear matter power spectra relative to the linear matter power spectrum, for the gRPT perturbative model, the \citet{mead15} \textsc{HMCode}, the \citet{takahashi12} Halofit prescription, and the CosmicEmu emulated power spectrum \citep{heitmann14}. \textit{Centre}: Non-linear matter power spectra relative to the CosmicEmu model. The blue shaded region covers the power spectrum range within the prior range of our AGN feedback description. \textit{Bottom}: Full galaxy-matter power spectrum $P_{\rm gm}$ relative to the \citet{mead15} power spectrum multiplied by the linear galaxy bias, shown for the gRPT perturbation theory model (orange) and the fit formula of Eq.$\,$(\ref{eq:pgm}), using gRPT (purple) or \citet{mead15} (grey) for the non-linear matter power spectrum term. The dotted grey curve includes an additional $r(k)$ term obtained from a semi-analytic model of a galaxy sample similar to the one used in our analysis. The vertical grey line indicates the smallest scales used in the galaxy clustering modelling.}
	\label{fig:3dps}
\end{figure}

We compare the two non-linear power spectrum models with the alternative dark-matter only Halofit prescription by \citet{takahashi12}, as well as the CosmicEmu emulated power spectrum by \citet{heitmann14} in the top two panels of Fig.~\ref{fig:3dps}. The comparison is at $z=0.38$, which is the mean redshift of one of our two \lq lens\rq\ galaxy samples. We treat CosmicEmu as the truth in this case although the emulator itself is limited to $2\,\%$ accuracy on relevant non-linear scales. All non-linear recipes capture the deviation from the linear matter power spectrum well, with the perturbative model starting to break down beyond $k=0.25\,h\,{\rm Mpc}^{-1}$, in line with the S17 results. Our default \citet{mead15} model is systematically higher than CosmicEmu in the mildly non-linear regime\footnote{We note that this excess is parameter-dependent. For different cosmological parameter values deviations can also be negative, but they tend to be at a similar level and on the same scales.} around $k \approx 0.2\,h\,{\rm Mpc}^{-1}$, while Halofit shows a similar trend around $k=1\,h\,{\rm Mpc}^{-1}$. Systematic trends at high $k$ are to some degree absorbed by the freedom in the baryon feedback amplitude, with the allowed prior range shown as the shaded region in Fig.~\ref{fig:3dps}. However, this would not be the case for the excess power of the \citet{takahashi12} Halofit model as we limit the feedback model to only account for suppression of power.

Recently, \citet{knabenhans19} introduced the EuclidEmulator and demonstrated it to be in excellent agreement with CosmicEmu to $\sim \!1\,\%$ at $z \lesssim 1$ and over the majority of relevant $k$-scales, despite using completely independent simulations and emulation methods\footnote{We cannot use either emulator in the main analysis because they do not include baryonic feedback. Furthermore, CosmicEmu is limited to a maximum redshift of $z=2$, while the real-data redshift distributions extend to higher redshifts. The EuclidEmulator only covers a restricted parameter space, e.g. it does not model our fiducial value of $\sigma_8$, so that it cannot be added to Fig.~\ref{fig:3dps}.}. The emulators agree that the \citet{takahashi12} Halofit model consistently over-predicts power over a broad range of redshifts and cosmological parameters by typically $3\,\%$ at $k \sim 1\,h\,{\rm Mpc}^{-1}$ \citep{heitmann14,mead15,knabenhans19}. We can therefore reject Halofit as an accurate description of matter power at highly non-linear scales.

Nevertheless, it is relevant to include Halofit in comparisons as it is still widely used, for instance in the recent DES Year 1 \citep{troxel18} and HSC \citep{hikage19} cosmic shear analyses. However, \citet{troxel18} discarded highly non-linear scales that are affected by baryonic feedback at $2\,\%$ or more, which also suppresses potential inaccuracies in matter power spectrum modelling to negligible levels. \citet{hikage19} repeated their analysis with \textsc{HMCode} in a similar setup to ours and found a tolerable $0.2\sigma$ shift in the parameter combination $\sigma_8\,(\Omega_{\rm m}/0.3)^{0.45}$.

To assess the impact of modelling inaccuracies in the matter power spectrum on KiDS-1000, we perform a mock likelihood analysis of the joint clustering and weak lensing probes with our default \textsc{HMCode} model (for details see Sect.~\ref{sec:parameter_constraints}), constructing a noiseless mock data vector with (1) CosmicEmu and (2) \citet{takahashi12} Halofit as the non-perturbative non-linear matter power spectrum. This results in a shift of the best-fit $S_8$ from the input value by $0.1\sigma$ for CosmicEmu, with the best-fit $A_{\rm bary}$ very close to the dark-matter only value of 3.13. This demonstrates that our \textsc{HMCode} model accurately captures the state-of-the-art emulator predictions. The corresponding $S_8$ shift in the case of Halofit is $0.3\sigma$; if we had used this as our model, it would have rendered the non-linear matter power spectrum prescription the dominant systematic in KiDS cosmological modelling and inference.

\subsection{Galaxy power spectra}
\label{sec:pgm}

The galaxy power spectrum $P_{\rm gg}(k,z)$ underlying the galaxy clustering signals is adopted from S17 and based on the perturbation theory approach developed by \citet{chan12}. It is of the form
\eq{
\label{eq:pgg}
P_{\rm gg}(k,z) = \sum_{\alpha,\beta} \alpha\, \beta\, P_{\alpha
  \beta}(k,z) + b_1 \gamma_3^-\, P_{b_1 \gamma_3^-}(k,z) + P_{\rm noise}(k,z)\;,
}
with $\alpha,\beta \in \bb{b_1, b_2, \gamma_2}$. We have introduced the linear and quadratic bias parameters $b_1$ and $b_2$, and the non-local bias parameters $ \gamma_2$ and $\gamma_3^-$. The different power spectrum terms on the right-hand side of Eq.~(\ref{eq:pgg}) are convolutions of the linear power spectrum and given explicitly in Appendix A of S17, with the identification $P_{b_1 b_1} \equiv P_{\rm m, nl-pt}$.

In the same formalism the cross-power spectrum between the galaxy and matter distribution required for GGL is given by
\eqa{
\label{eq:pgm_pt}
P_{\rm gm, pt}(k,z) &= b_1 P_{\rm m, nl-pt}(k,z) + b_2 P_{b_2}(k,z) + \gamma_2 P_{\gamma_2}(k,z) \\ \nn
& \hspace*{0.3cm} +  \gamma_3^- P_{ \gamma_3^-}(k,z)\;,
}
where the right-hand side power spectrum terms are again provided in S17. To extend the model to smaller scales while ensuring consistency with the non-linear model for the cosmic shear signal, we undertake the following empirical modifications to this model: first, we replace the perturbative non-linear power spectrum in the linear bias term with the halo model-based version, $P_{\rm m, nl}$. We also add an additional function $r(k)$ that will be discussed further below.

Secondly, we re-formulate the remaining terms relative to the square of the linear matter power spectrum, which results in smoothly varying functions ${\cal F}$ that are robustly extrapolated into the deeply non-linear regime. To a very good approximation, these functions can also be assumed redshift-independent (an approach also adopted by S17). This leads to the following model:
\eqa{
\label{eq:pgm}
P_{\rm gm}(k,z) &= b_1 r(k) P_{\rm m, nl}(k,z) + \left\{ b_2\, {\cal F}_{b_2}(k)
  - \gamma_2\, {\cal F}_{\gamma_2}(k) \right.\\ \nn
& \hspace*{0.3cm}  \left. -\, \gamma_3^-\, {\cal F}_{\gamma_3^-}(k)
 \right\} P_{\rm m, lin}^{2}(k,z)\;,
}
with the definition
\eq{
\label{eq:pgm_fit}
{\cal F}_\alpha(k) := \frac{P_\alpha(k,z_{\rm ref})}{P_{\rm m, lin}^{2}(k,z_{\rm ref})} = \frac{h^3}{{\rm Mpc}^3} \exp \bc{  \sum_{i=0}^2 f_{\alpha,i}\, \bb{ \ln \br{ k \frac{\rm Mpc}{h} } }^i }\;
}
for each higher-order bias term, that is $\alpha \in [b_2, \gamma_2, \gamma_3^-]$. We express the logarithm of the functions ${\cal F}$ as second-order polynomials in $\ln k$ and fit the coefficients $f_{\alpha,i}$ to the perturbation theory model within its range of validity. We find that including quadratic terms in Eq.~(\ref{eq:pgm_fit}) is sufficient to capture the deviations in scale dependence from $P_{\rm m, lin}^{2}$. Since the non-local bias terms in Eq.~(\ref{eq:pgm_pt}) are consistently negative, we extract an overall minus sign and model the absolute values in ${\cal F}_{\gamma_2}$ and ${\cal F}_{\gamma_3^-}$.

Most of the cosmology dependence of the power spectrum terms in Eq.~(\ref{eq:pgm}) is captured by the linear matter power spectrum. Due to the convolutions in these terms there is some mode mixing which induces sensitivity to cosmological parameters that modify the shape of the power spectrum. We find it is sufficient to account for $n_{\rm s}$, the power-law slope of the primordial power spectrum, and the combination $\omega_{\rm c}/h$, where $\omega_{\rm c} = \Omega_{\rm c} h^2$ is the physical cold dark matter density. The latter governs the position of the peak in the matter power spectrum (strictly speaking, $\Omega_{\rm m} h$ should capture this dependence even better, but we find using the baseline parameter $\omega_{\rm c}$ sufficient). Again, a polynomial including terms up to second order is sufficient to capture the dependence, yielding
\eq{
\label{eq:pgm_fitterms}
f_{\alpha,i} := \sum_{m,n=0}^{m+n \leq 2} g_{\alpha,i}^{mn} \left( \frac{\omega_{\rm c}}{h}
\right)^m n_{\rm s}^n\;.
}
The complete set of best-fit coefficients $g_{\alpha,i}^{mn}$ is provided in Table~\ref{tab:pgm_fitcoeff}. We note that the ansatz of Eq.~(\ref{eq:pgm}) has the additional benefit of a speed-up over the perturbative calculations by several orders of magnitude in some parts of parameter space, removing a major bottleneck in the weak lensing likelihood evaluations.

Non-linear galaxy bias and stochasticity, mostly driven by satellite galaxies, both cause a de-correlation between the matter and galaxy density fields \citep{dekel99,cacciato12}. This can be expressed in the form of deviations of the correlation coefficient $r(k,z) = P_{\rm gm}(k,z) / \sqrt{ P_{\rm gg}(k,z) P_{\rm m}(k,z) }$ from unity. For $k < 0.25\,h\,{\rm Mpc}^{-1}$, gRPT predicts $r$ to be very close to 1, in good agreement with recent observational and simulation studies (\citealp{blake11,simon18,dvornik18}; see also \citealp{asgari20b}). However, these works also see a more pronounced deviation from $r=1$ beginning on scales just below the maximum $k$ that can be modelled perturbatively. Therefore, we have considered including an explicit function $r(k)$ into the first term of Eq.~(\ref{eq:pgm}) to enable additional flexibility in galaxy biasing on scales beyond where perturbation theory is valid. For illustration we adopt the functional form presented in \citet{asgari20b}, which was derived from semi-analytic modelling in N-body simulations for a BOSS CMASS-like sample (as originally developed in \citealp{simon18}).

The bottom panel of Fig.~\ref{fig:3dps} illustrates the non-linear bias modelling in $P_{\rm gm}$. gRPT predicts an increase in power of up to $10\,\%$ relative to linear galaxy bias alone at $k =0.25\,h\,{\rm Mpc}^{-1}$. Its scale dependence is accurately reproduced by the fit formula of Eq.~(\ref{eq:pgm}) when using the gRPT non-linear matter power spectrum for the $b_1$ term. Switching to the \citet{mead15} matter power spectrum mildly increases power with respect to gRPT where the latter is valid, in line with the small excess seen in $P_{\rm m, nl}$. The non-perturbative matter power spectrum smoothly extends the non-linear biasing trend to smaller scales, with the excess power levelling off above $k \approx 0.5\,h\,{\rm Mpc}^{-1}$. Incorporating the \citet{asgari20b} expression for $r(k)$ enhances power significantly further beyond those scales.

Since realistic simulations to highly non-linear scales of both the lens and source galaxy samples are currently not available to us for validation, we consider the approach to galaxy biasing in the highly non-linear regime outlined here as currently too speculative to adopt in our cosmological analysis. Therefore, we set $r \equiv 1$ and limit the GGL signals to scales that only have significant contributions from $k < 0.3\,h\,{\rm Mpc}^{-1}$, where the perturbative approach holds and our empirical modifications are minor. While we thus discard high signal-to-noise GGL measurements, the impact on the overall constraining power of our data sets is minimal, largely because we make use of clustering signals measured over a much larger survey area than GGL (see Sect.~\ref{sec:data}). We note however that forthcoming analyses where the clustering and GGL areas are compatible would suffer from such a conservative approach, making the development of a robust non-linear biasing model a priority. The hybrid approach outlined above, or a more comprehensive halo model of all astrophysical contributions to the large-scale structure probes involved \citep[see e.g.][]{fortuna20}, are promising Ans\"atze in this regard.

\subsection{Clustering summary statistics}
\label{sec:clustering_stats}

The translation from three-dimensional galaxy power spectrum to the clustering observable follows S17, with the key steps summarised here. As a first step, redshift-space effects are taken into account. If $\mu$ denotes the cosine of the angle between the separation of galaxy pairs and the line of sight, the redshift-space power spectrum is given by
\eqa{
\label{eq:rsd}%
P_{\rm gg, s}(k,\mu,z) &= W_\infty(f_{\rm g}\, k\, \mu) \left\{ P_{\rm gg}(k,z) + 2 f_{\rm
  g}(a)\, \mu^2 P_{\rm g \theta}(k,z) \right. \\ \nn
& ~~ \left. + f_{\rm g}(a)^2 \mu^4 P_{\theta \theta}(k,z) + \mbox{higher-order terms} \right \}\;,
}
where $P_{\rm gg}$ is taken from Eq.~(\ref{eq:pgg}) and 
\eq{
\label{eq:growthrate}
f_{\rm g} (a) = \frac{\dd \ln D(a)}{\dd \ln a}\;
}
is the structure growth rate, with $D(a)$ the linear growth factor and scale factor $a=(1+z)^{-1}$. The galaxy-velocity cross-power spectrum reads
\eq{
\label{eq:pgvel}
P_{\rm g \theta }(k,z) = b_1 P_{{\rm m} \theta}(k,z) + b_2\, P_{b_2}(k,z)
  + \gamma_2\, P_{\gamma_2}(k,z) + \gamma_3^-\, P_{\gamma_3^-}(k,z)\;,
}
where $P_{{\rm m} \theta}(k,z)$ and $P_{\theta \theta}(k,z)$ in Eq.~(\ref{eq:rsd}) are the matter density contrast cross-power spectrum, and the auto-power spectrum, of the velocity divergence $\theta$, respectively\footnote{In this subsection $\theta$ denotes velocity divergence in accordance with the perturbation theory literature, but in the remainder of the paper we use $\theta$ as a variable for angular separation.}. These terms are also derived from gRPT, using the same prescription as for the density spectra \citep{crocce12}, and do not contain additional dependencies on our bias parameters. The remaining power spectrum terms match those in Eq.~(\ref{eq:pgm_pt}). We refer the reader to S17 for the explicit form of the higher-order terms in Eq.~(\ref{eq:rsd}). Fingers of god generated by small-scale virial motion are modelled by the term
\eq{
\label{eq:velocitydispersion}
W_\infty(x) = \frac{1}{\sqrt{1 + x^2 a_{\rm vir}^2}}\, \exp \br{ - \frac{x^2 \sigma_v^2}{1 + x^2 a_{\rm vir}^2} }\;,
}
where $\sigma_v$ is the one-dimensional linear velocity dispersion, which is calculated from $P_{\theta \theta}$ (cf. Eq.~39 in \citealp{scoccimarro04}). The quantity $a_{\rm vir}$ is a free parameter in the analysis that accounts for small-scale, non-linear contributions to the velocities and therefore deviations from Gaussianity in the velocity distribution.

The redshift-space galaxy power spectrum is transformed to correlation functions as a function of comoving lag $s$, $\mu$, and redshift $z$ as follows,
\eqa{
\label{eq:bosswedgefouriertrafo}
 \xi_{\rm gg} \br{s,\mu,z} &= \sum_{l=0}^{2} {\rm L}_{2l}(\mu) \frac{(-1)^l
   (4l+1)}{(2 \pi)^2} \int_0^\infty \dd k\, k^2\, {\rm j}_{2l}(ks) \\ \nn
& \times \int_{-1}^1
 \dd \mu_1 {\rm L}_{2l}(\mu_1) P_{\rm gg, s}(k,\mu_1,z)\;,
}
where ${\rm L}_i$ denotes the Legendre polynomial of degree $i$, and ${\rm j}_i$ is the spherical Bessel function of order $i$. The expansion in Legendre polynomials is truncated at $l=2$ as higher-order contributions are small on the scales considered\footnote{Once the Alcock-Paczynski distortions are taken into account, the final clustering wedges may have contributions from higher multipoles.}. The correlation function is then cast into three redshift-space wedges, equidistant in $\mu$. The three-dimensional galaxy pair separation and $\mu$ are calculated from the observed redshift assuming a fiducial cosmology, which leads to Alcock-Paczynski distortions \citep{alcock79}. The final prediction for the clustering observable is hence
\eq{
\label{eq:bosswedgecorr}
\xi^\prime_{{\rm gg,} i}(s^\prime,z) = \frac{1}{\mu^\prime_{i, \rm up} -
  \mu^\prime_{i, \rm lo}} \int_{\mu^\prime_{i, \rm
    lo}}^{\mu^\prime_{i, \rm
    up}} \dd \mu^\prime  \xi_{\rm gg} \br{s \bb{\mu^\prime,
    s^\prime},\mu \bb{\mu^\prime, s^\prime},z}\;
}
for wedges between $\mu_{\rm lo}$ and $\mu_{\rm up}$ and indexed by $i$, where primes denote quantities evaluated in the fiducial cosmology. The translation for $s$ and $\mu$ is carried out according to
\eq{
\label{eq:ap}
s_\perp = \frac{f_{\rm K}(\chi[z_{\rm mean}])}{f_{\rm K}^\prime(\chi[z_{\rm mean}])}\, s_\perp^\prime\;; ~~~
s_\parallel = \frac{H ^\prime(z_{\rm mean})}{H(z_{\rm mean})}\, s_\parallel^\prime\; 
}
for the perpendicular and line-of-sight components of the separation vector, respectively. We use the comoving angular diameter distance $f_{\rm K}$, the comoving (radial) distance $\chi$, and the Hubble parameter $H(z)$ evaluated at the mean redshifts $z_{\rm mean}$ of the lens galaxy samples. 

As in S17, we divide the BOSS galaxies into two redshift bins, each modelled with an individual set of galaxy bias and other non-cosmological parameters; cf. Table~\ref{tab:fiducialpars}. The correlation functions overall are modelled at the mean redshift in each bin, neglecting evolution within the bin. S17 showed this approximation to be accurate on simulations that include redshift evolution. Moreover, an analysis of the same data set with angular correlation functions measured in thin redshift shells produced fully consistent results \citep{salazar17}. Overall, a likelihood analysis of mock BOSS clustering signals derived from an N-body simulation revealed subdominant biases of the posterior maximum with respect to the input cosmology of less than $0.3\sigma$ for $S_8$ and less than $0.4\sigma$ for $\Omega_{\rm m}$ (S17; \citealp{troester19}).

\subsection{Cosmic shear summary statistics}
\label{sec:cosmicshear_stats}

Cosmic shear two-point statistics can be expressed as linear functionals of angular power spectra, which in turn are line-of-sight projections of the three-dimensional matter power spectrum. We perform tomography, that is to say we consider all unique combinations of a number of source galaxy sub-samples, denoted by indices $i$ and $j$ in the following. Statistics derived from galaxy ellipticity measurements receive two contributions, one from the gravitational lensing effect (subscript \lq G\rq) and one from intrinsic alignments (subscript \lq I\rq; referred to as IA henceforth) of galaxy ellipticities. This results in \lq observed shear\rq\ power spectra
\eq{
\label{eq:cl_cosmicshear}
C^{(ij)}_{\epsilon \epsilon}(\ell) = C^{(ij)}_{\rm GG}(\ell) +
C^{(ij)}_{\rm GI}(\ell) + C^{(ij)}_{\rm IG}(\ell) + C^{(ij)}_{\rm II}(\ell)\;,
}
where the right-hand side power spectra are given by Limber-approximated projections of the form \citep{kaiser92,loverde08}
\eq{
\label{eq:generallimber}
C^{(ij)}_{\rm ab}(\ell) = \int^{\chi_{\rm hor}}_0 \!\!\! \dd \chi\;
\frac{W^{(i)}_{\rm a} (\chi)\; W^{(j)}_{\rm b} (\chi)}{f^2_{\rm
    K}(\chi)}\; P_{\rm m, nl} \br{\frac{\ell+1/2}{f_{\rm K}(\chi)},z(\chi)}\;,
}
with ${\rm a,b} \in \bc{\rm I,G}$. The integral runs over the entire line of sight to the horizon $\chi_{\rm hor}$. The weak lensing kernel is
\eq{
\label{eq:kernel_lensing}
W^{(i)}_{\rm G} (\chi) = \frac{3 H_0^2 \Omega_{\rm m}}{2\, c^2}
\frac{f_{\rm K}(\chi)}{a(\chi)} \int_{\chi}^{\chi_{\rm hor}} \dd
\chi'\; n^{(i)}_{\rm S}(\chi')\; \frac{f_{\rm K}(\chi' - \chi)}{f_{\rm K}(\chi')}\;,
}
where $\Omega_{\rm m}$ is the total matter density parameter and $n^{(i)}_{\rm S}(\chi)$ is the probability density distribution of comoving distances of galaxies in source sample $i$, which in practice we express in terms of redshift\footnote{We employ the same symbol for the comoving distance distribution and the redshift distribution to keep notation simple. They can be distinguished by their function argument.}, that is $n_{\rm S}(z)=n_{\rm S}(\chi)\, \dd \chi / \dd z$. We choose an IA kernel that produces the so-called NLA model \citep{bridle07},
\eq{
\label{eq:kernel_IA}
W^{(i)}_{\rm I} (\chi) = - A_{\rm IA}\; \br{\frac{1+z(\chi)}{1+z_{\rm pivot}}}^{\eta_{\rm IA}}  \frac{C_1 \rho_{\rm
  cr}\, \Omega_{\rm m}}{D\br{a[\chi]}}\; n^{(i)}_{\rm S}(\chi)\;,
}
where $C_1 \rho_{\rm cr } \approx 0.0134$ is constant, and $z_{\rm pivot}$ is an arbitrary pivot which we set to 0.3 for compatibility with earlier KiDS and IA analyses \citep{joachimi11}. The IA kernel introduces two additional parameters: the dimensionless IA amplitude $A_{\rm IA}$ and additional freedom in the redshift dependence of the IA strength via $\eta_{\rm IA}$. As there is currently no evidence for significant extra redshift evolution in our source samples (see e.g. KV450), we only consider $\eta_{\rm IA}$ in extended analyses and set $\eta_{\rm IA}=0$ by default.

More flexibility in the IA model to account for the still rather limited physical understanding of galaxy alignments may seem prudent \citep{blazek19,samuroff19}, but it enables residual systematics in the source redshift distributions to disguise as a physical signal, which subsequently risks biasing constraints on primary cosmological parameters \citep{efstathiou18}. This concern would not necessarily extend to IA parameters that do not modify the overall amplitude of weak lensing two-point statistics, or the scaling of their amplitude with redshift (such as additional freedom on highly non-linear scales). However, it is this property that also makes these parameters less relevant for the cosmological inference; their absence would cause a poor goodness of fit rather than biases in the cosmological parameters if the data preferred the more complex IA model.

There is growing evidence that IA model parameters pick up residual discrepancies in the scaling of cosmic shear signals with source redshifts, rather than solely the IA contribution: IA amplitudes in both the most recent KiDS and DES analyses tend to be higher than the consensus of direct IA measurements predicts \citep{fortuna20}. \citet{hikage19} do not detect IA, but also see a dependence of the IA amplitude constraint on the choice of photometric redshift algorithm employed. Recently, \citet{wright20b} showed that the fiducial KV450 constraint of $A_{\rm IA} \sim 1$ is reduced to values fluctuating around zero for a conservative selection of source galaxies with more secure redshift calibration (we adopt this approach; see Sect.~\ref{sec:photo-z} for details). In conclusion, we continue to follow the philosophy of earlier KiDS analyses to employ the simplest, most predictive IA model as long as it yields a good fit to the data.

Band powers are defined as angular averages over the angular power spectra,
\eq{
\label{eq:bpdef}
{\cal C}_{{\rm E/B},l}^{(ij)} := \frac{1}{{\cal N}_l} \int_0^\infty \dd \ell\, \ell\; S_l(\ell)\; C_{\epsilon \epsilon, {\rm E/B}}^{(ij)}(\ell)\;,
}
where $S_l$ is the band power response function for an angular bin $l$. The normalisation is given by
\eq{
\label{eq:normalisation}
{\cal N}_l = \int_0^\infty \dd \ell\, \ell^{-1}\; S_l(\ell)\;,
}
which is designed such that the band power traces $\ell^2 C_{\epsilon \epsilon, {\rm E/B}}$ at the log-centre of bin $l$. Here, the spin-2 gravitational shear field has been decomposed into curl-free E-modes and divergence-free B-modes. The former carry the signal of interest, that is $C_{\epsilon \epsilon, \rm E} \equiv C_{\epsilon \epsilon}$, whereas we assume that none of our cosmological and astrophysical signals produce B-modes intrinsically, so $C_{\epsilon \epsilon, \rm B} \equiv 0$. For both gravitational lensing and the II-term prediction in the NLA model, B-modes are only generated at second order and therefore safely negligible in KiDS \citep{bernardeau97,schneider98,hirata04}, as are other potential sources of B-modes like source redshift clustering \citep{schneider02b} or exotic physics \citep[e.g.][]{thomas17}. B-mode statistics are therefore used as null tests for residual systematics in the data.

The band powers are derived from the two-point correlation functions, which are related to the angular power spectra via Hankel transformations of the form
\eqa{
\label{eq:hankelshear}
\xi_+^{(ij)}(\theta) &= \int_0^\infty \frac{\dd \ell \ell }{2 \pi} \; \bc{ C_{\epsilon \epsilon, {\rm E}}^{(ij)}(\ell) + C_{\epsilon \epsilon, {\rm B}}^{(ij)}(\ell)} \; {\rm J}_0(\ell \theta)\;; \\ \nn
\xi_-^{(ij)}(\theta) &= \int_0^\infty \frac{\dd \ell \ell }{2 \pi} \; \bc{ C_{\epsilon \epsilon, {\rm E}}^{(ij)}(\ell) - C_{\epsilon \epsilon, {\rm B}}^{(ij)}(\ell)} \; {\rm J}_4(\ell \theta)\;,
}
where ${\rm J}_\mu$ is a cylindrical Bessel function of the first kind of order $\mu$. Due to the orthogonality of the Bessel functions, these equations are readily inverted to express power spectra as a functional of $\xi_\pm$. Inserting the result into Eq.~(\ref{eq:bpdef}) yields
\eq{
\label{eq:bpestimate_ideal}
{\cal C}_{{\rm E/B},l}^{(ij)} =  \frac{\pi}{{\cal N}_l}\; \int_0^\infty \dd \theta\, \theta\; T(\theta) \bc { \xi_+^{(ij)}(\theta)\; g_+^l(\theta) \pm \xi_-^{(ij)}(\theta)\; g_-^l(\theta)  }\;,
}
where we defined the kernels
\eqa{
\label{eq:gplusminus}
g_\pm^l(\theta) &=  \int_0^\infty \dd \ell\, \ell\, S_l(\ell)\; {\rm J}_{0/4}(\ell \theta)\;.
}
Formally, Eq.~(\ref{eq:bpestimate_ideal}) is exact for the weight function $T \equiv 1$. However, in practice the correlation functions can only be measured over a finite range of angular separations, so $T$ has to vanish outside this range. To avoid ringing in the band power kernel due to this cut-off, we apodise the angular scales entering the band power computation in the form of a Hann window,
\eq{
\label{eq:apodisation}
T(\theta) = \left\{ \begin{aligned}  0\,; &\; x < x_{\rm lo} - \frac{\Delta_x}{2} \\ \cos^2 \bb{\frac{\pi}{2} \frac{x- (x_{\rm lo}+\Delta_x/2)}{\Delta_x} }\,;  &\; x_{\rm lo} -\frac{\Delta_x}{2} \leq x <  x_{\rm lo} +\frac{\Delta_x}{2} \\ 1 \,; &\;  x_{\rm lo} +\frac{\Delta_x}{2} \leq x <  x_{\rm up} -\frac{\Delta_x}{2} \\  \cos^2 \bb{\frac{\pi}{2} \frac{x-(x_{\rm up} - \Delta_x/2)}{\Delta_x} } \,; &\;  x_{\rm up} -\frac{\Delta_x}{2}  \leq x <  x_{\rm up} +\frac{\Delta_x}{2}\\ 0 \,; &\; x \geq  x_{\rm up} +\frac{\Delta_x}{2}\;. \end{aligned} \right. 
}
where $x = \log \theta$ and $\Delta_x$ is the log-width of the apodisation, which is a free parameter. The apodisation at the lower and upper angular limits is centred on the scales $x_{\rm lo} = \log \theta_{\rm lo}$ and $x_{\rm up} = \log \theta_{\rm up}$, respectively. The apodisation suppresses out-of-band sensitivity in Fourier space at the price of somewhat extending the range of angular scales used in the input correlation functions at both ends.

In this work we opt for a band power response function that is a top hat between $\ell_{{\rm lo},l}$ and $\ell_{{\rm up},l}$ (for other choices, see \citealp{becker16b}). The normalisation is then given by ${\cal N}_l = \ln \br{\ell_{{\rm up},i}/\ell_{{\rm lo},i}}$. The functions $g_\pm$ can be expressed in closed form as \citep{schneider02,uitert18}
\eqa{
\label{eq:bp_kernel_cosmicshear}
g_+^l(\theta) &= \frac{1}{\theta^2} \bb{ \theta \ell_{{\rm up},l}\;
  {\rm J}_1(\theta \ell_{{\rm up},l}) - \theta \ell_{{\rm lo},l}\;   {\rm J}_1(\theta \ell_{{\rm lo},l}) }\;; \\ \nn
g_-^l(\theta) &= \frac{1}{\theta^2} \bb{ {\cal G}_-(\theta \ell_{{\rm
      up},l})  - {\cal G}_-(\theta \ell_{{\rm lo},l}) }\;,
}
with
\eq{
{\cal G}_-(x) = \br{x - \frac{8}{x}} {\rm J}_1(x) - 8 {\rm J}_2(x)\;.
}

While Eq.~(\ref{eq:bpestimate_ideal}) forms the basis of our estimator, it is convenient for the modelling to link the band powers directly to the angular power spectra via 
\eqa{
\label{eq:bp_cosmicshear}
{\cal C}_{{\rm E},l}^{(ij)} &=  \frac{1}{2 {\cal N}_l} \int_0^\infty \dd
\ell\, \ell \bc{ W^l_{\rm EE}(\ell)\; C^{(ij)}_{\epsilon \epsilon, \rm
    E}(\ell) + W^l_{\rm EB}(\ell)\; C^{(ij)}_{\epsilon \epsilon, \rm B}(\ell) }\;; \\ \nn
{\cal C}_{{\rm B},l}^{(ij)} &=  \frac{1}{2 {\cal N}_l} \int_0^\infty \dd
\ell\, \ell \bc{ W^l_{\rm BE}(\ell)\; C^{(ij)}_{\epsilon \epsilon, \rm
    E}(\ell) + W^l_{\rm BB}(\ell)\; C^{(ij)}_{\epsilon \epsilon, \rm B}(\ell) }\;,
}
with kernels given by 
\eqa{
\label{eq:kernelshear}
\nn
W^l_{\rm EE}(\ell) &= W^l_{\rm BB}(\ell) = \!\! \int_0^\infty \!\! \dd \theta\, \theta\; T(\theta) \bc{ {\rm J}_0(\ell \theta)\; g_+^l(\theta) + {\rm J}_4(\ell \theta)\; g_-^l(\theta) }; \\
W^l_{\rm EB}(\ell) &= W^l_{\rm BE}(\ell) = \!\! \int_0^\infty \!\! \dd \theta\, \theta\; T(\theta) \bc{ {\rm J}_0(\ell \theta)\; g_+^l(\theta) - {\rm J}_4(\ell \theta)\; g_-^l(\theta) }.
}

\begin{figure}
	\includegraphics[width=\columnwidth]{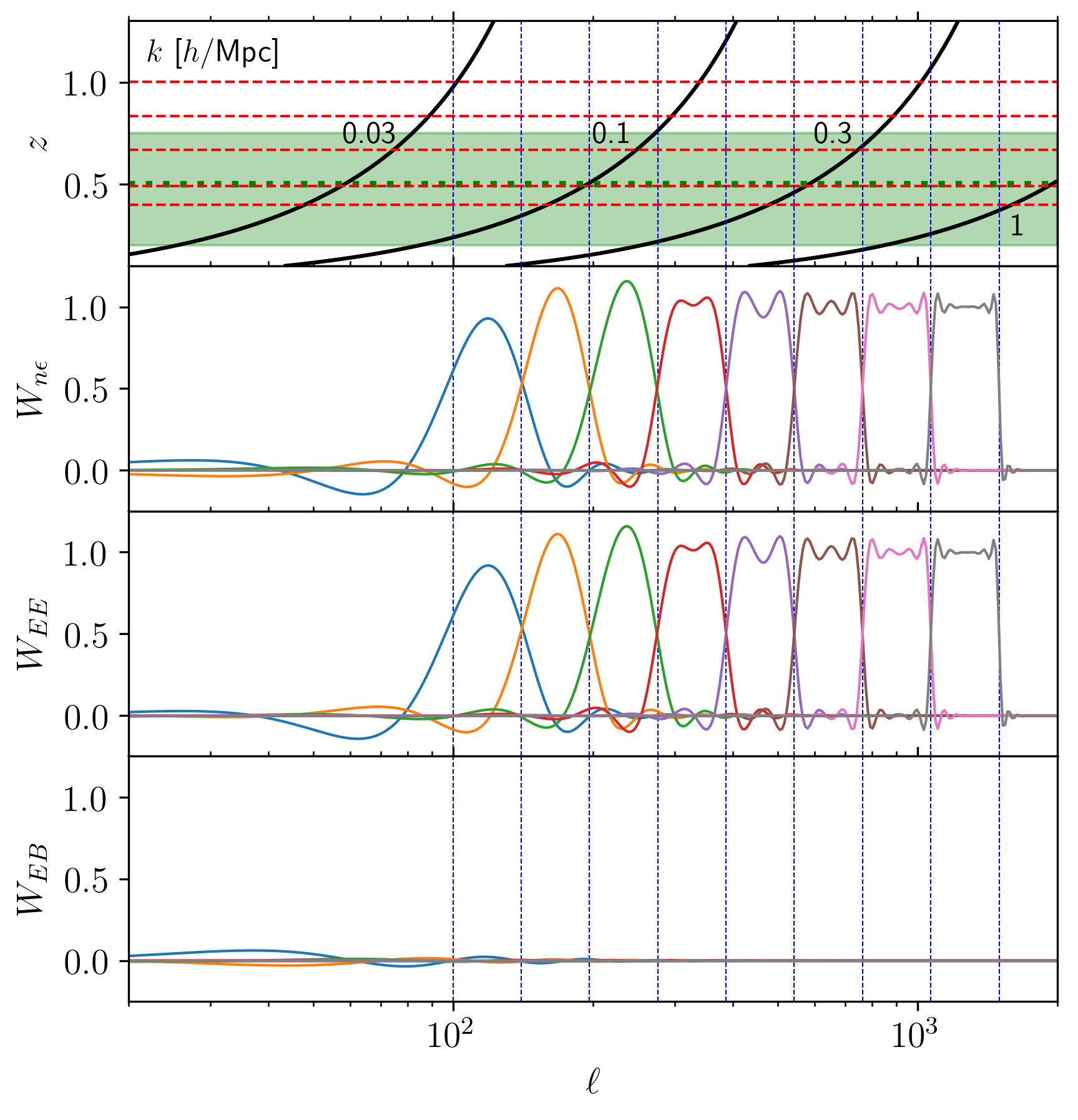}
	\caption{Bandpower Fourier-space filters for 8 bands logarithmically spaced $\ell \in [100,\,1500]$, as indicated by the vertical lines. The second to fourth panels from the top show the galaxy-galaxy lensing, cosmic shear, and EB-mode mixing kernels; see Eqs. (\ref{eq:kernelshear}), and (\ref{eq:kernelggl}). The top panel shows redshift $z$ as a function of $\ell$ for a selection of wavenumbers $k$ given in units of $h\,{\rm Mpc}^{-1}$. Red horizontal lines mark the mean redshifts of the source redshift bins, and the green band shows the redshift range of the two lens samples with a split at $z=0.5$. The apodisation width is chosen as $\Delta_x=0.5$.}
	\label{fig:bandpower_filters}
\end{figure}

\begin{figure}
	\includegraphics[width=\columnwidth]{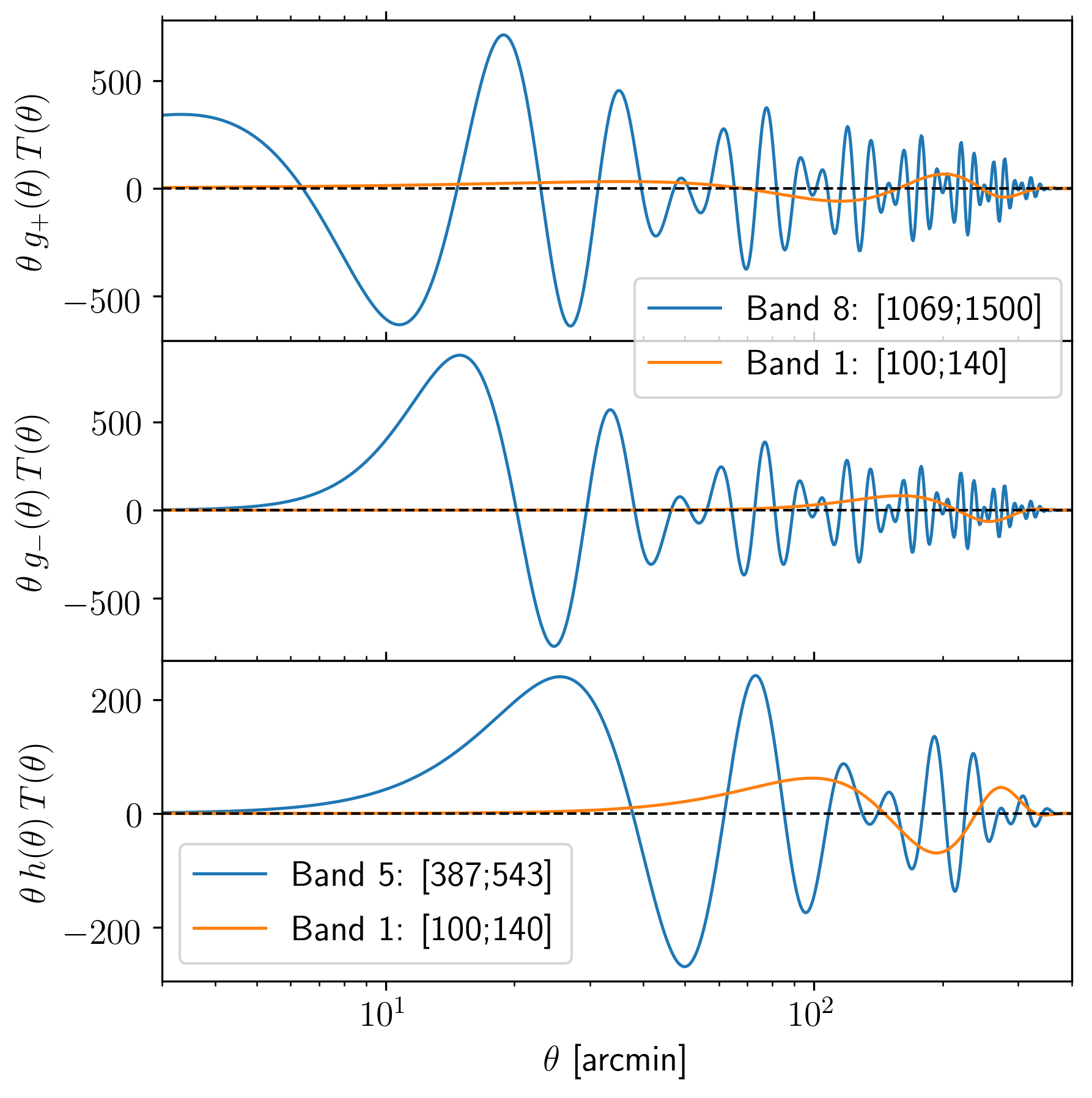}
	\caption{Bandpower real-space filters for different bands with the angular frequency range given in square brackets. For each case we show the lowest and highest band used in the fiducial analysis setup. The default correlation function binning has been assumed, with apodisation centred on the angular range boundaries of $0.5\,{\rm arcmin}$ and $300\,{\rm arcmin}$, using a log-width of $\Delta_x=0.5$. \textit{Top}: filter for $\xi_+$; \textit{centre}: filter for $\xi_-$; \textit{bottom}: filter for $\gamma_{\rm t}$; cf. Eqs.~(\ref{eq:apodisation}), (\ref{eq:bp_kernel_cosmicshear}), and (\ref{eq:bp_kernel_ggl}).}
	\label{fig:bandpower_filters_real}
\end{figure}

The band power filters in Fourier space (Eq.~\ref{eq:kernelshear}) and in configuration space ($T(\theta)\, g_\pm^l(\theta)$, cf. Eq.~\ref{eq:bpestimate_ideal}) are shown in Figs.~\ref{fig:bandpower_filters} and \ref{fig:bandpower_filters_real}, respectively, for our default correlation function binning and band power definitions. In particular, we choose eight bands in the range $100 < \ell < 1500$, uniformly spaced in the log. At higher angular frequencies, where many modes are available for a given band, our band powers have a very clean selection in Fourier space, with $W^l_{\rm EE}(\ell)$ approaching a top hat to very good approximation. Equation~(\ref{eq:bp_cosmicshear}) shows that band power B-modes could in principle still be generated even if $C_{\epsilon \epsilon, \rm B} \equiv 0$, but the mode mixing via $W^l_{\rm EB}(\ell)$ is negligible.

For the low-$\ell$ bands these windows become broader while still clearly peaking in the centre of the defined band, with some out-of-band sensitivity especially at $\ell<100$ (see also \citealp{asgari15} for a detailed analysis of band power filter dependencies). These inherently linear scales are still accurately modelled; the gradual decline in the accuracy of the flat-sky and Limber approximations does not impact our analysis due to the strongly suppressed sensitivity. The corresponding real-space filters shown in Fig.~\ref{fig:bandpower_filters_real} are highly oscillatory and therefore require fine binning of the input correlation functions. For the scales used in our analysis the apodisation only significantly affects large angular scales, reaching $50\,\%$ signal suppression at $300\,{\rm arcmin}$ in this case.

Figure~\ref{fig:bandpower_signal} demonstrates that the recovery of the underlying angular power spectra by the band powers is excellent, so that we can use the band powers directly as a faithful representation of the binned angular power spectra in the data. We emphasise however that the translation from true power spectrum to band power according to Eq.~(\ref{eq:bp_cosmicshear}) and the associated mode mixing is fully taken into account in all quantitative modelling of both the signal and its covariance (in contrast to the earlier work by \citealp{uitert18}).

Figure~\ref{fig:bandpower_signal} also translates the residual uncertainties in matter power spectrum modelling discussed in Sect.~\ref{sec:matterps} to band powers. Due to the line-of-sight projection, effects at certain $k$-scales are spread out over a wider range of angular frequencies, such that all band powers are impacted to some degree. While the limited accuracy in non-linear power spectrum modelling as indicated by the difference between the \citet{takahashi12} and \citet{mead15} non-linear prescriptions is at the few per-cent level and close to the limit of tolerable inaccuracy for the current constraining power of data (see the discussion in Sect.~\ref{sec:matterps}), the potentially large suppression of power by AGN feedback warrants the inclusion of at least one nuisance parameter in the likelihood analysis.

\subsection{Galaxy-galaxy lensing summary statistics}
\label{sec:ggl_stats}

\begin{figure}
	\includegraphics[width=\columnwidth]{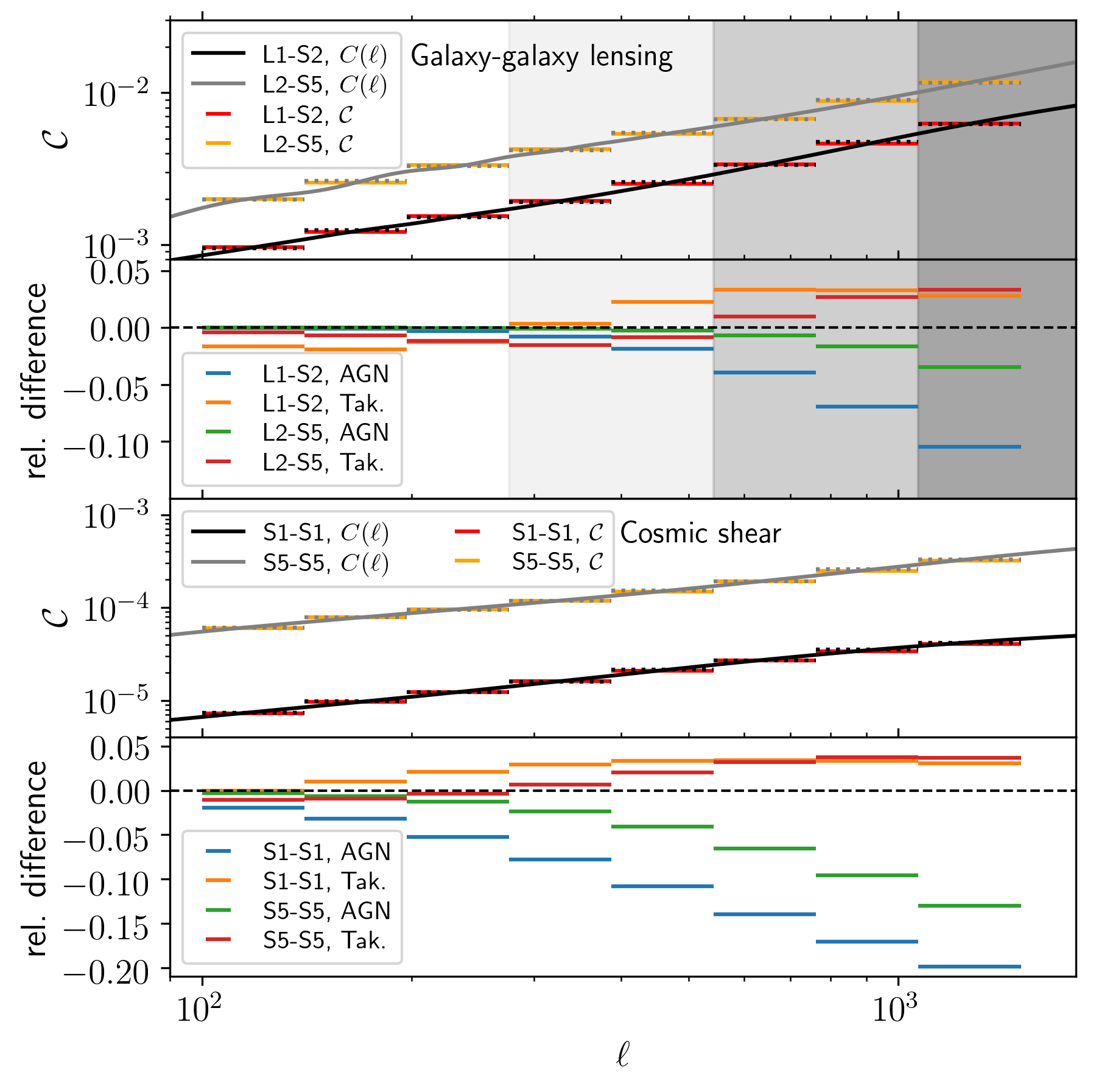}
	\caption{Bandpowers for galaxy-galaxy lensing (top panels) and cosmic shear (bottom panels) for a selection of redshift bin combinations (indicated by \lq L\rq\ and \lq S\rq\ for lens and source galaxy samples, respectively; see Sect.~\ref{sec:data} for details). Grey and black lines show the true underlying angular power spectrum; the associated dotted lines the average of these power spectra over the angular frequency band. The bottom panels show the relative deviations in the cases of maximum AGN feedback (blue/green) and using the \citet{takahashi12} Halofit version (orange/red) with respect to \citet{mead15} without AGN feedback. The grey shaded regions in the top panels show the conservative limits applied to the low- and high-redshift lens samples (excluding all and light/medium grey regions, respectively), as well as the more optimistic cut if galaxy biasing were well understood (only excluding the dark grey region).}
	\label{fig:bandpower_signal}
\end{figure}

We proceed analogously to the cosmic shear case for GGL, writing the angular power spectra as
\eq{
\label{eq:cl_ggl}
C^{(ij)}_{\rm n \epsilon}(\ell) = C^{(ij)}_{\rm gG}(\ell) +
C^{(ij)}_{\rm gI}(\ell) + C^{(ij)}_{\rm mG}(\ell)  \;,
}
where $i$ now indexes lens galaxy samples, and $j$ source samples. We include three contributions: the cross-correlation between the lens galaxy distribution and the source gravitational shear (\lq gG\rq) which contains most of the cosmologically relevant information, the alignment of source galaxies physically close to foreground lenses (\lq gI\rq), and the correlation between gravitational shear and the lensing-induced magnification bias in the lens sample (\lq mG\rq). The latter term is discussed in detail in Appendix$\,$\ref{sec:magbias}. We have ignored potential contributions from IA-magnification bias correlations, which would not exceed the per-cent level \citep{joachimi10}.

The line-of-sight projections read
\eq{
\label{eq:ggl_limber}
C^{(ij)}_{\rm g\, a}(\ell) = \int^{\chi_{\rm hor}}_0 \!\!\! \dd \chi\;
\frac{n^{(i)}_{\rm L} (\chi)\; W^{(j)}_{\rm a} (\chi)}{f^2_{\rm
    K}(\chi)}\; P_{\rm gm} \br{\frac{\ell+1/2}{f_{\rm K}(\chi)},\chi}\;
}
for ${\rm a} \in \bc{\rm I,G}$, where $n^{(i)}_{\rm L} (\chi)$ is the comoving distance distribution of lens sample $i$. The three-dimensional galaxy-matter power spectrum is given by Eq.~(\ref{eq:pgm}) and the source sample kernels by Eqs.~(\ref{eq:kernel_lensing}) and (\ref{eq:kernel_IA}). Since we project along the line of sight over broad line-of-sight galaxy distributions, we do not have to take redshift-space distortions into account. The magnification bias term is modelled as
\eq{
\label{eq:ggl_magbias}
C^{(ij)}_{\rm mG}(\ell) = 2 \br{\alpha_{\rm mag}^{(i)} - 1}\, C^{(ij)}_{\rm GG}(\ell) \;,
}
where $\alpha_{\rm mag}^{(i)}$ is usually understood as the power-law slope at the faint end of the luminosity function of a flux-limited sample $i$ \citep{bartelmann01}. However, the lens samples that we employ have a complex selection function which deviates substantially from a simple flux limit in a single band. As detailed in Appendix$\,$\ref{sec:magbias}, we have developed and validated an approach to measure an \emph{effective} luminosity function slope $\alpha_{\rm mag}$, which can be employed in the standard formalism.

Assuming a top-hat band power response as in the cosmic shear case, GGL band powers can be expressed in terms of the tangential shear correlation function,
\eq{
\label{eq:bpestimate_ggl_ideal}
{\cal C}^{(ij)}_{{\rm n\epsilon},l} =  \frac{2 \pi}{{\cal N}_l}\; \int_0^\infty \dd \theta\, \theta\; \ba{\gamma_{\rm t}}^{(ij)}(\theta)\; T(\theta)\; h^l(\theta)\;,
}
with
\eq{
\label{eq:hankelggl}
\ba{\gamma_{\rm t}}^{(ij)}(\theta) = \int_0^\infty \frac{\dd \ell \ell }{2 \pi} \; C^{(ij)}_{\rm n\epsilon}(\ell)\; {\rm J}_2(\ell \theta)\;.
}
Likewise, they can be directly calculated from the angular power spectrum,
\eq{
\label{eq:bpobsggl}
{\cal C}^{(ij)}_{{\rm n\epsilon},l} = \frac{1}{{\cal N}_l}
\int_0^\infty \dd \ell\, \ell\; W^l_{\rm n \epsilon}(\ell)\; C^{(ij)}_{\rm n\epsilon}(\ell)\;,
}
with the kernel
\eq{
\label{eq:kernelggl}
W^l_{\rm n\epsilon}(\ell) = \int_{\theta_{\rm min}}^{\theta_{\rm max}} \dd \theta\, \theta\; T(\theta)\; {\rm J}_2(\ell \theta)\; h^l(\theta)\;.
}
Here, we have followed \citet{uitert18} in defining the function
\eqa{
\label{eq:bp_kernel_ggl}
h^l(\theta) &= - \frac{1}{\theta^2} \left[ \theta \ell_{{\rm up},l}
  {\rm J}_1(\theta \ell_{{\rm up},l}) - \theta \ell_{{\rm lo},l}
  {\rm J}_1(\theta \ell_{{\rm lo},l}) \right.\\ \nn
&~~~~~~ + \left. 2 {\rm J}_0(\theta \ell_{{\rm up},l}) - 2 {\rm J}_0(\theta \ell_{{\rm lo},l}) \right].
}
Figure~\ref{fig:bandpower_filters} also displays the $W^l_{\rm n\epsilon}$ kernel whose form is very similar to the mode-preserving kernel (first equality of Eq.~\ref{eq:kernelshear}) for cosmic shear. The top panel illustrates that wavenumbers above $k=1\,h\,{\rm Mpc}^{-1}$ do not contribute significantly to the band power signals except for Band 8. We therefore study an optimistic scenario that includes Bands 1-7 ($\ell < 1069$) for GGL in the inference. However, by default we restrict the analysis to Bands 1-3 ($\ell < 276$) in the low-redshift lens bin (L1) and Bands 1-5 ($\ell < 543$) in the high-redshift bin (L2), which strongly suppresses all highly non-linear scales beyond $k=0.3\,h\,{\rm Mpc}^{-1}$ (cf. the grey shading in Fig.~\ref{fig:bandpower_signal}). For completeness the GGL real-space filter is shown in the bottom panel of Fig.~\ref{fig:bandpower_filters_real}. The conclusions from Fig.~\ref{fig:bandpower_signal} are analogous to the cosmic shear case: excellent recovery of the underlying power spectrum by the band powers and similar impact of residual matter power spectrum uncertainties on the bands.

\section{Data and measurements}
\label{sec:data}

\begin{table}
\caption{Effective survey areas and sky overlap. The overlap is also illustrated in Fig. \ref{fig:Survey_footprint}.}              
\label{tab:surveyareas}      
\centering                                      
\begin{tabular}{llll}          
\hline\hline                        
 & KiDS-1000 & BOSS & 2dFLenS \\    
\hline                                   
survey area & 773.3$^a$ & 9329.0$^b$ & 510.8$^c$ \\
KiDS overlap & - & 319.5 & 341.9 \\
\hline
\end{tabular}
\tablefoot{All survey areas given in ${\rm deg}^2$. Areas involving BOSS are extracted from a \textsc{Healpix} map with $N_{\rm side}=2048$ (due to the limited size of the random catalogue used to construct the footprint); all other areas are obtained from $N_{\rm side}=4096$ maps. The KiDS overlap is limited to full nine-band KiDS+VIKING coverage. $a)$ Unmasked area with nine-band coverage. $b)$ From \citet{alam17}. $c)$ Using only survey area adjacent to KiDS-S.}
\end{table}

\begin{figure*}
	\centering
	\includegraphics[width=17cm]{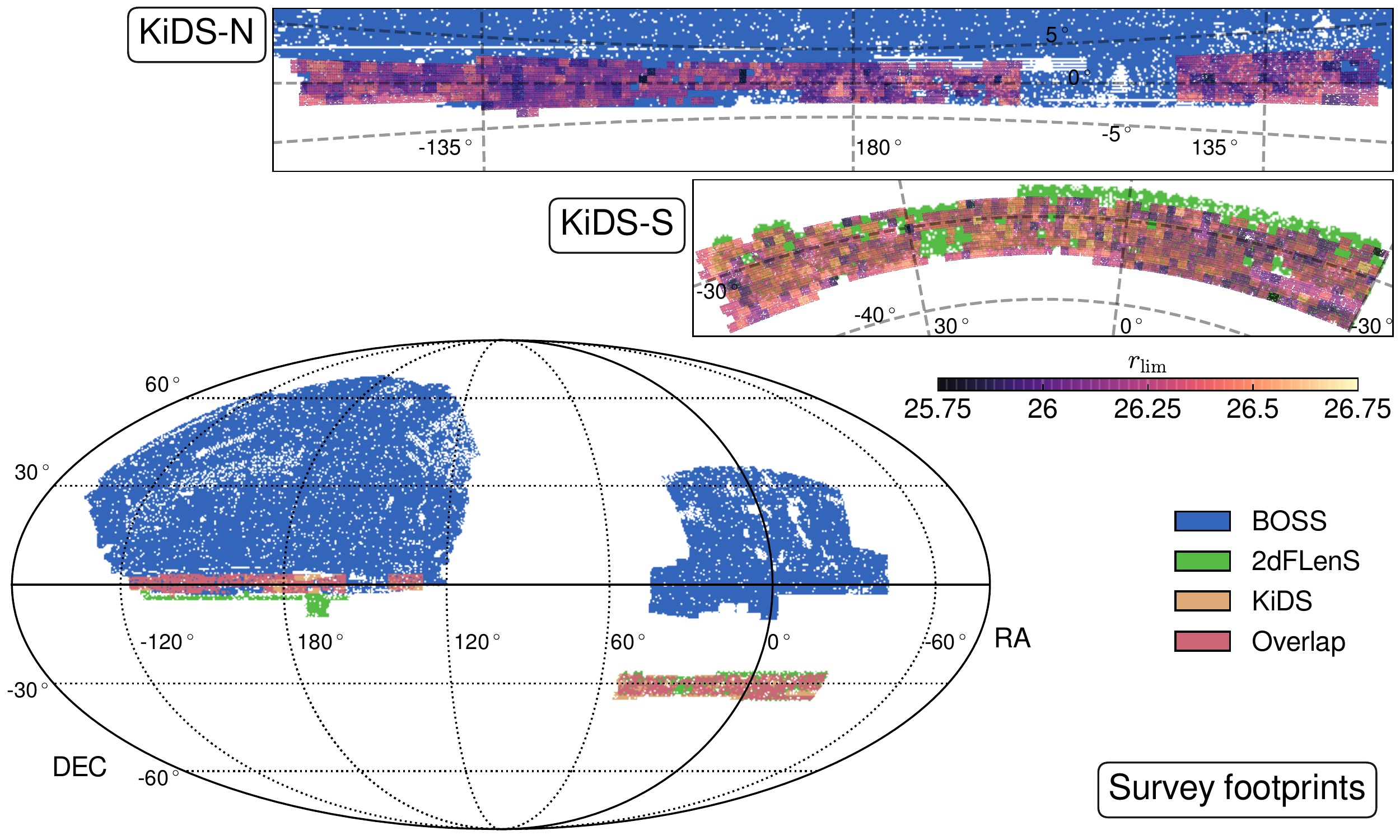}
	\caption{\textit{Bottom}: Survey footprints of three data sets used in this analysis: BOSS (blue), 2dFLenS (green), and KiDS (orange). Overlapping regions of KiDS with either BOSS or 2dFLenS are shown in pink. \textit{Top}: Cut-outs of the KiDS areas showing the variable depth patterns of our source samples as indicated by the $r$-band magnitude limit, $r_\mathrm{lim}$ (at $1\sigma$ for adaptive aperture size; see \citealp{kuijken19} for details).}
	\label{fig:Survey_footprint}
\end{figure*}

In this section we summarise the characteristics of the surveys involved and provide an overview of the data processing up to the level of the summary statistics that enter the likelihood analysis.  Figure~\ref{fig:Survey_footprint} shows the footprints of the KiDS, BOSS, and 2dFLenS surveys. Table~\ref{tab:surveyareas} provides the corresponding survey areas and the respective overlaps.

\subsection{BOSS and 2dFLenS}

Our analysis relies on two spectroscopic galaxy surveys, the Sloan Digital Sky Survey (SDSS)-III BOSS \citep{eisenstein11} and the 2dFLenS \citep{blake16} surveys, which cover disjoint areas of sky and were conducted from the northern and southern hemispheres, respectively. The clustering measurements used in our analysis were made over the full BOSS area but do not employ 2dFLenS, whereas the GGL measurements use lenses from both surveys with roughly the same contribution in terms of sky area to maximise the overlap with KiDS.

\begin{figure}
	\includegraphics[width=\columnwidth]{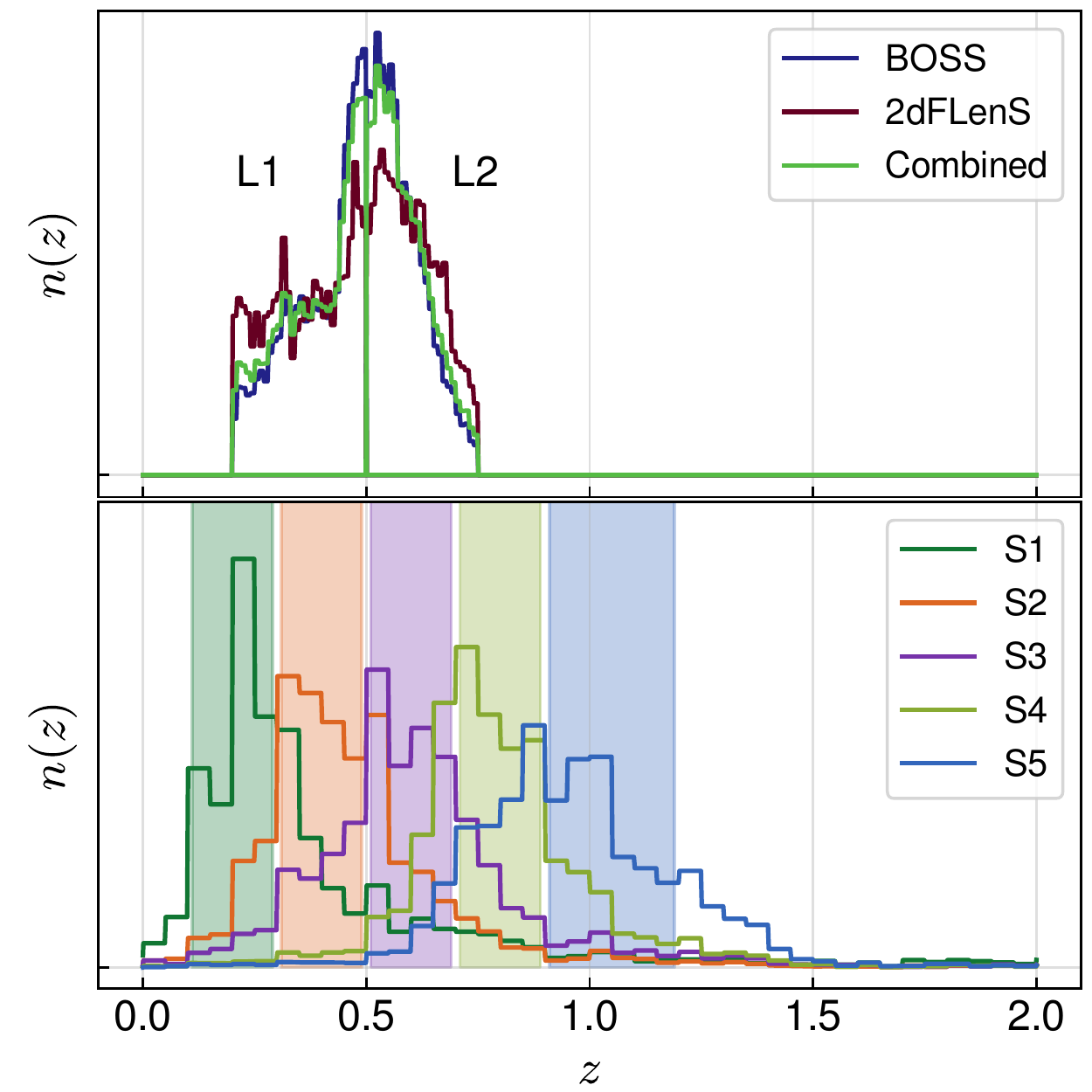}
	\caption{Redshift distributions of the lens (upper panel) and source (lower panel) galaxy samples. Lens galaxies are split at $z=0.5$ into two redshift bins for both the BOSS and 2dFLenS surveys. Modelling is carried out with a weighted combination of their redshift distributions (in green). Lens distributions are constructed for galaxies in the KiDS overlap regions, respectively. Source galaxies are split into five tomographic bins based on their best photometric redshift estimates, as indicated by the shaded bands. The source redshift distributions are KiDS-1000-like and built from KV450 data.}
	\label{fig:nOfZ}
\end{figure}

We use the same galaxy samples as S17 who based their analysis on the final data release of BOSS, DR12 \citep{alam15}, which combined the two main BOSS galaxy samples, LOWZ and CMASS, to obtain contiguous redshift coverage between $z=0.2$ and $z=0.75$\footnote{Data access: \texttt{data.sdss.org/sas/dr12/boss/lss}}; see Fig.~\ref{fig:nOfZ}, upper panel. As in the direct measurements of baryon acoustic oscillations and redshift-space distortions in the final BOSS data, S17 incorporated an additional $\sim 1000\,{\rm deg}^2$ of LOWZ in the Northern Galactic Cap with modified target selection that yielded a lower number density of successful redshift measurements, bringing the effective survey area of both the LOWZ and CMASS samples to ca. $9300\,{\rm deg}^2$ \citep{reid16}. The LOWZ and CMASS samples are dominated by massive early-type galaxies, but unlike other sample selections of luminous red galaxies (LRGs), they contain a significant fraction of spiral galaxies as well as of order $10\,\%$ satellites \citep{white11,masters11}.

\begin{table}
\caption{Galaxy sample properties for BOSS/2dFLenS lens (\lq L\rq) samples and KiDS-1000 source (\lq S\rq) samples.}              
\label{tab:sampleprops}      
\centering                                      
\begin{tabular}{lllll}          
\hline\hline                        
Bin ID & $z$ range & mean $z$ & $n_{\rm eff}$ [${\rm arcmin}^{-2}$] & $\sigma_{\epsilon,i}$ \\    
\hline                                   
L1 & $\bb{0.2,\,0.5}$ & 0.38/0.36 & 14.5/5.8 $\!\times 10^{-3}$ & - \\
L2 & $\bb{0.5,\,0.75}$ & 0.61/0.60 & 16.6/6.1 $\!\times 10^{-3}$ & - \\
\hline
S1 & $\bb{0.1,\,0.3}$ & 0.39 & 0.85 &  0.28\\
S2 & $\bb{0.3,\,0.5}$ & 0.49 & 1.56 &  0.27\\
S3 & $\bb{0.5,\,0.7}$ & 0.67 & 2.23 &  0.28\\
S4 & $\bb{0.7,\,0.9}$ & 0.83 & 1.52 &  0.27\\
S5 & $\bb{0.9,\,1.2}$ & 1.00 & 1.39 &  0.28\\
\hline
\end{tabular}
\tablefoot{Redshift ranges for the source samples are in terms of the best photometric redshift estimate $z_{\rm B}$. The first (second) entries for lens mean redshifts and $n_{\rm eff}$ are for BOSS (2dFLenS). The galaxy number density $n_{\rm eff}$ is given in units of ${\rm arcmin}^{-2}$; $\sigma_{\epsilon,i}=\sigma_\epsilon/\sqrt{2}$ is the dispersion per ellipticity component. We note that the source sample properties were extracted from KV450 as well as blinded KiDS-1000 data but are expected to be representative of the final KiDS-1000 data. The mean redshift of bin S1 lies outside its $z_{\rm B}$ range, which is due to an extended high-redshift tail; see Fig.~\ref{fig:nOfZ}.}
\end{table}

Corrections for observational systematics in the BOSS samples \citep{reid16,ross17} are applied. Potential residual systematics would primarily affect large scales, but \citet{troester20} showed that removing these from the analysis does not impact on the cosmological constraints or the goodness of fit. The range over which the clustering model is valid was studied in detail by S17, leading to the choice of $20\,h^{-1}{\rm Mpc}$ as a conservative minimum scale. \citet{troester19} confirmed that this scale remains appropriate when analysing the data in the more constrained $\Lambda$CDM framework as opposed to the generic summary parameters employed by S17. The addition of weak lensing data could in principle further tighten the accuracy requirement. However, the mock results in Fig.~1 of \citet{troester19} show that any systematic trends would shift the posterior orthogonally to $S_8$ and thus the direction in parameter space impacted by cosmic shear. Hence, we conclude the same clustering scales as in S17 can also be employed in the joint analysis with KiDS-1000.

2dFLenS\footnote{Data access: \texttt{2dflens.swin.edu.au/data.html}} was conducted at the Anglo-Australian Telescope, designed to complement BOSS in southern regions of deep imaging surveys used for weak lensing studies. The survey totals $731\,{\rm deg}^2$ of which we use the largest contiguous portion that overlaps the southern KiDS field. The target selection for its LRG-like samples follows that of BOSS but relies on the VST ATLAS survey \citep{shanks15} for target imaging. The resulting galaxy samples are similar to their BOSS counterparts, but sparsely sampled at $\sim \! 40\,\%$ of BOSS's galaxy number density (see Table~\ref{tab:sampleprops} and Fig.~\ref{fig:nOfZ}).

We combine the GGL measurements in the BOSS and 2dFLenS overlaps weighted by the respective galaxy pair counts. This corresponds to inverse variance weighting in the limit that sample variance is negligible. The redshift distributions are combined consistently with that approach (see the green histogram in Fig.~\ref{fig:nOfZ}), and the combined statistics are treated as a single measurement in the subsequent analysis. We refrain from deriving potentially slightly more optimal weights from the full covariance (see Sect.~\ref{sec:covariances}) as this would complicate the data processing and introduce a cosmology dependence in these weights through sample variance contributions. We considered re-weighting the 2dFLenS samples to obtain a closer match to the BOSS redshift distribution, but decided not to do so as this would have further reduced the effective number density of 2dFLenS. Re-weighting 2dFLenS was found to have no impact on the likelihood analysis. We define two lens galaxy samples, denoted by L1 and L2, cutting at $z=0.5$. The third redshift bin in the range $\bb{0.4;0.6}$ used by S17 is not considered as it is almost fully correlated with L1 and L2.

\subsection{KiDS-1000}

The Kilo-Degree Survey (KiDS) was conducted from 2011 to 2019 at the $2.6\,$m VLT Survey Telescope (VST) using OmegaCAM \citep{kuijken11}. This analysis makes use of Data Release 4 (DR4; \citealp{kuijken19}), comprising essentially all tiles with $ugri$-band data taken up to January 2018 and covering a raw survey area of $1006\,{\rm deg}^2$. By design, KiDS has the same footprint as the VISTA Kilo-degree INfrared Galaxy survey (VIKING; \citealp{edge13}), together providing deep nine-band photometry in the $ugriZYJHK_{\rm s}$ bands. DR4 is a complete re-reduction of the KiDS data set with the VIKING reduction of \citet{wright19} fully incorporated, and more than doubles the area with respect to DR3.

Earlier KiDS data releases, which cosmological analyses so far have been based on, still had quite fragmented sky coverage while DR4 is fairly homogeneous in three larger patches, two of which are equatorial with small separation (KiDS-N) and one that straddles ${\rm Dec}=-30\,{\rm deg}$ (KiDS-S); cf. Fig.~\ref{fig:Survey_footprint}. Correlation function measurements from these patches are combined through  galaxy pair weighting and further processed as a single statistic, with redshift distributions, shear calibration, galaxy number densities, etc. averaged accordingly.

KiDS data is processed with two pipelines: \textsc{Astro-WISE} \citep{mcfarland13} and dedicated software for joint flux measurements in KiDS and VIKING images \citep[see][]{wright19} produce coherent nine-band photometry particularly relevant for photometric redshift estimation (see Sect.~\ref{sec:photo-z}); \textsc{Theli} \citep{erben05} is optimised for weak lensing shear measurement and run on the $r$-band images only. KiDS benefits from a survey and instrument design that is optimised for weak lensing studies, with benign PSF variations across the $1\,{\rm deg}^2$ field of view and a mean $r$-band seeing of $0.7\,{\rm arcsec}$ in DR4. Improvements in the more recent observations mean that DR4 has overall lower PSF variability across the survey than DR3. Gravitational shear estimates are obtained with the model-based \textit{lens}fit \citep{miller07,miller13,fenech17} software, with only minor changes compared to previous KiDS analyses (see KV450). For details of the image processing, nine-band photometry, and shear catalogue production we refer the reader to \citet{kuijken19,wright19} and \citet{giblin20}, respectively.

\subsection{Photometric redshifts and tomographic binning}
\label{sec:photo-z}

\begin{figure}
        \centering
	\includegraphics[width=0.65\columnwidth]{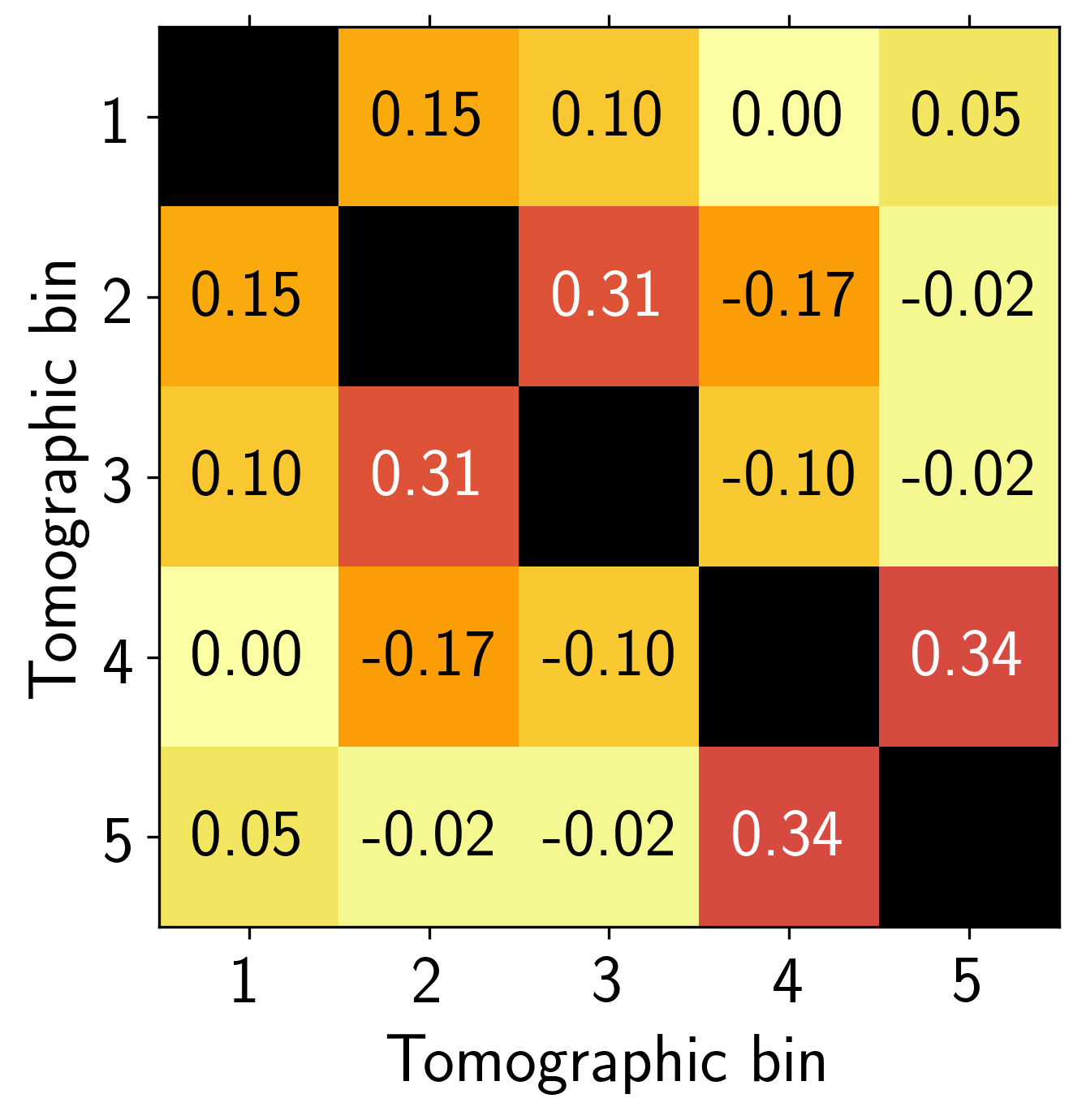}
	\caption{Correlations between the shifts of the source redshift bins S1--5 arising from the SOM calibration. The covariance was estimated from multiple mock realisations as described in \citet{wright20}.}
	\label{fig:nz_shifts}
\end{figure}

Photometric redshifts are determined from the joint KiDS and VIKING photometry with the Bayesian template-fitting code \textsc{BPZ} \citep{benitez00} as detailed in \citet{wright19}. The output maximum posterior redshift $z_{\rm B}$ is used to divide the source galaxy sample into five tomographic bins with boundaries $\bc{0.1; 0.3; 0.5; 0.7; 0.9; 1.2}$, as in KV450. The photometric redshifts are not used for any other purpose in the KiDS analysis. 

The source redshift distributions $n_{\rm S}(z)$ of the tomographic bins, shown in the lower panel of Fig.~\ref{fig:nOfZ}, are determined from a compilation of spectroscopic redshift data sets that overlap with KiDS-1000 and dedicated calibration observations with the VST in additional fields with deep VISTA and spectroscopic survey coverage \citep{wright19}. Totalling more than $25\,000$ objects, the spectroscopic redshift collection is then re-weighted to become representative of the KiDS source galaxy distribution. In a major update from previous analyses, we adopt the methodology presented in \citet{wright20} applied to DR4, which means that, rather than defining spectroscopic representation via a nearest-neighbour search in the nine-dimensional magnitude space of KiDS+VIKING (referred to as the DIR approach), we now use an unsupervised clustering approach employing a self-organising map (SOM, \citealp{kohonen82}). The SOM creates a two-dimensional representation of colour space onto which both the spectroscopic and source galaxy samples are mapped. The relative abundance of the two samples in a given cell of the SOM space is then used to derive the re-weighting. 

The procedure is applied to each tomographic bin individually to reduce the risk of misrepresenting multiple galaxy populations that overlap in colour space by a small number of spectra. Contrary to the previous DIR method, regions without any spectroscopic representation can now be removed from the source sample altogether to create a \lq gold\rq\ sample with secure calibration. For KV450, this step reduces the effective number density of galaxies by $15-17\,\%$ in S1--4 and $6\,\%$ in S5, which however is partially offset by a reduced uncertainty in the centring of the $n(z)$ \citep[cf.][]{wright20b}.

As detailed in \citet{wright20}, 100 mocks based on quasi-independent lines of sight in the MICE2 simulations \citep{carretero15,crocce15,fosalba15a,fosalba15b,hoffmann15} are used to assess the fidelity of the SOM approach outlined above. Biases in the mean of the redshift distributions are small, $\delta_z \lesssim 0.01$, with standard deviations of 
\eq{
\label{eq:sigma_zshift}
\sigma_z = \bc{0.005;0.006;0.006;0.004;0.005}\,
}
in S1-5, respectively. Moreover, the simulations reveal substantial correlations between the calibration in each bin, caused by the same spectroscopic redshifts being used in the calibration of multiple tomographic bins. While these correlations are strong in the previous nearest-neighbour approach with correlation coefficients around 0.9 for adjacent bins, they are mild for the SOM algorithm; see Fig.~\ref{fig:nz_shifts}. 

We account for the residual uncertainty in the tomographic redshift distributions by allowing shifts in the mean redshift of each bin with a multivariate Gaussian prior on the vector of $n(z)$ shifts $\vec{\delta}_z$ with the correlation structure determined by Fig.~\ref{fig:nz_shifts}. The standard deviations of the prior are given by the values in Eq.~(\ref{eq:sigma_zshift}) but multiplied by a factor 2 to account for additional uncertainty due to the limitations in the representativeness of our MICE mocks (see \citealp{hildebrandt20} for details). We note that those mocks contain the smaller survey area of the KV450 analysis, but as the method is insensitive to the source survey area, we do not expect any impact on our results. The calibration of the redshift distributions and the propagation of the residual uncertainty into the likelihood analysis focus on the mean only because biases in higher moments of the distributions are expected to be subdominant \citep{amara07}. In a separate publication we demonstrate this explicitly for a KiDS cosmic shear analysis by propagating the statistical uncertainty of the full redshift distribution functions \citep{stolzner21}.

In the KiDS-1000 analysis the $\vec{\delta}_z$ prior will be centred on the mock-based estimates of the bias \citep{wright20}, but for the demonstration of the methodology in this work we keep the mean of the Gaussian at zero. As the calibrated $n_{\rm S}(z)$ for KiDS-1000 are not produced until a fairly late stage of the data processing, we adopt in this methodology study the source sample properties as measured from KV450 (see Table~\ref{tab:sampleprops}), again with little change expected in the transition to KiDS-1000. For the redshift probability density functions themselves we adopt a KiDS-1000 distribution from one of the blinded shear catalogues (the $n_{\rm S}(z)$ vary due to the blinded shear estimation weights; see Appendix~\ref{sec:merrors}), calibrated with the approach described in \citet{wright20}. Further details on the redshift calibration for KiDS-1000, including additional validation via cross-correlations with spectroscopic reference samples, can be found in \citet{hildebrandt20}.

As Fig.~\ref{fig:nOfZ} shows, some source bins have substantial probability of galaxies being located at lower redshift than the lens samples L1 and/or L2. Sources in front of lenses do not generate any of the cosmological or astrophysical signals that we model, except for a very weak magnification bias contribution. To prevent any potential unaccounted systematics entering the data set via this route, we therefore exclude the lens-source bin combinations L1-S1, L2-S1, L2-S2, and L2-S3 from the GGL data vector entering the likelihood. We test for the impact of excluding bin combinations with large overlap between lenses and sources (L1-S2 and L2-S4; see Sect.~\ref{sec:parameter_constraints}), but include them by default as they contain valuable information about intrinsic alignments.

\subsection{Shear calibration}
\label{sec:shear_calibration}

We adopt the shear calibration approach of KV450, which builds on the image simulations presented in \citet{kannawadi19}. The simulations are generated by sampling from galaxy images emulated from superior Hubble Space Telescope observations in the COSMOS field \citep{griffith12}, including realistic distributions of colour, size, ellipticity, and their correlations. While the simulations were designed to accurately reproduce galaxy properties per tomographic redshift bin (rather than only for the overall survey), explicit cross-talk between shear measurement selection bias and photometric redshift estimates is not yet taken into account \citep[see e.g. the discussion in][]{asgari19}. This requires multi-band image simulations and will be addressed in forthcoming analyses as it is expected to be a second-order effect. Leading-order effects are captured because the COSMOS galaxies are selected based on their $z_{\rm B}$ value obtained from KiDS photometry. \citet{troxel18} find the residual biases to be more than an order of magnitude smaller than the response of shear to noisy galaxy images (see Appendix~\ref{sec:merrors} for details), which we can therefore assume to be negligible since our multiplicative shear bias corrections are small to begin with.

Both additive and multiplicative shear bias are generally significant enough to require calibration and uncertainty propagation into the likelihood analysis. It is demonstrated in \citet{giblin20} that biases proportional to PSF ellipticity ($\alpha_{\rm PSF}$) or residual PSF ellipticity after modelling ($\beta_{\rm PSF}$) introduce negligible bias on cosmology for KiDS-1000. \citet{kannawadi19} found the multiplicative biases $m$ for the KV450 analysis setup to be at the per-cent level. The $m$-correction is applied when computing correlation functions, and propagated accordingly into effective number densities $n_{\rm eff}$, observed ellipticity dispersions $\sigma_\epsilon$, and galaxy pair counts, as detailed in Appendix$\,$\ref{sec:merrors}.

No multiplicative bias corrections are applied to the mocks used in this work to demonstrate the methodology, but we do consider the uncertainty in this correction, which constitutes a significant contribution to the total error budget (see Sect.~\ref{sec:covariances}). As in KV450, we model the uncertainty as an additive contribution to the covariance of the two-point statistics, assumed to follow a Gaussian distribution with standard deviation $\sigma_m$. The only major update with respect to KV450 that could affect shear calibration significantly is the switch to the SOM-derived gold sample. However, re-running the \citet{kannawadi19} analysis with the new sample selection yields negligible changes in the shear bias. We therefore work with the $\sigma_m$ values for the KV450 setup.

We take a more nuanced approach to extracting $\sigma_m$ values from the \citet{kannawadi19} results (see in particular their Fig.~12) than KV450 who chose a blanket value of $\sigma_m=0.02$ which comfortably envelopes the various simulation configurations tested. We consider the two most extreme settings versus the fiducial case: the case with correlations between galaxy size and ellipticity switched off (as a conservative floor for this dependence), and the case that uses the global galaxy properties rather than those specific to each tomographic bin (as a conservative floor of accounting for the redshift dependence of galaxy properties). The resulting multiplicative biases vary systematically between these two cases as a function of redshift. In each tomographic bin we choose half of the maximum spread between the three cases as the standard deviation of the calibration uncertainty on the fiducial case, with a floor of 0.01 which corresponds to the statistical noise in the fiducial simulation results. This leads to 
\eq{
\sigma_m = \bc{0.019;0.020;0.017;0.012;0.010}\,
}
for S1-5. As in KV450, we assume that the multiplicative bias corrections are fully correlated between the tomographic bins since the $\sigma_m$ values are determined by systematic trends that smoothly vary across redshift. \citet{kannawadi19} identified additional dependencies on the clustering of source galaxies, stellar number density on the sky, and properties of the input COSMOS catalogue, which could act to de-correlate multiplicative biases, but these contributions remain subdominant. Thus, the covariance contribution due to multiplicative shear bias uncertainty reads
\eqa{
\label{eq:cov_mbias}
{\rm Cov}_{\rm mult} \bb{{\cal C}^{(ij)}_a;\, {\cal C}^{(kl)}_b} &= {\cal C}^{(ij)}_a\, {\cal C}^{(kl)}_b \\ \nn
& \hspace*{-1cm} \times \bc{ \sigma_m^{(i)} \sigma_m^{(k)} + \sigma_m^{(i)} \sigma_m^{(l)}  + \sigma_m^{(j)} \sigma_m^{(k)} + \sigma_m^{(j)} \sigma_m^{(l)}  } \;,
}
where $\sigma_m^{(i)}=0$ for any index $i$ indicating a lens sample in a GGL measurement. The subscripts  $a$ and $b$ could either be E/B for cosmic shear or ${\rm n} \epsilon$ for GGL. To arrive at this expression, we have assumed that the expectation of $m$ is zero (after correction), and that $\sigma_m^{(i)} \ll 1$ (see \citealp{blake20} for a more general equation).

Additive shear bias can be directly estimated from the data with some residual statistical uncertainty. Overall, additive shear bias terms are at the few times $10^{-4}$ level and corrected individually for the two ellipticity components and for KiDS-N/S\footnote{The two main KiDS patches are calibrated separately as they differ substantially in sky position as well as Galactic latitude, which leads to differences in their observational and astrophysical characteristics, such as the distribution of seeing and the number density of stars.} before further processing of the data. The residual uncertainty in KV450 on the additive bias measurement was $2 \times 10^{-4}$ which was propagated into the likelihood analysis via a nuisance parameter with a Gaussian prior with scatter given by this error. The global additive bias only enters $\xi_+$ and particularly affects large angular scales where the cosmological signal is smallest. Ideal band powers derived from correlation functions available over the entire positive real numbers are insensitive to additive shear contributions as those would only affect the monopole. For the finite range of angular scales used for the band powers in this work, we find the additive bias contribution to be negligibly small, reaching at most $\sim 1\,\%$ on large angular scales in the lowest-redshift tomographic bin combination. We therefore do not use an additive bias nuisance parameter ($\delta c$ in KV450) in the band power analysis.

We also identified a spatially varying additive bias pattern linked to three problematic CCD chips in the OmegaCAM field of view (see Fig.~2 in KV450), which was measured via stellar PSF ellipticities and then propagated into $\xi_\pm$ with a free overall amplitude that was treated as a nuisance parameter. KV450 showed this nuisance parameter to have no effect on cosmological constraints. Since the amplitude of the pattern has not increased in significance in KiDS-1000 and impacts even less on the band powers, we opt to neither correct for it nor include the related nuisance parameter ($A_c$ in KV450). The pattern could in principle generate a characteristic (weak) signal in the B-modes \citep{asgari19}, which however are found to be consistent with zero in \citet{giblin20}.

While our shear calibration procedure takes the impact of blended galaxy images into account, strongly blended objects are likely to be rejected by the shear measurement algorithm or remain undetected if obscured by a foreground object. Since this rejection is more likely to occur along over-dense lines of sight, weak lensing two-point statistics are biased low as a result \citep{hartlap11}. However, for the current generation of surveys this effect only causes biases at the per-cent level over relevant angular scales (\citealp{harnois18}; see also \citealp{samuroff18}) and can therefore be neglected.

\subsection{Correlation function measurements}

All summary statistics in this analysis are based on configuration-space correlation function measurements, which have the major advantage that their expectation values are independent of the survey geometries. We adopt the clustering wedge correlation function measurements \citep{kazin12} from S17 whose salient features we summarise below. The consensus analysis in \citet{alam17} showed that the S17 results are in good agreement with other \lq full-shape\rq measurements on the same data, using real-space multipole as well as Fourier-space statistics. The wedge-based analyses generally produced tighter parameter constraints, for example the standard deviation on distance was about a third smaller than for the multipole statistics.

Two-dimensional correlation functions $\xi_{\rm gg}(s,\mu)$ are measured with a Landy-Szalay estimator \citep{landy93} in linear bins of separation $s$ with size $5\,h^{-1}\,{\rm Mpc}$ and three equidistant bins in $\mu$, totalling 84 data points per redshift bin. The angle entering $\mu$ is measured between the separation vector of a galaxy pair and the line of sight at the midpoint of $s$. As in related BOSS DR12 analyses, $s$ and $\mu$ are calculated for a fiducial, spatially flat $\Lambda$CDM model with $\Omega_{\rm m}=0.31$. For the Landy-Szalay estimator, we adopt the random catalogues used in S17 with an oversampling factor of 50. The correlation functions include a combination of weights that minimise variance \citep{feldman94}, account for redshift failures and fibre collisions, and correct for spatial patterns induced by a number of observational systematics \citep{ross17}.

Weak lensing correlation functions are computed using the public \textsc{TreeCorr}\footnote{\texttt{https://github.com/rmjarvis/TreeCorr}} \citep{jarvis04} tree code. Its key tuning parameter is \texttt{bin\_slop}, which sets the accuracy of placing galaxy pairs into the correct angular separation bin measured relative to the bin size. We optimise \texttt{bin\_slop} to minimise computation time without measurably impacting on the correlation function. Since we measure finely binned correlation functions with more than a hundred bins per decade of angular separation, fairly large values of \texttt{bin\_slop} (1.5 for cosmic shear and 1.2 for GGL) can be used. Parts of the KiDS-1000 analysis also employ coarsely binned correlation functions; these are obtained by merging the finely binned measurements.

We use the standard estimator for $\xi_\pm$ \citep{schneider02}; see Appendix~\ref{sec:merrors} for its explicit form including our choice of weights. GGL is measured through the average tangential shear around lenses, estimated via \citep{mandelbaum2005}
\eq{
\label{eq:ggl_estimator}
\widehat{\ba{\gamma_{\rm t}}}(\theta) = \frac{\sum_{ls} w_l\, w_s\, \epsilon_{{\rm t},l \rightarrow s}\, \Delta_{ls}(\theta)}{\sum_{rs} w_r\, w_s\, \Delta_{rs}(\theta)}  {\cal N}_{\rm rnd} - \frac{\sum_{rs} w_r\, w_s\, \epsilon_{{\rm t},r \rightarrow s}\, \Delta_{rs}(\theta)}{\sum_{rs} w_r\, w_s\, \Delta_{rs}(\theta)} \;,
}
where the sums with indices $l$, $s$, and $r$ run over all elements of the lens, source, and random catalogues, respectively. Here, we have used the bin selector function $\Delta_{ij}(\theta)$ which is unity if the angular separation of a galaxy pair indexed by $i$ and $j$ falls into a bin centred on $\theta$, and zero otherwise. The estimator also accommodates different weights $w$ for the lens, random, and source samples as indicated by their index, and $\epsilon_{{\rm t},l \rightarrow s}$ denotes the tangential ellipticity of object $s$ measured with respect to the position of object $l$. The term ${\cal N}_{\rm rnd} := \sum_r w_r/ \sum_l w_l$ reduces to the oversampling factor of the random catalogue with respect to the catalogue of lens galaxies for unit weights. We use 100 times more random points than lens galaxies, which we verify reduces any additional noise components due to the finite number of random points to negligible levels (see Appendix~\ref{sec:anacov}).

The second term in Eq.~(\ref{eq:ggl_estimator}) removes potential spurious contributions to the signal that do not average out due to survey boundaries and masks. It also suppresses non-Gaussian and additive noise contributions to the covariance of the estimator \citep{singh17}. The average tangential ellipticity in the first term is not normalised by the effective number of lens-source pairs, but by the number of pairs between the source and random catalogues. This modification is sometimes referred to as the boost correction and removes any signal suppression due to clustering between lenses and sources in case their redshift distributions overlap. We experiment with different weights in the GGL estimator \citep[see e.g.][]{shirasaki18,blake20}, but find that they do not improve signal-to-noise significantly over the basic form of Eq.~(\ref{eq:ggl_estimator}).

\subsection{Band power measurements}
\label{sec:bandpower_measurement}

Band powers can be derived via linear transformation of the correlation functions. For cosmic shear, our E- and B-mode estimators are given by a discretised version of Eq.~(\ref{eq:bpestimate_ideal}),
\eq{
\label{eq:bp_eb}
\hat{{\cal C}}^{(ij)}_{{\rm E/B},l} =  \frac{\pi}{{\cal N}_l}\; \sum_k \Delta \theta_k\, \theta_k\; T(\theta_k) \bc { \hat{\xi}_+^{(ij)}(\theta_k)\; g_+^l(\theta_k) \pm \hat{\xi}_-^{(ij)}(\theta_k)\; g_-^l(\theta_k)  }\;,
}
where the sum runs over all angular bins of the correlation function, each with bin width $\Delta \theta_k$. The normalisation ${\cal N}_l$, the apodisation $T(\theta_k)$, and the kernels $g_\pm$ were defined in Sect.~\ref{sec:cosmicshear_stats}. These expressions would be unbiased estimators of the angular power spectrum in the limit that the sum asymptotes into an integral over the non-negative real numbers and $T \equiv 1$ \citep{schneider02,uitert18}. In practice the angular extent over which correlation functions can be measured is limited, which is fully accounted for in the modelling of the band powers. Effects of the discretisation in Eq.~(\ref{eq:bp_eb}) are negligible as long as the binning is chosen finely enough to smoothly sample the highly oscillatory kernels (see Fig.~\ref{fig:bandpower_filters_real}). The corresponding GGL estimator is
\eq{
\label{eq:bp_ne}
\hat{{\cal C}}^{(ij)}_{{\rm n\epsilon},l} =  \frac{2 \pi}{{\cal N}_l}\; \sum_k \Delta \theta_k\, \theta_k\; T(\theta_k)\; \widehat{\ba{\gamma_{\rm t}}}^{(ij)}(\theta_k)\; h^l(\theta_k)\;,
}
derived from Eq.~(\ref{eq:bpestimate_ggl_ideal}). Throughout, the superscripts $(ij)$ indicate redshift bin combinations.

We find that 300 logarithmically spaced bins between $0.5\,{\rm arcmin}$ and $300\,{\rm arcmin}$ (the default range used in KV450) avoids any discretisation effects. This binning scheme is continued for both smaller and larger angular separation to allow for apodisation. An apodisation width of $\Delta_x=0.5$ (cf. Eq.~\ref{eq:apodisation}) is found to suppress sensitivity of the band powers far from their designated band width while only mildly expanding the angular range entering the estimators. Consequently, correlation functions in the range $\bb{0.6;234}\,{\rm arcmin}$ enter at full sensitivity, contributions reduce to half at the nominal bin edges of $0.5/300\,{\rm arcmin}$, and scales outside the interval $\bb{0.4;385}\,{\rm arcmin}$ do not contribute at all. The minimum scale is still well above the regime where galaxy blending and the image cutout size with which \textit{lens}fit works ($\lesssim 10\,{\rm arcsec}$) could impact on the correlation functions. Conversely, the maximum scale is of the same order as the extent in declination of the KiDS patches (cf. Fig.~\ref{fig:Survey_footprint}) beyond which galaxy pair counts would rapidly decline and isotropy-breaking systematics in the shear measurement not fully average out.

We choose 8 logarithmically spaced bands with a lower boundary of $\ell=100$ and an upper limit of $\ell=1500$. The band power response to lower angular frequencies than $\ell \sim 100$ becomes very broad (see Fig.~\ref{fig:bandpower_filters}) and would thus necessitate more expensive modelling of very large-scale modes. On scales smaller than $\ell=1500$ the modelling uncertainty of non-linear matter power and baryonic feedback prevents us from exploiting measurements efficiently. All 8 bands are used in the cosmic shear signals. Our conservative approach to modelling GGL means we only use the lowest three bands for correlations involving L1 and the lowest five bands for those involving L2 (see the discussion in Sect.~\ref{sec:pgm}). We also explore a more progressive setting that only discards the highest $\ell$-band for GGL and thus scales of $k \approx 1\,h\,{\rm Mpc}^{-1}$ and smaller.

\section{Simulations}
\label{sec:mocks}

We create suites of dedicated simulations with the primary purpose of assessing the accuracy of the two-point statistics we measure and of their covariance. A key question we address is whether the previously used analytic covariance prescription is still good enough for the increasingly constraining data sets of the current survey generation, despite its necessarily idealising assumptions about survey geometry and survey homogeneity. Non-Gaussian contributions to the covariance are subdominant and have been assessed before by comparison with a suite of N-body simulations \citep{hildebrandt17,harnois18}, so that we now opt for a fast approximate simulation approach that can accommodate the full survey volume and allows for a large number of mock realisations of the data.  The resulting simulation suite has the additional benefit of providing us with accurate sampling distributions of various summary statistics that we generate from the data (see Sect.~\ref{sec:inference}).

\subsection{Scope}
\label{sec:mocks_scope}

A number of challenges related to survey characteristics arise in the joint galaxy clustering and weak lensing analysis that necessitate careful validation of signals and covariances with mocks. First, lens and source galaxies occupy vastly different footprints in the sky (see Fig.~\ref{fig:Survey_footprint} and Table~\ref{tab:surveyareas}). This complicates the modelling of cross-variances in particular, and the strong assumptions used in analytic modelling to date \citep[cf.][]{uitert18} are yet to be tested. Secondly, while our choice of two-point statistics are by design insensitive to the survey geometry, it does impact on their covariance. The KiDS footprint with its two disjoint and elongated fields, whose extent in declination is of the same order as the largest scales we measure, could violate the sweeping approximations currently made in the most widely used covariance models.

It is also unclear from first principles how to measure the effective area of a survey that to good approximation scales the variances of all data. One could obtain an accurate estimate of all unmasked area or instead focus more on the extent of the footprint, that is, exclude features like the numerous star masks which are on scales much smaller than those that are considered in the likelihood analysis. This then becomes a question of adequate spatial resolution in the survey mask (either of a pixelated map or a random catalogue). Different choices here can lead to tens of per cent differences in variances and thus outweigh the impact of other contributions like non-Gaussian terms which have received much attention in the literature.

Finally, virtually all analytic modelling has hitherto relied on the assumption of homogeneity, thereby ignoring the spatial variability in the data induced by observing conditions \citep[see][though]{heydenreich20}. However, the depth of an imaging survey varies from pointing to pointing due to changes in conditions such as seeing and sky transparency. Larger-scale patterns can be generated through a combination of seasonal weather patterns and visibility restrictions of targets. The resulting variable survey depths will be more pronounced in surveys like KiDS that visit sky areas only once during the lifetime of the programme. The impact is visualised in the lower part of Fig.~\ref{fig:Survey_footprint} where colours ranging from yellow to purple indicate one magnitude of per-pointing variation in the $r$-band limiting magnitude $r_\mathrm{lim}$ ($1\sigma$ detection in adaptive apertures), which is sensitive to both the seeing and the background noise level in the images.

In order to evaluate the impact of multiple survey footprints, mask geometry, and variable depth on our analysis, we require simulations that allow us to impose the realistic footprints and overlaps of the surveys, as well as galaxy samples with precisely the same observational characteristics as in the data. Validation is greatly aided by having a-priori knowledge of the input two-point statistic that is to be recovered, as well as by generating a large number of realisations of the Universe to beat down noise in the estimates of large covariance matrices. These conditions lead us to make use of fast, full-sky random field simulations as described in the following.

We opt for lognormal, rather than Gaussian, random fields because the latter are unable to recover the input two-point statistics at the level of accuracy that we require. At the spatial resolution of our mocks, the KiDS-1000 area is covered by of order $5 \times 10^6$ pixels which define volume elements in shells of the matter distribution. If the probability distribution of the matter density contrast $\delta$ in these volume elements is Gaussian, values of $\delta$ are not bounded from below. Hence, unphysical values of $\delta < -1$ in the tail of the distribution are realised in a small fraction of the pixels. Any modification of these problematic cases, for example by setting them to an empty pixel ($\delta=-1$), leads to significant biases in the two-point statistic of the matter density field at the few per cent accuracy that we aim for. Lognormal statistics of the matter density contrast have the additional benefit of being close to the realistic case \citep{kayo01}, which enables us to at least qualitatively explore the non-Gaussian properties in the covariance (see Sect.~\ref{sec:covcomp}).

We emphasise that, by design, our mock suite returns the two-point statistics that we put in. It is therefore not suitable to validate our baseline models of these statistics. This validation was performed in previous works using a range of N-body and hydro-dynamical simulations; see the discussion and references in Sect.~\ref{sec:modelling}. In our mocks we do not couple the shear estimates to the foreground galaxy population, so that we do not incorporate neighbour-exclusion bias, which is expected to vary with the local survey depth. However, this effect was shown to be insignificant for the current generation of surveys \citep{harnois18}. We note that the impact of variable depth on the shear calibration is taken into account in the data analysis; see the discussion in Appendix~\ref{sec:merrors}.

\subsection{Construction of mocks}
\label{sec:mocks_construction}

\begin{figure}
	\centering
	\includegraphics[width=\columnwidth]{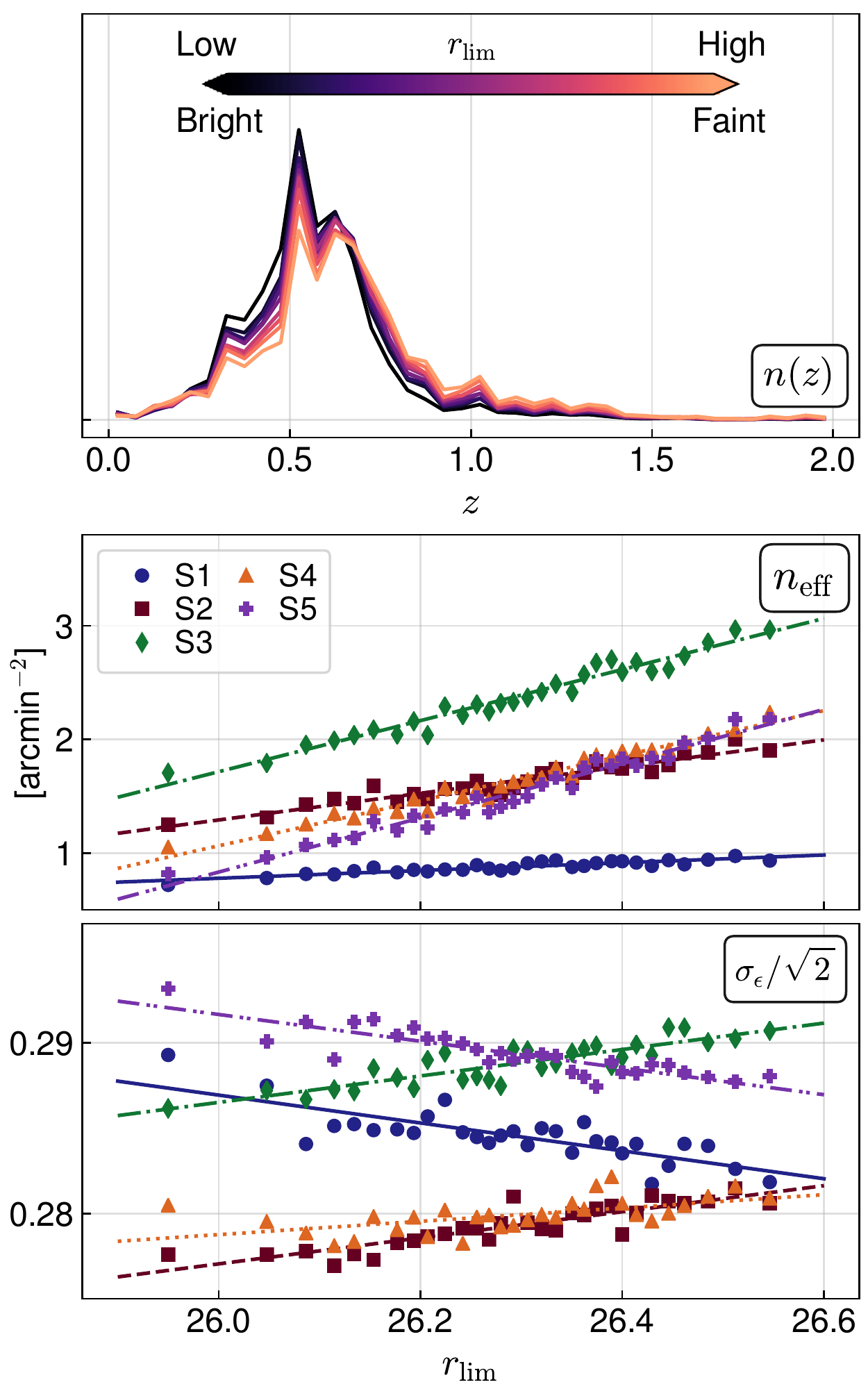}
	\caption{Characteristic of variable depth effects measured from early KiDS-1000 data. 
	\textit{Top panel:} Redshift distributions of source galaxies in the third tomographic bin (S3) in 10 equi-populated bins of $r_\mathrm{lim}$ (1$\sigma$ adaptive aperture size). \textit{Centre panel:} Effective galaxy number density as a function of $r_\mathrm{lim}$ in the five source galaxy bins, together with linear fits. \textit{Lower panel:} Same as centre panel but for the observed ellipticity dispersion $\sigma_\epsilon$.}
	\label{fig:KiDS_varDepth}
\end{figure}

The matter distribution of our fast mocks is generated with \textsc{Flask}\footnote{\texttt{http://www.astro.iag.usp.br/$\sim$flask}} \citep{xavier16}, building concentric shells of lognormal random fields with known angular matter power spectra and a minimum density contrast of $-1$ (corresponding to an empty line of sight). We note that the same approach to validating covariance models was also taken by \citet{des_y1_methods}. The input cosmology for the mocks is specified in Table~\ref{tab:fiducialpars}. The matter shells are created in 18 redshift bins of roughly equal intervals in comoving distance between $150 - 200\,h^{-1}\,{\rm Mpc}$. The bin edges closest to the redshift limits of the lens samples are moved to exactly 0.2, 0.5, and 0.75. The correlations between matter shells are fully included via their cross-power spectra. Since the matter slices are fairly thin, the angular matter power spectra are computed using the full curved-sky, non-Limber expressions. Weak lensing shear and convergence fields are then obtained from integrating over the matter shells along the line-of-sight with the weighting of Eq.~(\ref{eq:kernel_lensing}).

We create a new software, \textsc{Salmo}\footnote{Speedy Acquisition of Lensing and Matter Observables; the code will be made public on acceptance of the KiDS-1000 analysis papers.}, to populate the resulting matter distribution with galaxies based on the visibility masks and redshift distributions provided by the surveys. In contrast to the standard \textsc{Flask} functionality, \textsc{Salmo} allows us to implement pointing-dependent selection functions and thus spatially varying redshift distributions, galaxy densities, and ellipticity dispersions. Galaxies are sampled following a Poisson process with constant galaxy bias corresponding to the fiducial values of $b_1$ in the lens samples and then assigned weak lensing signals and shape noise. Tracers with different angular and radial selection functions can thus be simulated based on the same realisation of the underlying matter distribution. Masks for the source galaxy samples are taken from KiDS-1000; those for the lens samples are made from areas of BOSS and 2dFLenS that overlap KiDS (with the exception of the study presented in Appendix~\ref{sec:bosskidscorr}, which is based on a set of simulations with the full BOSS footprint). Both masks and the \textsc{Flask}-generated maps have a \textsc{Healpix}\footnote{\texttt{http://healpix.sourceforge.net}} \citep{gorski05} resolution of $N_\mathrm{side} = 4096$, which corresponds to a characteristic angular size of 0.86~arcmin below which the power of density fluctuations realised in the simulations is suppressed. 

Lens redshift distributions are either adopted individually from the BOSS and 2dFLenS data, or an inverse variance-weighted combination is applied uniformly to all lenses (cf. Fig.~\ref{fig:nOfZ}, top panel). Since the final KiDS-1000 redshift solutions were still under construction when the mocks were created, we proceed as follows, making use of the default KV450 DIR approach (using a weighting of spectroscopically observed galaxies in the KiDS-VIKING colour space). We divide the $1\,{\rm deg}^2$ pointings in KV450 into 10 equi-populated bins of $r_{\rm lim}$ and repeat the redshift distribution calibration on each subset for all five tomographic bins with the full set of spectroscopic sources. The resulting distributions are then assigned to each of the pointings within the respective bins and the mock galaxies are sampled from them accordingly. To limit the size of simulations, we truncate all redshift distributions at $z=2$ beyond which only $0.3\,\%$ of source galaxies are located. The average of these individual redshift distributions is shown in Fig.~\ref{fig:nOfZ} and used in the modelling of the data vector.

Figure~\ref{fig:KiDS_varDepth} illustrates the characteristics of variable depth effects for KiDS, as measured from an early version of KiDS-1000 data with blinded ellipticity measurements (which will affect $\sigma_\epsilon$ only at the per-cent level). The upper panel shows the redshift distributions for the different $r_{\rm lim}$ bins in the third tomographic bin (S3). We can see that, as expected, for faint $r_\mathrm{lim}$ the distribution is shifted towards higher redshifts as better seeing and transparency allow for more faint galaxies to enter the sample, which tend to be at higher redshifts. The number of low-redshift galaxies is only weakly affected, but after normalisation their proportion effectively decreases.

The centre and lower panels of Fig.~\ref{fig:KiDS_varDepth} show the dependence of the effective number density $n_\mathrm{eff}$ and the observed ellipticity dispersion $\sigma_\epsilon$ on $r_\mathrm{lim}$. In this case galaxies from all five tomographic bins are divided into 30 equi-populated $r_\mathrm{lim}$ bins selected from the KiDS-1000 footprint. We observe a very clear linear dependence for both quantities in all tomographic bins. The galaxy number density increases with deeper data as expected, while there is no clear trend for $\sigma_\epsilon$, as the sign of the slope reverses for the lowest and highest tomographic redshift bins. We surmise that there is a complex interplay between faint-end galaxies altering the galaxy population and hence the intrinsic ellipticity distribution on the one hand and those galaxies increasing the image level noise which broadens $\sigma_\epsilon$ on the other (one of the reasons why multiplicative shear bias is calibrated per tomographic bin in KiDS; cf. \citealp{kannawadi19}). Based on the linear fits to these trends, we incorporate a per-pointing variation for $n_\mathrm{eff}$ and $\sigma_\epsilon$ into the mocks as well.

With the methodology above, we produce four sets of mocks\footnote{We choose to name the different mock setups after birds in order to easily distinguish between them.}, following identical underlying dark matter field realisations but differing in their level of realism regarding survey footprints and spatial variability in the redshift distributions: \textbf{Buceros} -- a single rectangular mask with fully overlapping surveys, spatially uniform lens redshift distributions, and spatially uniform source redshift distributions; \textbf{Cygnus} -- realistic survey masks, spatially uniform lens redshift distributions, and spatially uniform source redshift distributions; \textbf{Diomedea} -- realistic survey masks, different lens redshift distributions in KiDS-N/S from BOSS/2dFLenS, and spatially uniform source redshift distributions; and \textbf{Egretta} -- realistic survey masks, different lens redshift distributions in KiDS-N/S, and variable depth source redshift distributions and galaxy sample properties.

Contrary to full N-body or other particle- and/or mesh-based runs, lognormal random field simulations are extremely fast. Using the setting described above, a set of full-sky tomographic density and lensing maps can be generated within less than an hour of wall-clock time using four CPUs in parallel. Catalogue generation with two BOSS-like lens bins and five KiDS-like source bins requires about 20 minutes with one CPU. Parallel computation of many survey realisations is then trivial to execute on distributed high-performance computing systems. The speed of the simulation generation shifts the computational bottleneck to the processing of the mock data, in our case driven by the calculation of correlation functions. We create $5000$ realisations of each survey setup which we find to reduce the sampling noise in covariance estimates to negligible levels for our comparison.

The fidelity of our mocks is assessed by applying the measurement pipeline to the mock catalogues in the case of uniform galaxy distributions and comparing the resulting two-point statistics to the theoretical predictions. The mock mean agrees to $3\,\%$ or better for both cosmic shear and GGL signals over relevant angular scales, with the exception of cosmic shear correlation functions involving bin S1 for which we see an excess signal around $5\,\%$ over the theoretical model. The latter is due to the small number of redshift shells in our \textsc{Flask} simulations which has the strongest impact on the low-redshift source sample with the smallest redshift baseline in the line-of-sight integration to obtain the lensing signal. This limitation would be easy to remedy, but as the runtime of mocks scales approximately quadratically with the number of shells, we prioritise the ability to create a larger number of mock realisations. While the small excess power in S1 should be borne in mind in the comparison with analytic covariances (see Sect.~\ref{sec:covcomp}), it is of little consequence for the likelihood analysis since the cosmic shear signals including S1 have low signal-to-noise.

\subsection{Impact of spatial variability}
\label{sec:spatial_variability}

\begin{figure}
	\includegraphics[width=\columnwidth]{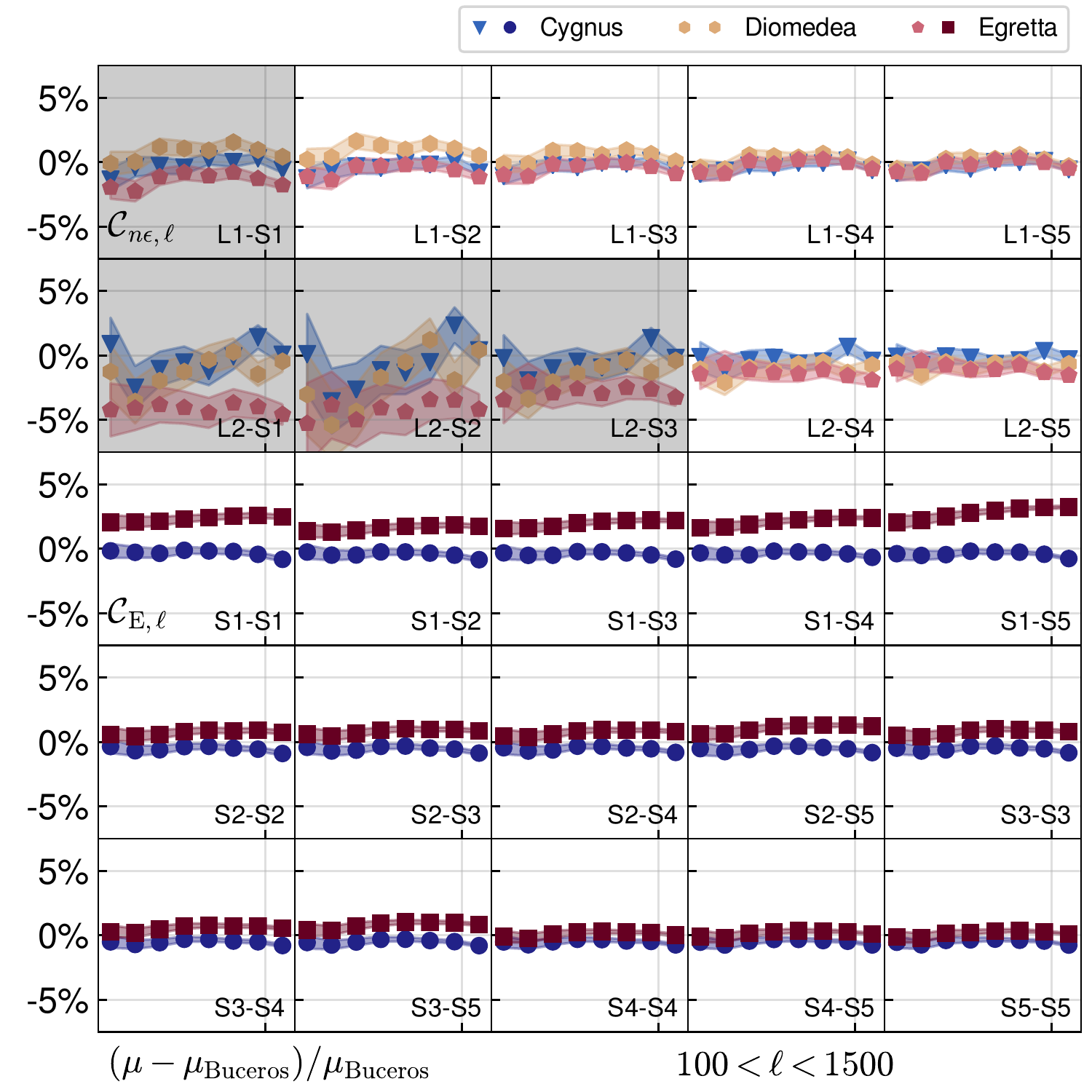}
	\caption{Comparison of mean estimates of band power between different mock setups. Plotted is the relative deviation from the Buceros case (rectangular mask, uniform depth). Blue, yellow, and red colours correspond to the Cygnus (realistic mask, uniform depth), Diomedea (like Cygnus but using individual lens sample properties for BOSS and 2dFLenS, hence only shown for GGL), and Egretta (realistic mask and depth variations) cases, respectively. The top two rows show GGL signals; the bottom three the cosmic shear signals, with bin combinations indicated in the panels (those not used in the likelihood analysis are greyed out). Bands around the data points indicate the standard error determined from a jackknife estimate of variance on the ratios shown, based on mean signals calculated from $1000$ realisations.}
	\label{fig:mean_diff_egretta}
\end{figure}

We run our measurement pipeline on the four variants of mock catalogues and study the relative difference between the resulting band power signals in Fig.~\ref{fig:mean_diff_egretta} (differences in their covariance will be discussed in Sect.~\ref{sec:covcomp}). To suppress noise, we have averaged signals over $1000$ realisations and switched off shape noise. Switching from a simple to the realistic survey geometry (Buceros to Cygnus) has no impact on correlation functions and their derived statistics -- a major advantage of our choice of harmonic space measure.

Comparing Diomedea to Cygnus shows the effect of averaging GGL over the BOSS and 2dFLenS overlap and modelling the resulting signal with the averaged redshift distribution. This approximation is accurate to within $2\,\%$ and thus fully sufficient for our analysis. We caution however that this conclusion is to a large degree due to the factor 2.5 lower number density in 2dFLenS, which means that the contribution to both the signal and the shape of the redshift distributions is dominated by BOSS (cf. Fig.~\ref{fig:nOfZ}). Averaging over spatial variability in the lens samples, as ubiquitously done when calculating summary statistics, has the potential to cause significant biases in GGL if the parts of the survey(s) that vary in depth carry a more evenly balanced statistical weight than in our case.

The Egretta case additionally incorporates variable depth in the source galaxy samples. Its impact on GGL is small and even partially reverses the small trends due to variability in the lens samples in correlations with L1. The effect on cosmic shear is more pronounced, especially for correlations involving the low-redshift S1 for which deviations of up to $4\,\%$ can be observed at the highest angular frequencies considered. The cosmic shear case was recently studied in depth by \citet{heydenreich20} who developed a semi-analytic model and demonstrated very good recovery of \textsc{Flask/Salmo}-generated mock signals. They concluded that on small scales the trend is well accounted for by the pointing-to-pointing variation of survey depth, whereas above degree scales correlations between the depths of adjacent pointings contribute as well.

The mean redshifts of S1 vary the most, ranging between 0.35 and 0.65 (see Fig.~2 of \citealp{heydenreich20}), leading to the most pronounced variable depth effect. This is likely driven by the extended high-redshift tail of the S1 redshift distribution (Fig.~\ref{fig:nOfZ}) which is strongly affected by the limiting magnitude and drives up the S1 mean redshift well beyond the photometric redshift boundary.

The cosmic shear band powers are consistently enhanced by the variable depth effect. As mean redshifts vary roughly linearly with depth, the amplitude of two-point statistics acquires a quadratic dependence to first order via the lensing efficiencies (cf. Eq.~\ref{eq:kernel_lensing}), so will tend towards higher amplitudes on average\footnote{Variable depth also generates B-mode cosmic shear signals \citep{vale04}, but they are unlikely to be detectable with KiDS.}. The source redshift distributions enter the GGL signals linearly, so as long as most of the sources are located behind the lens galaxies, the contributions from the different lines of sight average to the signal based on the mean redshift distribution used for Buceros. However, if the sources are mostly positioned in front of the lenses, it is the high-redshift tails of $n_{\rm S}^{(i)}(z)$ that determine the signal amplitude, and these have a non-linear dependence on depth (cf. Fig.~\ref{fig:KiDS_varDepth}, top panel). This leads to the $\sim\!5\,\%$ negative trend in the L2-S1 and L2-S2 GGL signals, which however are not used further in our analysis.

Since the high signal-to-noise cosmic shear signals involving S3-5 are only affected at the per-cent level, we can expect a mild impact on cosmological parameter constraints overall, with $S_8$ somewhat over-estimated. Parameters that are sensitive to the small modifications of redshift scaling and angular scale dependence, like $A_{\rm IA}$ and $A_{\rm bary}$, could in principle be more strongly affected. We add the relative difference between the means of the Egretta and Buceros mocks seen in Fig.~\ref{fig:mean_diff_egretta} to a noiseless mock data vector and carry out a likelihood analysis of the joint weak lensing signals (see Sect.~\ref{sec:parameter_constraints}), which results in insignificant changes to all model parameters, for instance a $0.12\sigma$ shift in $S_8$.

\section{Error modelling}
\label{sec:covariances}

We develop separate covariance models for galaxy clustering and the weak lensing signals, as detailed in the following. This is possible because these parts of the overall data vector are uncorrelated to very good approximation, primarily due to the much larger survey area of BOSS outside the KiDS footprint; see Appendix~\ref{sec:bosskidscorr} and in particular Fig.~\ref{fig:corr_paper} for a detailed analysis.

\subsection{Clustering covariance}

Since we can treat the clustering signal as fully independent of the weak lensing signals, we simply adopt the public S17 covariance that was estimated from 2045 \textsc{Multidark-Patchy} mocks of BOSS \citep{kitaura16}. Bias in the mean of the noisy inverse of the covariance estimate entering the likelihood is corrected \citep{hartlap07}. S17 also rescaled the posterior to account for noise bias in the second moments of the inverse covariance distribution \citep{percival14}, which we do not adopt because it only amounts to a $1.6\,\%$ change in parameter uncertainties which is within the numerical uncertainty of posterior sampling. Moreover, this ad-hoc modification would introduce inconsistencies in the constraints jointly with weak lensing as the latter do not suffer noise biases due to using a fully analytic covariance.

\subsection{Analytic weak lensing covariance}

\begin{figure}[h!!!]
	\includegraphics[width=\columnwidth]{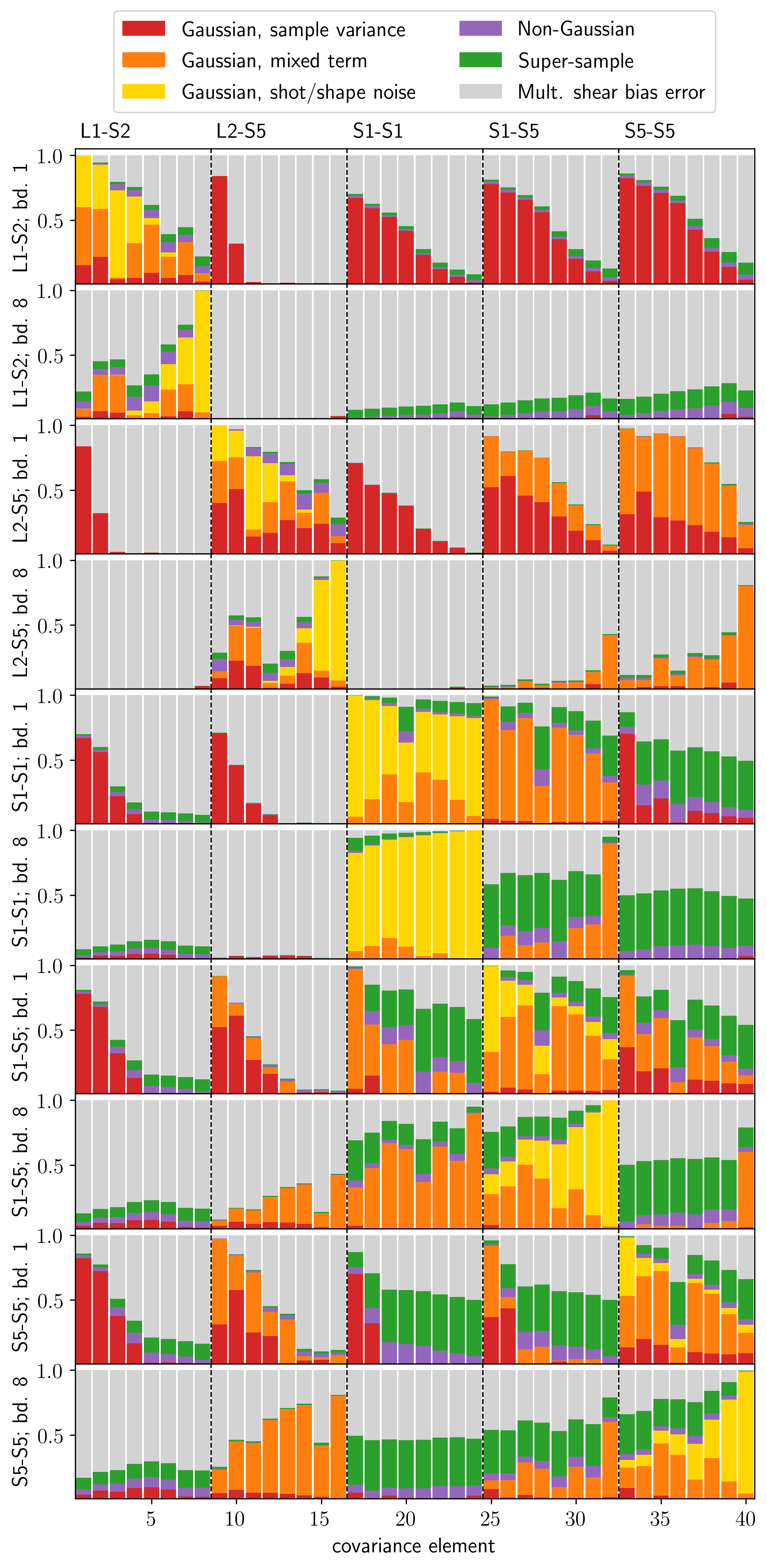}
	\caption{Contributions to selected elements of the analytic band power covariance. Shown are the relative contributions of the absolute values of the six covariance terms: sample variance (red), mixed term (orange), and shot/shape noise (yellow) in the Gaussian covariance; non-Gaussian in-survey covariance (purple) and super-sample covariance (green); and the multiplicative shear bias uncertainty (grey). Each panel corresponds to a single row in the covariance matrix of the full GGL and cosmic variance data vector, for the signal and angular frequency band indicated in the label; columns in each panel have the same ordering of signals but show all 8 bands, respectively. A representative subset of signals was chosen: two GGL signals, as well as a low- and high-redshift cosmic shear signal, and the corresponding redshift bin cross-correlation.}
	\label{fig:cov_contribution}
\end{figure}

\begin{figure}
	\includegraphics[width=\columnwidth]{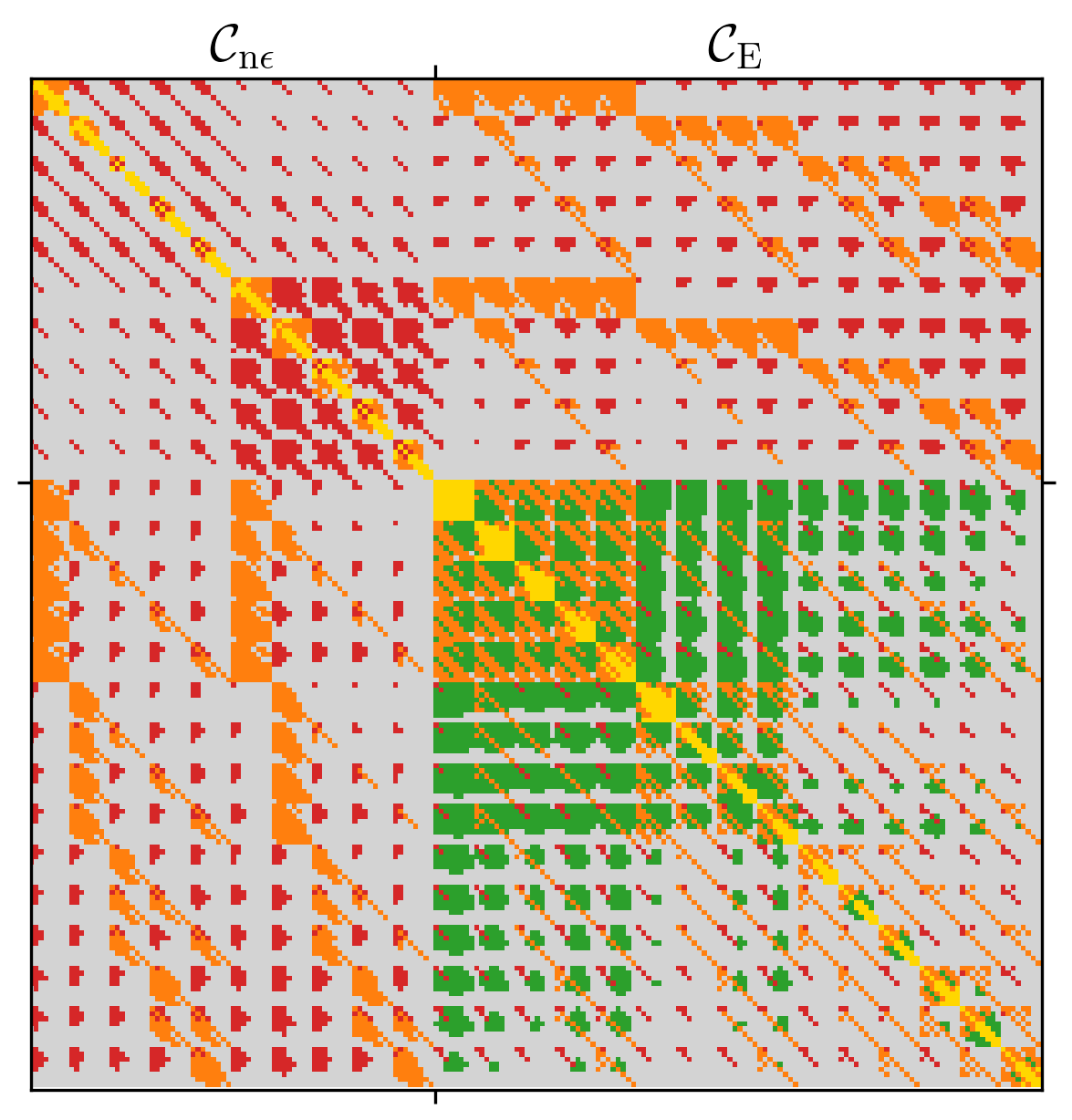}
	\caption{Dominant contribution (absolute value) to the band power covariance. The colour coding is identical to Fig.~\ref{fig:cov_contribution}, i.e. sample variance (red), mixed term (orange), and shot/shape noise (yellow) in the Gaussian covariance; non-Gaussian in-survey covariance (purple, always sub-dominant) and super-sample covariance (green); and the multiplicative shear bias uncertainty (grey). The top left block contains the covariance of the GGL signals (${\cal C}_{\rm n \epsilon}$; without cuts in tomographic bins or angular scales), the bottom right block the cosmic shear covariance (${\cal C}_{\rm E}$). Smaller blocks discernible correspond to the different bin combinations, and within these, individual $8 \times 8$ blocks correspond to the eight band powers of the signal for a single tomographic bin combination.}
	\label{fig:cov_contribution_matrix}
\end{figure}

Analytic approaches to the weak lensing covariance have now been widely adopted by the community as they provide sufficiently accurate, computationally fast, and noiseless models \citep{dessv16,hildebrandt17,hikage19}. We adopt the model used for cosmic shear in KV450 with minor updates and extend its use to GGL. A detailed description of the implementation is provided in  Appendix~\ref{sec:anacov}; here, we summarise the salient points.

The model is composed of five terms of cosmological origin,
\eq{
{\rm Cov} = {\rm Cov}_{\rm G, sva} +  {\rm Cov}_{\rm G, mix} +  {\rm Cov}_{\rm G, sn} +  {\rm Cov}_{\rm NG} +  {\rm Cov}_{\rm SSC}\;,
}
where \lq G\rq\ stands for the Gaussian contributions, that is terms that would also be present if summary statistics of a Gaussian random field were considered. The terms split further into a pure noise component (\lq sn\rq; covering shot noise due to the finite sampling of the matter density distribution by galaxies and shape noise due to the randomly oriented intrinsic ellipticities of galaxies), a pure sampling variance contribution due to observing a finite volume of the Universe (\lq sva\rq), and a component that mixes the two (\lq mix\rq). Additional terms arise due to the non-Gaussianity of the underlying fields, which generate higher-order correlations between modes measured by the survey (captured in the \lq NG\rq\ term), as well as between modes within the survey and those on scales larger than the survey footprint (known as super-sample covariance; \lq SSC\rq). A sixth, non-cosmological contribution accounting for statistical uncertainty due to the multiplicative shear bias calibration (see Eq.~\ref{eq:cov_mbias}) completes our analytic description.

We calculate real-space correlation function covariances which are subsequently transformed to band powers analogously to the two-point statistics measurements (see Appendix~\ref{sec:cov_model_bp} for the explicit formulae). The modelling of the Gaussian terms follows the approach of \citet{joachimi08}, with the exception of the noise term which takes the actual galaxy pair counts measured from the data (cf. Appendix~\ref{sec:cov_model_gaussian}) as opposed to working with the expectation for a uniform survey \citep[cf.][]{troxel18b}. The non-Gaussian components are constructed through a halo model in line with earlier approaches (\citealp{takada13,lin14}; see also \citealp{krause17}). We refrain from implementing full halo occupation statistics for the galaxy clustering covariance terms entering the GGL covariance and its cross-correlations, and only include linear bias via $b_1$. Higher-order bias terms modify the power spectra by less than $10\,\%$, and the strongest deviations occur on scales where noise terms increasingly mitigate inaccuracies in the sample variance terms.
 
Figure~\ref{fig:cov_contribution} illustrates the contributions of the six model components to a representative subset of covariance elements. Since absolute values are plotted, the components may act to partially cancel each other. Diagonal elements are generally dominated by noise, especially on small scales. Only the GGL variances at low $\ell$ have an almost equal contribution from noise, Gaussian sample variance, and the mixed term. Away from the diagonal, the mixed term and super-sample covariance have strong impact on the level of correlation for cosmic shear. The multiplicative bias term can also reach contributions at the $50\,\%$ level, while for cross-correlations involving GGL signals it mostly dominates the budget. Only on large scales does sample variance add to correlations in this case; the non-Gaussian terms are generally of minor importance for all GGL covariance elements.

In Fig.~\ref{fig:cov_contribution_matrix} the full covariance is plotted with each element coloured according to the term making the largest contribution by absolute value. It is interesting to see that all components dominate the covariance in some parts, with the exception of the in-survey non-Gaussian term which never contributes more than $\sim 10\,\%$ to any element (the same conclusion was also reached by \citealp{barreira18}). While super-sample covariance is critical to include in the cosmic shear covariance, especially for correlations involving low-redshift source samples (top left part of the cosmic shear block) where the survey volume is smaller, it is the three Gaussian components whose accuracy drives the overall fidelity of the model. These terms are primarily assessed through the comparison with our mocks (see below). The importance of precise shear calibration is also evident: smaller uncertainty on the multiplicative bias has the potential to de-correlate most elements of the data vector and thereby improve constraining power.

\subsection{Comparison with simulated covariances}
\label{sec:covcomp}

\begin{figure}
	\includegraphics[width=\columnwidth]{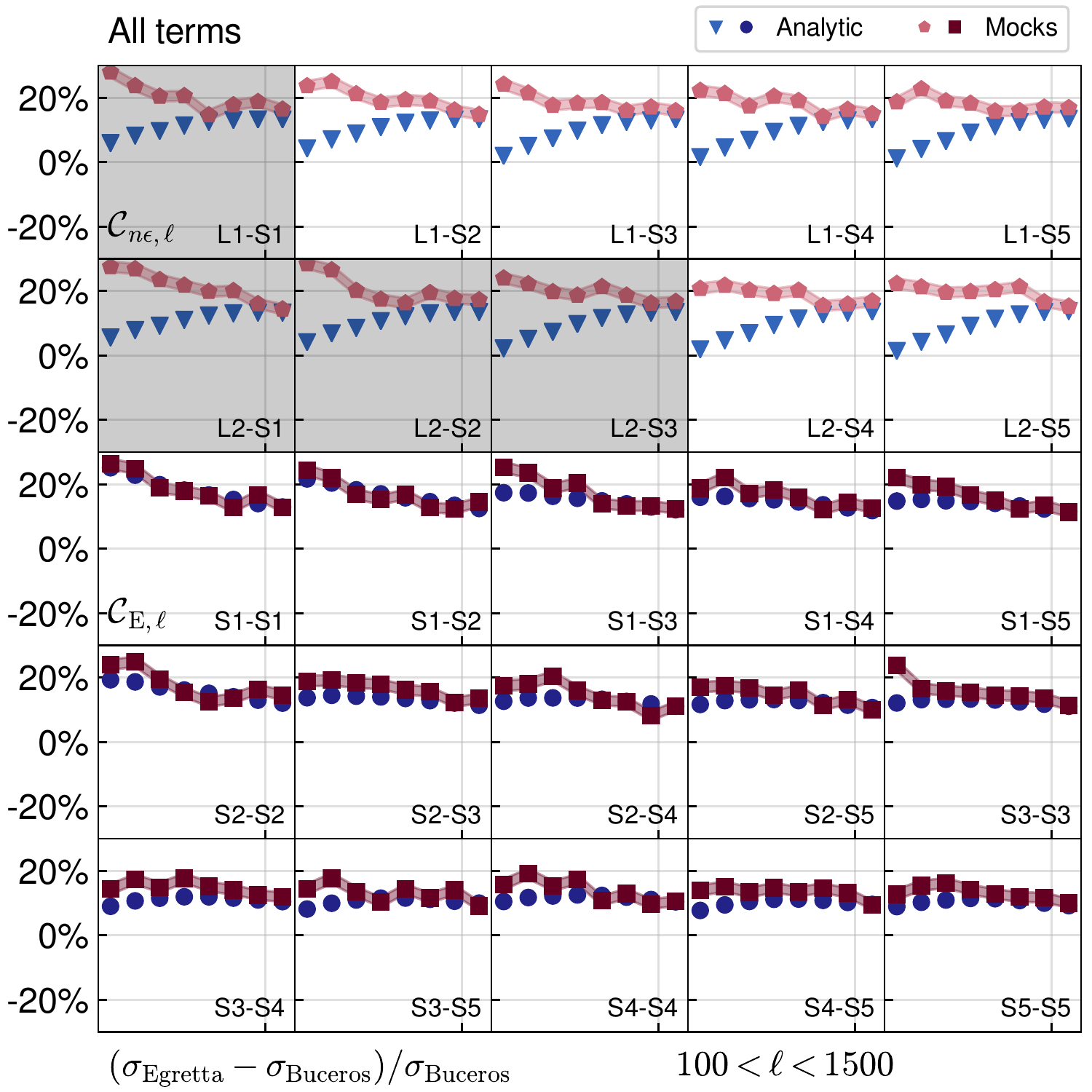}
	\caption{Relative difference between the standard deviations of different mock setups for the covariance of the weak lensing band power signals. The relative deviation of the Egretta case (realistic footprint and depth variations) from the Buceros case (rectangular mask, uniform depth) is displayed. Red (blue) symbols show results for the mock (analytic) covariance. The top two rows show GGL signals; the bottom three cosmic shear signals, with bin combinations indicated in the panels. GGL signals that are not used in the analysis have been greyed out. Bands around the mock data points indicate the standard error determined from a jackknife estimate of variance.}
	\label{fig:std_diff_egretta_footprint}
\end{figure}

\begin{figure}
	\includegraphics[width=\columnwidth]{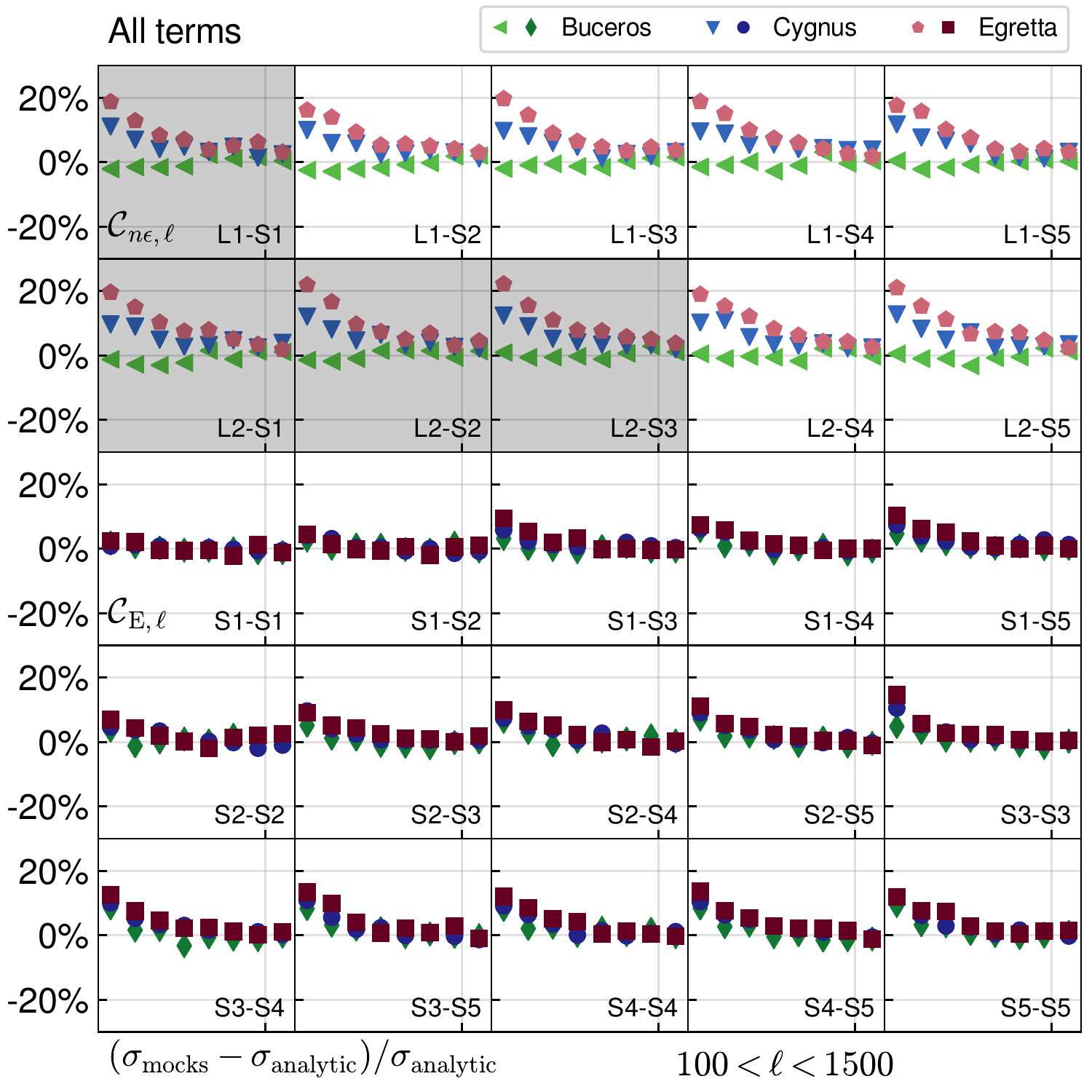}
	\caption{Relative difference between the square root of the diagonals of the mock and analytic covariances for the weak lensing band power signals. The top two rows show GGL signals; the bottom three cosmic shear signals, with bin combinations indicated in the panels. Three cases are shown: spatially uniform galaxy samples in a simple footprint (Buceros, green);  spatially uniform galaxy samples in the realistic survey footprints (Cygnus, blue); and spatially varying depth in the realistic footprints (Egretta, red). GGL signals that are not used in the analysis have been greyed out.}
	\label{fig:std_diff_egretta}
\end{figure}

The \textsc{Flask + Salmo} mocks are designed for high accuracy in the galaxy sample population properties as well as the two-point statistics of the weak lensing signal, so that they are well suited to validate the Gaussian terms in the analytic covariance. Due to the lognormal approximation of the matter density distribution in the simulations, the connected four-point function underlying the non-Gaussian covariance contributions is less accurate, which is tolerable as these terms are overall subdominant. Nonetheless, the lognormal assumption is capable of reproducing all salient features in the sample covariance of weak lensing correlation functions \citep{hilbert11}. It should be noted that, while \citet{hilbert11} assumed lognormality for the weak lensing convergence, we make the more accurate assumption of a lognormally distributed matter density field, which is known to be a very good approximation to the one-point probability density, as well as the third and fourth moments, of the matter density contrast \citep{kayo01}.

The first point we address is how the combination of the realistic, patchy survey footprint of KiDS-1000 and the spatial variations in the survey depth affect the covariance. This is shown in Fig.~\ref{fig:std_diff_egretta_footprint} where we plot the relative difference between the Egretta and Buceros setups\footnote{In all comparisons of this section we only include the covariance contributions of cosmological origin, i.e. there is no multiplicative shear bias calibration uncertainty.}. Compared to a simple, uniform survey of identical area, the square root of the diagonal of the covariance increases by about $20\,\%$, decreasing smoothly to $\sim \! 10\,\%$ excess in the highest source bin combinations for cosmic shear. This trend is driven by shape and shot noise through modifications of the number of galaxy pairs available in the correlation function measurement; super-sample covariance modifies the ratio in the opposite direction but remains subdominant on the diagonal. The analytic model accurately captures the majority of the survey property effects as it employs the pair counts measured from the real data. The only substantial deviation occurs for GGL on large scales where the analytic approach predicts much smaller changes. We find that this discrepancy is caused by the mixed noise-sample variance term ${\rm Cov}_{\rm G, mix}$ which in the analytic model disregards survey geometry effects altogether.

An alternative view is provided by Fig.~\ref{fig:std_diff_egretta} in which we directly compare mock and analytic covariances for the three survey setups Buceros, Cygnus, and Egretta. The agreement for cosmic shear is generally very good and independent of the survey setup. At small angular frequency the analytic model under-predicts the standard deviation by typically $10\,\%$, especially at higher redshift, which again originates from the idealised treatment of survey geometry in the Gaussian sample variance terms. We find these limitations to also be behind the interesting trends that emerge in the GGL parts of the covariance: mocks and theory match nearly perfectly in the case of Buceros, but deviate on large scales when switching to the realistic footprint (Cygnus) and a little further when introducing variable depth (Egretta), reaching $+20\,\%$.

\begin{figure}
	\includegraphics[width=\columnwidth]{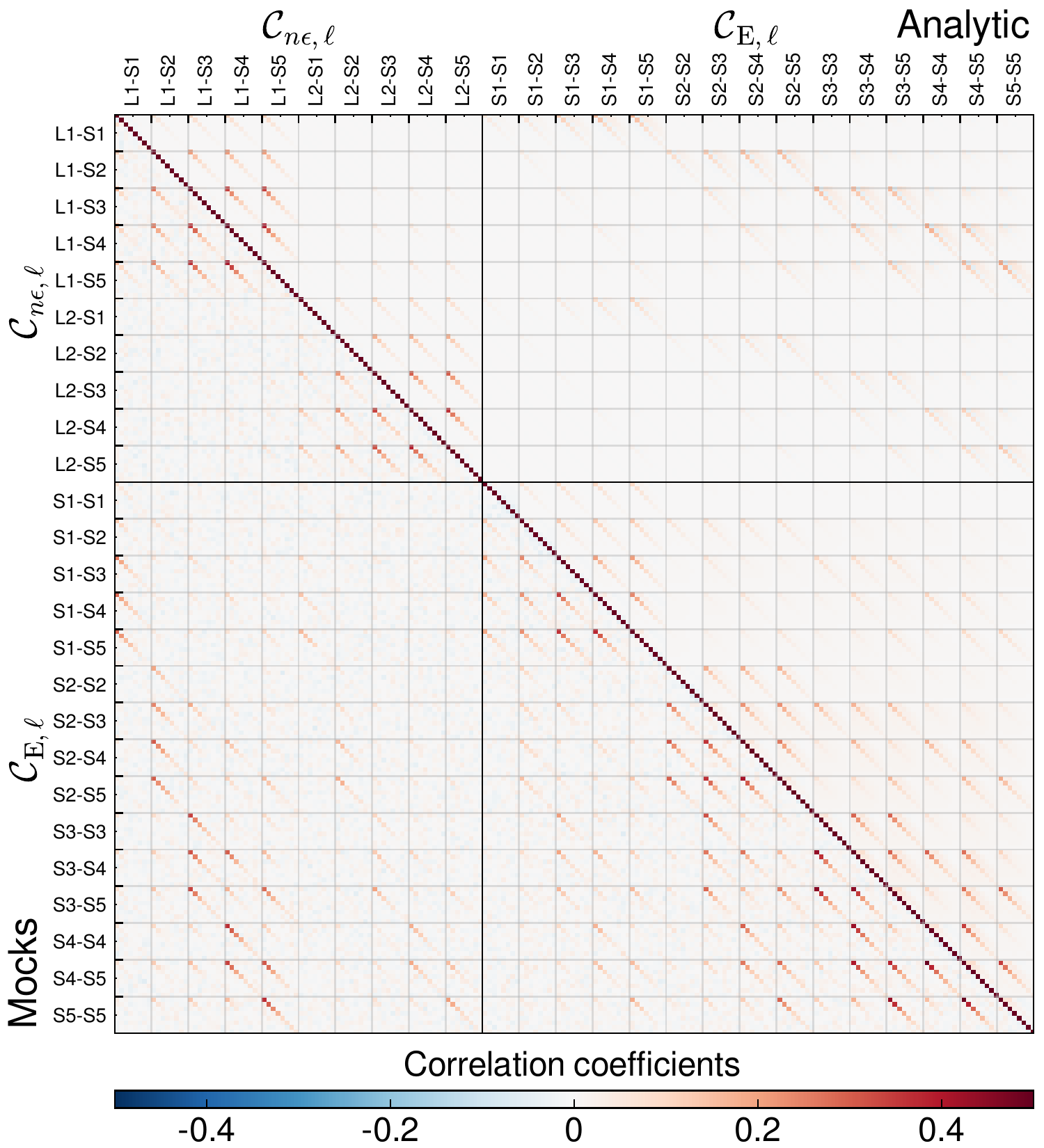}
	\caption{Comparison of correlation coefficients of the weak lensing band power covariance between mocks (lower left) and analytic covariance (upper right) in the most realistic setup (Egretta). The top left block is for GGL; the bottom right block for cosmic shear. Galaxy sample combinations are indicated on the axes.}
	\label{fig:corr_comp_egretta}
\end{figure}

We compare the off-diagonal terms via the correlation matrix shown in Fig.~\ref{fig:corr_comp_egretta}, for the most realistic (Egretta) survey setup. Generally, the correlations between the weak lensing band powers are small, not exceeding 0.5. Mock and analytic model agree to a high degree in the structure of the correlation matrix with the vast majority of elements differing in agreement with the statistical uncertainty due to the finite number of mocks. As the only clear systematic pattern, the sub-diagonals in the cross-covariance between GGL and cosmic shear are more pronounced in the mock covariance. Consulting Fig.~\ref{fig:cov_contribution_matrix}, these are terms that are also dominated by the mixed term, thus pointing to the aforementioned limitations once again.

In Appendix~\ref{sec:covcomp_app} we provide a more detailed discussion of mock and analytical covariances, including the performance of individual terms and results for the corresponding configuration-space signals. In Sect.~\ref{sec:parameter_constraints} we investigate the impact on the width of posterior parameter constraints of residual differences in our covariance models, due to different levels of realism in accounting for survey properties as well as due to limiting assumptions in our analytic model.

However, one could also incur a bias in the central value of the posterior if the covariance in the likelihood does not accurately describe the noise properties of the data vector. For a rough estimate of the magnitude of such bias in our analysis, we calculate the systematic difference between data vectors generated with noise from two different covariances labelled A and B. Assuming a Gaussian likelihood, these noisy data vectors of size $N_{\rm d}$ are given by $\vec{d}_i = \vec{\mu} + \tens{L}_i \vec{z}$ with $i \in \bc{\rm A,B}$, where $\vec{z}$ is a random variate drawn from a multivariate standard normal distribution (with zero mean and unit covariance). We denote the mean of the data vector by $\vec{\mu}$, and $\tens{L}$ is a triangular matrix obtained from the Cholesky decomposition of the covariance matrix, ${\rm Cov} = \tens{L} \tens{L}^\tau$. 

We now consider the difference between two data vectors generated with the same random variate but different covariances, $\Delta \vec{d} := \vec{d}_{\rm B} - \vec{d}_{\rm A}$. Since the mean is shared, it cancels in the difference $\Delta \vec{d}$. The expected variance of this difference is then given by
\eq{
\sigma^2_{\Delta {\rm Cov},\alpha} := \ba{ \Delta d_{\alpha}^2 } = \sum_{i=1}^{N_{\rm d}} \br{ L_{{\rm B},\alpha i} - L_{{\rm A},\alpha i}  }^2\;,
}
for an element of the data vector indexed by $\alpha$, where we exploited the relation $\ba{z_\alpha z_\beta}=\delta_{\alpha \beta}$. For the example of analytic covariances in the Buceros and Egretta setups as models A and B, we obtain for the relative shift in the weak lensing signal, that is $\sigma_{\Delta {\rm Cov},\alpha} / \mu_\alpha$, values between $4 \times 10^{-3}$ and $10^{-4}$ for GGL and values smaller than $5 \times 10^{-5}$ for cosmic shear. From this we conclude that biases due to remaining inaccuracies in the covariance model are well below the sensitivity of the current generation of surveys.

\section{Parameter inference methodology}
\label{sec:inference}

We build a dedicated inference pipeline\footnote{The KiDS Cosmology Analysis Pipeline (KCAP) will be made public on acceptance of the KiDS-1000 analysis papers.} using the \textsc{CosmoSIS}\footnote{\texttt{https://bitbucket.org/joezuntz/cosmosis/wiki}} \citep{zuntz15} analysis framework, assembling a mix of its standard components and new bespoke modules. We choose the nested sampling algorithm \textsc{MultiNest}\footnote{\texttt{https://github.com/farhanferoz/MultiNest}} \citep{feroz08,feroz09,feroz19} to explore the resulting posterior distributions, as we find it to strike a good balance between computational speed and robustness in navigating high-dimensional parameter spaces with several highly non-linear degeneracies.

\subsection{Parameters and priors}
\label{sec:parameters_priors}

\begin{table}
\caption{Choice of fiducial model parameters and priors.}              
\label{tab:priors}      
\centering                                      
\begin{tabular}{lll}          
\hline\hline                        
Parameter & Symbol & Prior \\    
\hline                                   
CDM density & $\omega_{\rm c}$ & $\bb{0.051,\,0.255}$ \\
Baryon density & $\omega_{\rm b}$ & $\bb{0.019,\,0.026}$ \\
Density fluctuation amp. & $S_8$ & $\bb{0.1,\,1.3}$ \\
Scalar spectral index & $n_{\rm s}$ & $\bb{0.84,\,1.1}$ \\
Hubble constant & $h$ & $\bb{0.64,\,0.82}$ \\
\hline
Linear galaxy bias & $b_1 \;[2]$ & $\bb{0.5,\,9}$ \\
Quadratic galaxy bias & $b_2 \;[2]$ & $\bb{-4,\,8}$ \\
Non-local galaxy bias & ${\gamma_3^-} \;[2]$ & $\bb{-8,\,8}$ \\
Virial velocity parameter & $a_{\rm vir} \;[2]$ & $\bb{0,\,12}$ \\
Intrinsic alignment amp. & $A_{\rm IA}$ & $\bb{-6,\,6}$ \\
Baryon feedback amp. & $A_{\rm bary}$ & $\bb{2,\,3.13}$ \\
\hline
Redshift offsets & $\delta_z \;[5]$ & ${\cal N}(\vec{0};C_{\delta z})$ \\
\hline
\end{tabular}
\tablefoot{The first section corresponds to the primary cosmological parameters. The second section contains astrophysical parameters related to galaxy bias, intrinsic alignments, and baryon feedback on the matter power spectrum. The third section lists observational nuisance parameters. Prior values in square brackets are the limits of top-hat priors, while ${\cal N}(\mu;C)$ corresponds to a normal prior with mean $\mu$ and (co-)variance $C$ (cf. Fig.~\ref{fig:nz_shifts}). The number in square brackets indicates how many instances of the parameter enter the analysis.}
\end{table}

In Table~\ref{tab:priors} we list our fiducial set of model parameters and their priors, with a total of 20 parameters varied for joint clustering and weak lensing analyses. The five free parameters describing a spatially flat $\Lambda$CDM cosmology are largely adopted from CMB analyses for which they are natural, which means that they are approximately independently constrained with clear links to features in the observables \citep{kosowsky02}. This is not necessarily true for low-redshift probes of large-scale structure. In particular, weak lensing is primarily constrained through the amplitudes of its otherwise smooth two-point signals, described by a non-linearly degenerate combination of $\Omega_{\rm m}$ and $\sigma_8$. It has therefore become customary to report constraints from cosmic shear in terms of the parameter $S_8 = \sigma_8\,(\Omega_{\rm m}/0.3)^{1/2}$, approximately measuring the lateral width of the posterior across the aforementioned degeneracy. We note however that the exact direction of the $\Omega_{\rm m}$ -- $\sigma_8$ degeneracy is affected by the source redshift range of the data set, as well as the range and weighting of angular scales entering the likelihood.

\begin{figure}
	\includegraphics[width=\columnwidth]{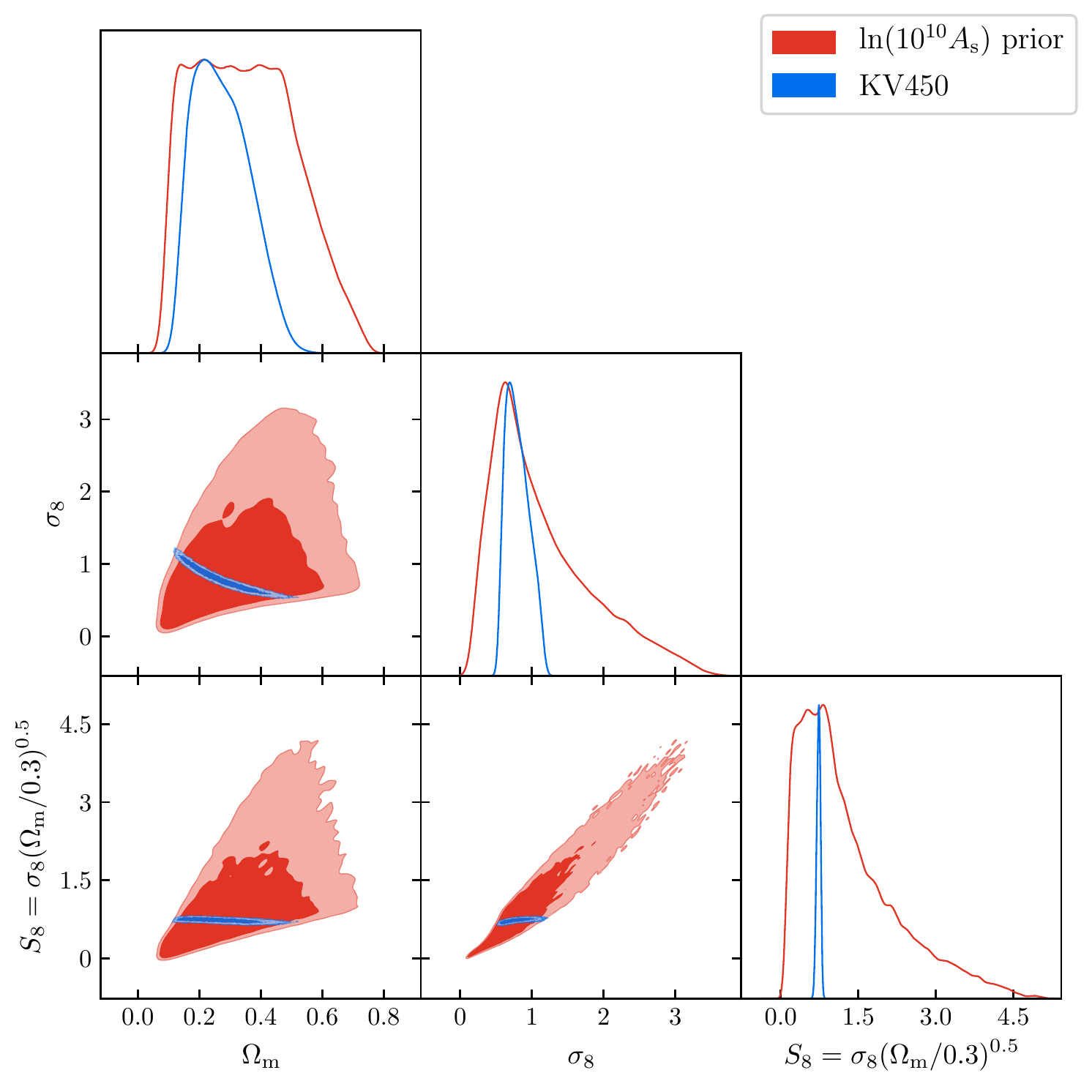}
	\caption{Prior volume resulting from using wide top-hat priors on the sampling parameters $\bc{\ln A_{\rm s},\omega_{\rm c},\omega_{\rm b},n_{\rm s},h}$ (in red), shown together with the KV450 posterior (in blue). Dark (light) shades correspond to the two-dimensional $1\sigma$ ($2\sigma$) credible regions. The prior does not impact on marginal $S_8$ constraints, but is informative in the $\Omega_{\rm m}$ -- $S_8$ and $\Omega_{\rm m}$ -- $\sigma_8$ planes.}
	\label{fig:priors_lnAs}
\end{figure}

\begin{figure}
	\includegraphics[width=\columnwidth]{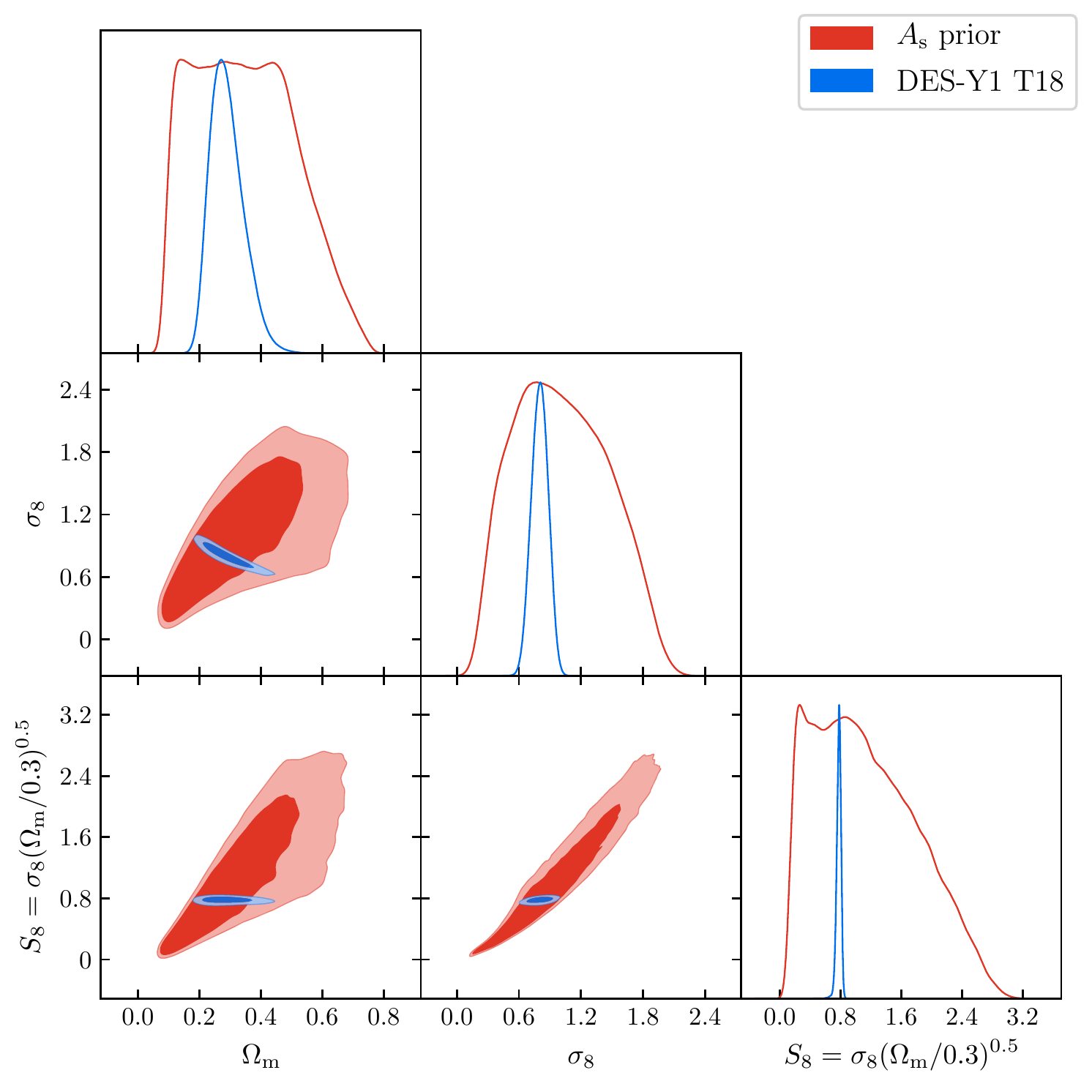}
	\caption{Same as Fig.~\ref{fig:priors_lnAs}, but now showing the prior volume using the sampling parameter $A_{\rm s}$ with a wide top-hat prior, alongside the DES Year 1 cosmic shear posterior. It should be noted that the axis range for the $\sigma_8$ and $S_8$ parameters is smaller than in Fig.~\ref{fig:priors_lnAs}.}
	\label{fig:priors_As}
\end{figure}

Standard sampling parameters for the CMB, like the amplitude of the primordial power spectrum of scalar density fluctuations, $A_{\rm s}$, and the physical cold dark matter density parameter, $\omega_{\rm c}$, remain largely unconstrained individually. The choice of priors for these parameters therefore directly impacts on the posteriors of weak lensing cosmological analyses, as illustrated in Figs.~\ref{fig:priors_lnAs} and \ref{fig:priors_As}. They show posteriors of the key cosmological parameters $\Omega_{\rm m}$, $\sigma_8$, and $S_8$ for the cosmic shear analyses of KV450 and DES Year 1 \citep{troxel18}, respectively. Alongside, we plot the corresponding prior volumes resulting from the cosmological sampling parameters chosen in each case, mapped by drawing $500\,000$ samples. The analyses assumed wide top-hat priors for the standard flat $\Lambda$CDM cosmological parameters, which we adopt here; in particular, KV450 used $\ln ( 10^{10} A_{\rm s})$ as a sampling parameter with prior range $[1.7,\,5]$, while DES Year 1 used $A_{\rm s}$ in the range $[ 5 \times 10^{-10},\, 5 \times 10^{-9} ]$. These assumptions translate into complex prior volumes for the derived low-redshift parameters. Figures~\ref{fig:priors_lnAs} and \ref{fig:priors_As} highlight that the extent of the $\Omega_{\rm m}$ -- $\sigma_8$ degeneracy, and thus the marginal posterior width for $\Omega_{\rm m}$, are almost entirely determined by the prior and not influenced significantly by the different constraining powers of the data sets (see also \citealp{joudaki20} for a discussion of the impact of the different $A_{\rm s}$ priors) .

In light of the observed tension in $S_8$ constraints from \textit{Planck} and galaxy weak lensing probes it is particularly important to avoid implicit informative priors on this parameter. Indeed, Figs.~\ref{fig:priors_lnAs} and \ref{fig:priors_As} highlight that both the KV450 and DES Year 1 setups disfavour high values of $S_8 \gtrsim 1$ a priori. However, the $S_8$ marginal posteriors remain unaffected by the prior as both data sets are highly constraining, and the $S_8$ prior is flat across the region with substantial posterior mass\footnote{By probability mass, we refer to the probability density function (PDF) integrated over certain regions of the PDF's domain.}. To guard against any undue impact by the prior choice and to simplify the parameter volume that needs to be explored by the sampler, we choose $S_8$ as our sampling parameter instead of $A_{\rm s}$ or $\ln A_{\rm s}$ and impose a top-hat prior that comfortably encompasses the full range of the previous, less constraining weak lensing analyses, as well as the \textit{Planck} $S_8$ posterior. We further assess the impact of the choice of fluctuation amplitude parameter for sampling in Sect.~\ref{sec:parameter_constraints}.

The other cosmological sampling parameters remain the same as in KV450, with unchanged wide priors for $\omega_{\rm b}$ and $h$. We tighten the prior range for $n_{\rm s}$ in order to avoid regions of very small expected posterior mass at large values of $n_{\rm s}$, where evaluations of the S17 clustering likelihood become prohibitively slow. To avoid prior volume artefacts in this weakly constrained parameter, we then symmetrise the top-hat prior around the theoretical expectation of values just below unity and set $n_{\rm s} \in \bb{0.84,\,1.1}$.

Since the prior for $\omega_{\rm c}$ directly impacts the inference on $\Omega_{\rm m}$ and $\sigma_8$, we seek to derive a more physically motivated prior range than the very wide $\omega_{\rm c} \in \bb{0.01,\, 0.99}$ chosen in KV450. We opt to motivate the prior from the independent measurement of luminosity distance via Supernova Type Ia measurements with the Pantheon sample \citep{scolnic18}. Their analysis yielded $\Omega_{\rm m} = 0.298 \pm 0.022$ for a flat $\Lambda$CDM model, which we translate into the top-hat prior range $\Omega_{\rm m} \in \bb{0.188,\,0.408}$ that encompasses the $\pm 5\,\sigma$ uncertainty of the marginal constraint including systematic errors. The boundaries of the $\Omega_{\rm m}$ prior are then translated into those for $\omega_{\rm c}$ using the extremes of the prior ranges of $\omega_{\rm b}$ and $h$. This choice also has the benefit of excluding parameter combinations for which the matter power spectrum model underlying all large-scale structure probe predictions is essentially untested -- for instance, the Coyote Universe \citep{heitmann14} suite of simulations used to calibrate the \citet{mead15} non-linear model is limited to the ranges $\Omega_{\rm m} h^2 \in \bb{0.12,\,0.155}$ and $\sigma_8 \in \bb{0.616,\,0.9}$.

The priors for the astrophysical parameters are adopted from KV450 and \citet{troester19} respectively for IA and galaxy bias. The intrinsic redshift dependence of the IA signal is by default kept fixed at $\eta_{\rm IA}=0$. We follow S17 in tying the otherwise poorly constrained non-local bias parameter $\gamma_2$ to the linear bias,
\eq{
\gamma_2 = - \frac{2}{7} \br{b_1 -1}\;.
}
This relation is physically motivated and holds to sufficient accuracy in realistic simulations. While a similar relation exists for $\gamma_3^-$, it is varied in the inference as a catch-all parameter for other non-linear power spectrum contributions not explicitly modelled in the perturbative approach. Finally, we include one offset parameter $\delta_z$ each for S1-5 that shifts the source redshift distributions within the multivariate Gaussian prior discussed in Sect.~\ref{sec:photo-z}.

\subsection{Gaussian likelihood assumption}
\label{sec:gaussian_likelihood}

\begin{figure}
	\includegraphics[width=\columnwidth]{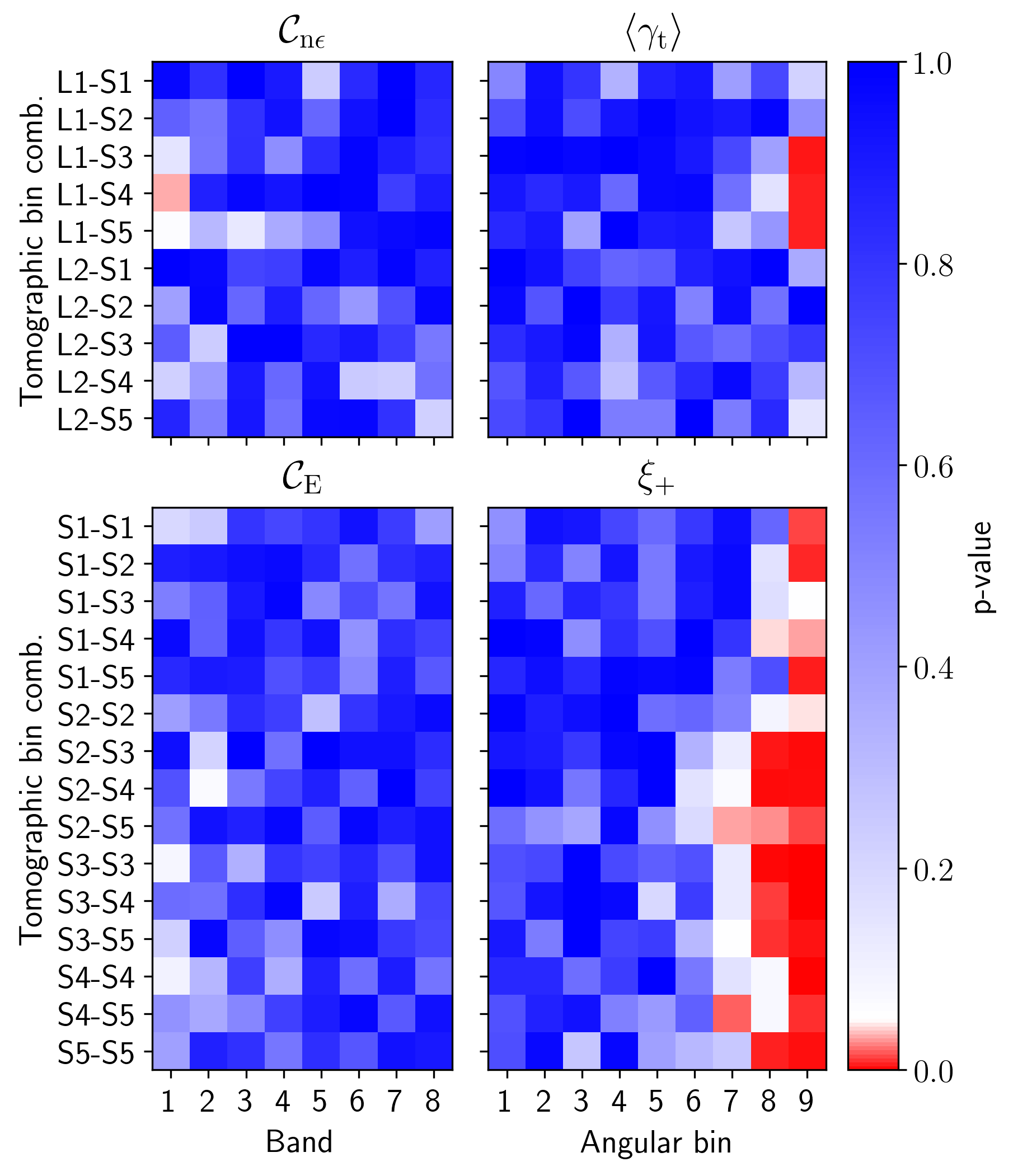}
	\caption{$p$-values for a KS-test of Gaussianity of the marginal likelihood distributions. Values above the typical threshold of 0.05 are blue; values below are red. Left panels show band powers for cosmic shear (bottom) and galaxy-galaxy lensing (top) for various tomographic bin combinations and angular frequency bands, with the largest scales on the left. Right panels display corresponding real-space correlation functions, with the largest scales on the right.}
	\label{fig:likelihood_pvalues}
\end{figure}

Along with the vast majority of large-scale structure cosmological analyses, we adopt a multivariate Gaussian likelihood. This is expected to be a generally excellent approximation if the summary statistics entering the likelihood have been averaged over many modes in the underlying fields. Exact likelihood expressions for two-point statistics of the non-Gaussian matter distribution and its tracers are unknown, and since only the central value and the covariance are usually readily calculable, a normal distribution is the least informative form of likelihood. Due to the small number of modes entering power spectra at low angular frequencies, significant deviations from Gaussianity occur \citep[e.g.][]{hamimeche08}, which propagate into configuration-space statistics that are sensitive to these modes, such as the cosmic shear correlation function $\xi_+$ at large angular separation \citep{schneider09,sellentin18,lin19}.

It is therefore an additional benefit of band powers that they have sensitivity over a compact range of angular frequencies, allowing us to reduce the impact of scales below $\ell \lesssim 50$ to a minimum; see Fig.~\ref{fig:bandpower_filters}. To demonstrate the impact, we create sampling distributions of both band power and configuration-space statistics measured from $1000$ mock realisations. We test the one-dimensional likelihoods of the elements of our data vector individually for Gaussianity, applying a Kolmogorov-Smirnov (KS) test to the standardised distributions and setting a threshold $p$-value of 0.05. As shown in Fig.~\ref{fig:likelihood_pvalues}, the band power likelihoods are consistent with being Gaussian, with a slight trend towards smaller $p$-values for the low-$\ell$ bands. In contrast, the correlation function likelihoods (in the standard KV450 setup with 9 logarithmically spaced bins in the range from $0.5$ -- $300\,{\rm arcmin}$) fail in the two to three largest scale bins for $\xi_+$ and to a lesser degree in $\gamma_{\rm t}$. We note in passing that our test is necessary but not sufficient because the multivariate distribution of band powers could be non-Gaussian despite Gaussian marginals. In future one could additionally apply tests that are sensitive to higher-order moments of the distribution \citep[see e.g.][]{sellentin18b}. 

As we can proceed with a Gaussian likelihood for the band-power statistics, it is important to emphasise that the covariance should not be varied alongside the mean when sampling the parameter space. It is a fundamental property of the normal distribution that its sample mean and covariance are statistically independent. The fact that in large-scale structure analyses both the mean (i.e. the model) and the covariance depend on cosmological and other parameters (primarily through sample variance) is a relic of the true, non-Gaussian sampling distributions that we approximate. Varying parameters in the covariance in the posterior sampling of a Gaussian likelihood would generate spurious, albeit small, extra constraining power because the parameter sensitivity of the covariance is treated erroneously as an independent source of information (see the discussion in \citealp{carron13}).

If the parameters are not varied in the covariance, the question arises which combination of parameters is the correct one to choose, as particularly amplitude-changing parameters like $\sigma_8$ or galaxy bias can  modify sample variance contributions considerably \citep[e.g.][]{harnois19}. \citet{uitert18} used an iterative approach to solve the issue, starting with a fiducial set of parameters and then repeating the likelihood analysis with the covariance re-calculated at the best-fit values of the previous iteration. In a forthcoming publication we will demonstrate that this procedure robustly converges to the correct posterior. Since we have best-fit values from the \citet{troester19} analysis for guidance, we expect that at most one iteration will be required for convergence on the key cosmological parameter constraints. For the mock analysis in this work, we evaluate the covariance at the fiducial parameter values.

If a mock-derived covariance is used in the likelihood, we correct for noise bias in the mean of the inverse covariance \citep{kaufman67,hartlap07}. Due to the large number of realisations, the correction is only $4\,\%$ and has negligible impact on the posterior.

\subsection{Goodness of fit}
\label{sec:gof}

The goodness of fit is a key statistic to judge how well the combination of the model and the assumed likelihood (including the covariance) describe the data. Traditionally, the weighted least squares, or $\chi^2$, statistic is evaluated at a choice of best fit of the sampling parameters and interpreted under the assumption that it follows a $\chi^2$ distribution with $k_{\rm dof} = N_{\rm d} - N_\Theta$ degrees of freedom. Here, $N_{\rm d}$ is again the size of the data vector $\vec{d}$ entering the likelihood, varying between 120 for cosmic shear only and 310 (adding 168 for clustering and 22 for GGL) for the joint analysis in their default setup, and $N_\Theta$ is the number of sampling parameters, where $N_\Theta = 12$ for cosmic shear and $N_\Theta = 20$ for the joint analysis. The goodness of fit is then quantified by the probability to exceed (PTE) the value of $\chi^2$ at the best fit, that is ${\rm Pr} ( \chi^2 > \chi^2_{\rm best} )$.

The assumption of the particular form of $\chi^2$ test statistic is only valid if (1) the underlying data is normally distributed, (2) the sampling parameters enter the model linearly, and (3) the parameter ranges are not significantly restricted by priors. Neither of these assumptions generally hold for cosmological analyses, and ours is no exception. We therefore need to establish a more accurate form of the sampling distribution of the best-fit $\chi^2$. This distribution can be built empirically from our mocks; if mocks are unavailable, realisations of $\chi^2_{\rm best}$ could analogously be generated directly from the data analysis via the posterior predictive distribution \cite[cf.][]{gelman92,gelman96}.

\begin{figure}
	\includegraphics[width=\columnwidth]{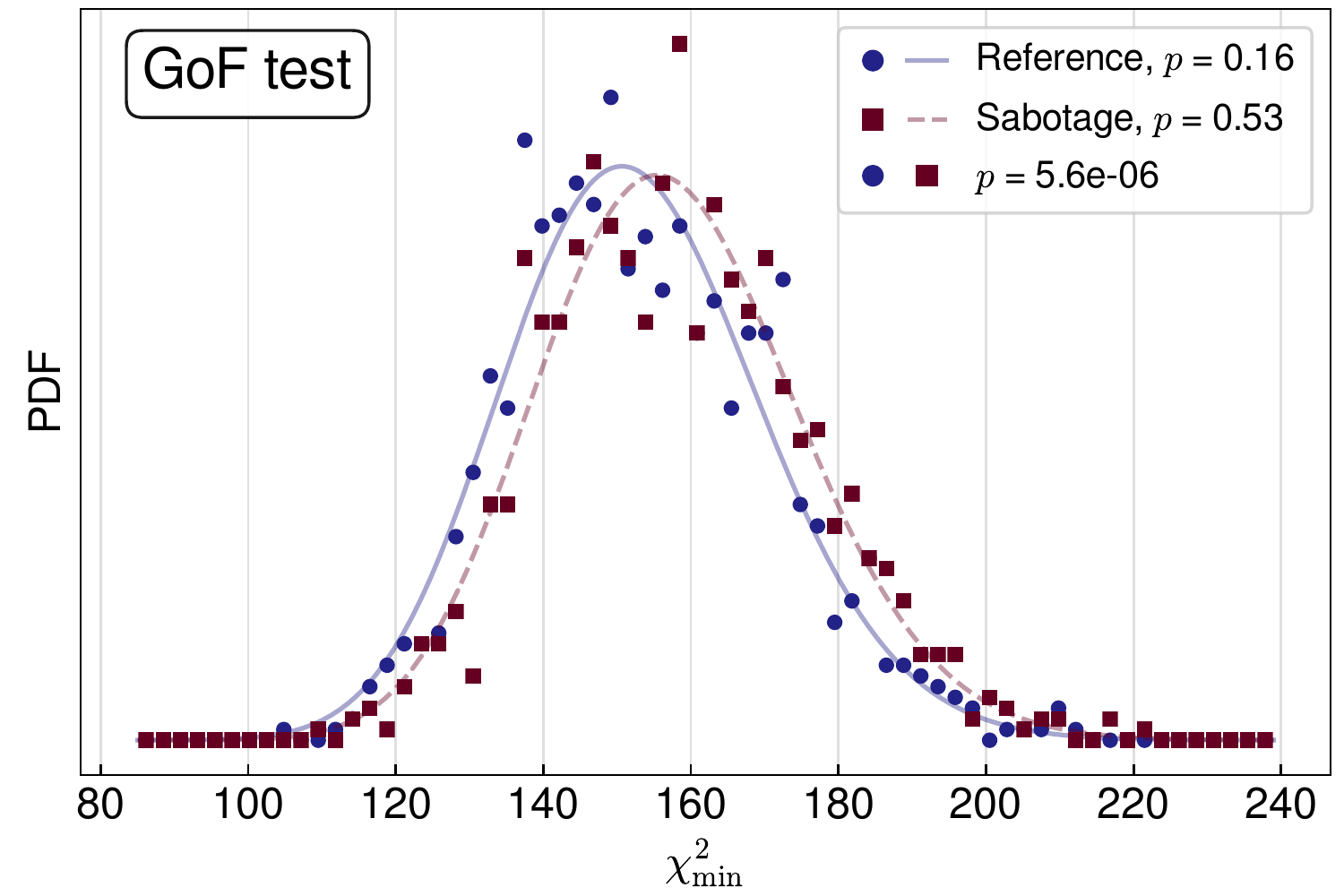}
	\caption{Goodness of fit sampling distributions. Shown are histograms of minimum $\chi^2$-values obtained from $1000$ mock realisations of data vectors analysed with our cosmic shear and GGL analysis pipeline. Blue points correspond to our baseline setup; red points to the result using a sabotaged data vector computed with the S3 and S4 redshift distributions systematically shifted upwards by $5\sigma$ of their Gaussian prior width. The curves are best fits of a $\chi^2$-distribution with its mean as free parameter. The solid blue curve is a good fit to the baseline results ($p$-value of 0.16), while the sabotaged distribution is still well fitted by a $\chi^2$-distribution (red dashed curve; $p=0.53$) which however is incompatible with the baseline ($p < 10^{-5}$).}
	\label{fig:chiSq_goodness}
\end{figure}

For every mock (or posterior predictive) realisation we maximise the posterior, or alternatively minimise the log-likelihood, to obtain an estimate of $\chi^2_{\rm best}$. An example of a resulting probability density function (PDF) created from $1000$ mocks is shown in Fig.~\ref{fig:chiSq_goodness} for the case of joint cosmic shear and GGL. We could now directly measure the compatibility of the $\chi^2_{\rm best}$ determined from the real-data analysis with being a sample drawn from this PDF via ${\rm Pr} ( \chi^2_{\rm best, mock} > \chi^2_{\rm best, obs} )$. However, for the joint clustering and weak lensing analysis in particular, we struggled to identify readily available optimisation algorithms\footnote{We did not explore algorithms that require function derivatives explicitly, such as conjugate gradient approaches, but recommend this for future work.} that ran fast enough to be applicable to a large number of mock realisations. Fortunately, in the cases where the $\chi^2_{\rm best}$ PDF is obtainable, it is still well described by a $\chi^2$-distribution; see Fig.~\ref{fig:chiSq_goodness}, and it is fair to assume that this remains the case for the most comprehensive analysis setup. The shape of the distribution is retained even for a substantially under-fitting model as shown by the red points and curve in Fig.~\ref{fig:chiSq_goodness}.

Therefore, the problem reduces to finding an estimate of the one free parameter of the $\chi^2_{\rm best}$ sampling distribution, which we denote by $k_{\rm dof, eff} := N_{\rm d} - N_{\Theta, {\rm eff}}$. This task can be intuitively understood as obtaining an effective number of constrained model parameters, $N_{\Theta, {\rm eff}}$, which will be smaller than the raw number of parameters due to parameter degeneracies and informative priors that severely restrict the freedom of a parameter to optimise the model.  We investigate a number of estimators for $N_{\Theta, {\rm eff}}$ that have seen applications in the astrophysical literature:
\eqa{
\label{eq:npar_eff_1}
N_{\Theta, \rm var} &= 2 \bb{ \ba{  \br{\chi^2}^2 }_{{\rm Pr} (\vec{\Theta} | \vec{d})} - \ba{ \chi^2 }^2_{{\rm Pr} (\vec{\Theta} | \vec{d})} }\;; \\
\label{eq:npar_eff_2}
N_{\Theta, \rm like} &= \ba{ \chi^2 }_{{\rm Pr} (\vec{\Theta} | \vec{d})} - \chi^2_{\rm min}\;; \\
\label{eq:npar_eff_3}
N_{\Theta, \rm post} &= \ba{ \chi^2 }_{{\rm Pr} (\vec{\Theta} | \vec{d})} - \chi^2\br{{\rm argmax}\bb{{\rm Pr} (\vec{\Theta} | \vec{d})}}\;,
}
where ${\rm Pr} (\vec{\Theta} | \vec{d})$ denotes the posterior, and angular brackets indicate an average over the posterior distribution (or samples thereof). Equation~(\ref{eq:npar_eff_1}) was recently proposed by \citet{handley19}. The other two expressions are variants of the basic estimator introduced by \citet{spiegelhalter02}, which employs a point estimate of $\chi^2$ at a point where \lq information is maximised\rq. We experiment with the maximum likelihood (Eq.~\ref{eq:npar_eff_2}) and the mode of the posterior (Eq.~\ref{eq:npar_eff_3}; see \citealp{raveri19} for a recent application), the former attaining the global minimum $\chi^2$ and the latter more attentive to the informative priors in our analysis. The posterior mean is a potential alternative \citep{kunz06}, but as it would always yield lower values of $N_{\Theta, {\rm eff}}$ than Eq.~(\ref{eq:npar_eff_3}), it does not perform favourably in our setup.

\begin{table}
\caption{Estimates of the effective number of model parameters.}              
\label{tab:gof}      
\centering                                      
\begin{tabular}{lllllll}          
\hline\hline                        
No. of pars. & \multicolumn{3}{l}{Cosmic shear} & \multicolumn{3}{l}{All WL} \\    
estimate & $\mu$ & $\sigma$ & $\alpha_p$ & $\mu$ & $\sigma$ & $\alpha_p$ \\
\hline                                   
$N_\Theta$ & 12 & - & 22.6 & 18 & - & 32.8 \\
$N_{\Theta, {\rm var}}$ & 3.0 & 0.5 & $-$4.0 & 7.4 & 1.3 & $-$4.3 \\
$N_{\Theta, {\rm like}}$ & 5.4 & 1.3 & 3.2 & 10.7 & 1.8 & 7.2 \\
$N_{\Theta, {\rm post}}$ & 3.5 & 0.6 & $-$2.3 & 7.6 & 1.1 & $-$3.9 \\
none & 0 & - & $-$12.6 & 0 & - & $-$29.4 \\
\hline
$N_{\Theta, {\rm eff}}$ & 4.5 & 0.8 & 0.4 & 9.1 & 1.3 & 1.6 \\
from $\langle \chi^2 \rangle$ & 4.5 & 0.6 & 0.4 & 8.7 & 0.4 & 0.2 \\
\hline
\end{tabular}
\tablefoot{Rows correspond to the total number of parameters varied in the analysis, $N_\Theta$, the three estimators for the effective number of parameters given in Eqs.~(\ref{eq:npar_eff_1}) to (\ref{eq:npar_eff_3}), and the assumption of no parameters varied. We also show our adopted estimator of $N_{\Theta, {\rm eff}} = (N_{\Theta, {\rm like}}+N_{\Theta, {\rm post}})/2$ and the effective number of parameters as inferred from the mean of the sampling distribution of minimum $\chi^2$-values (as in Fig.~\ref{fig:chiSq_goodness}). Columns denote the central value $\mu$, the standard deviation $\sigma$, and the quality of match to the $\chi^2_{\rm min}$ sampling distribution, $\alpha_p$ (cf. Eq.~\ref{eq:alphap}).}
\end{table}

\begin{figure*}
\includegraphics[width=\columnwidth]{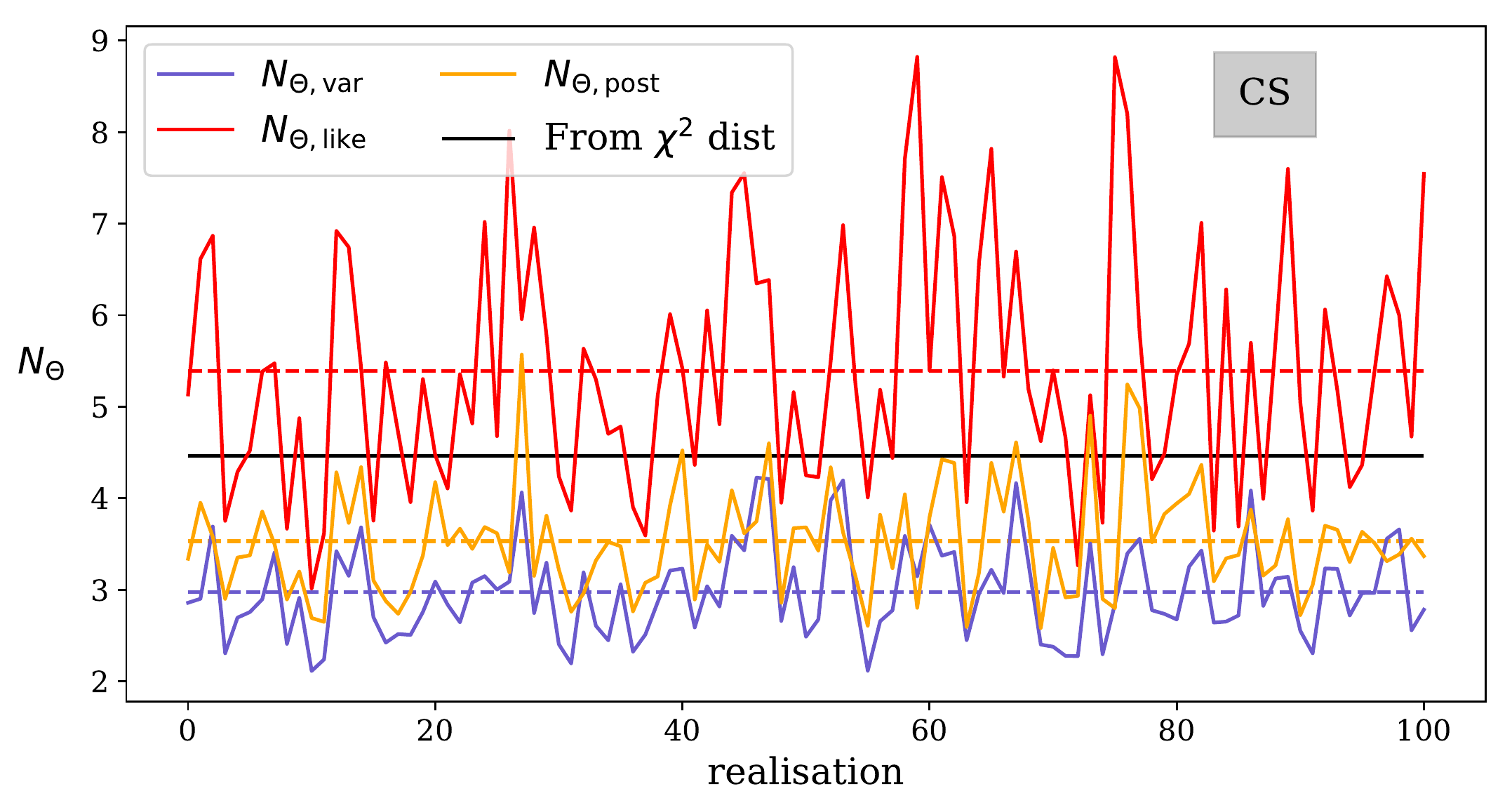}
\includegraphics[width=\columnwidth]{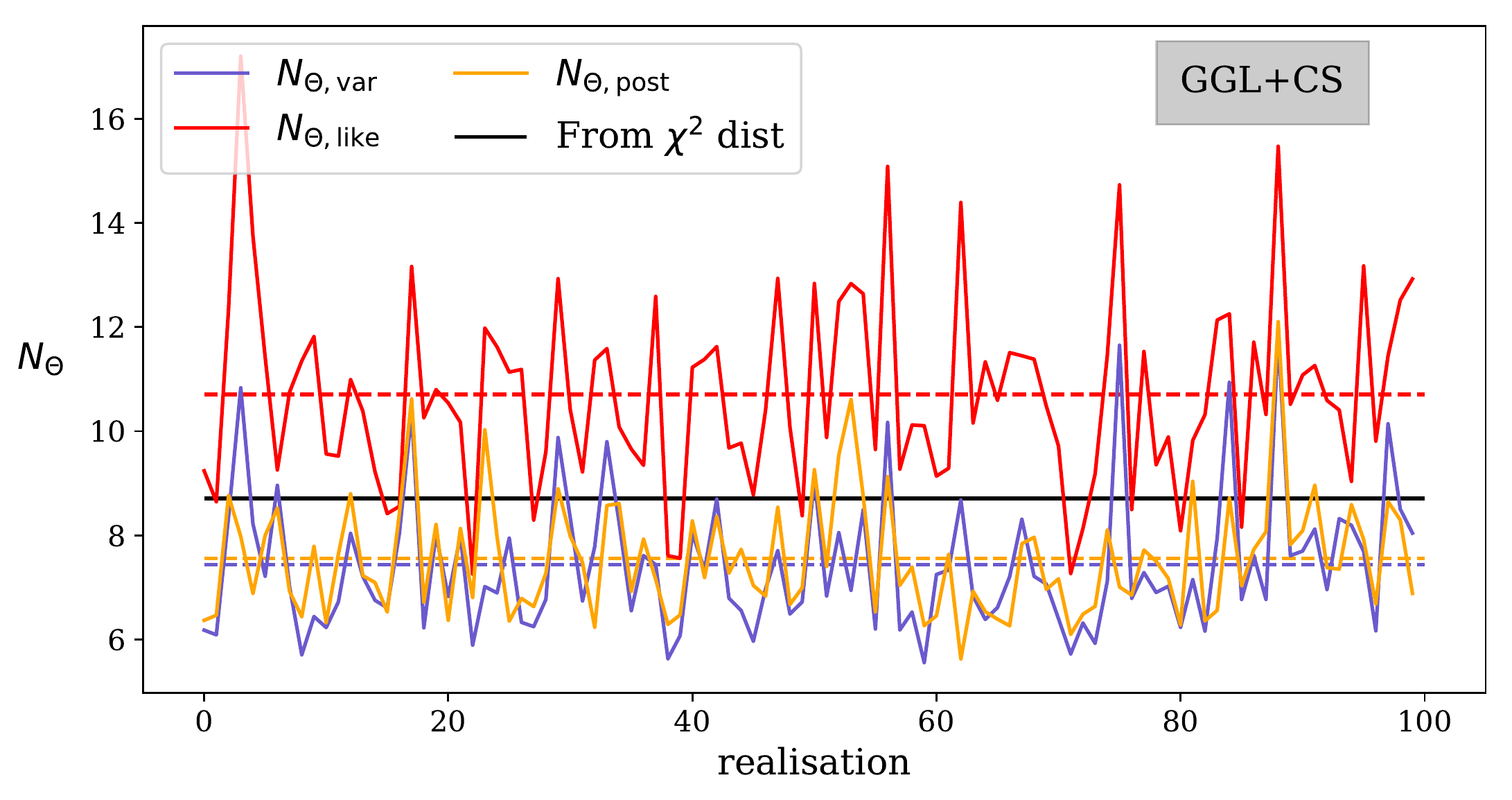}
\caption{Estimates of the effective number of parameters for 100 realisations of a cosmic shear \textit{(left)} and a joint cosmic shear+GGL \textit{(right)} data vector. The estimators defined in Eqs.~(\ref{eq:npar_eff_1}) to (\ref{eq:npar_eff_3}) are shown in blue, red, and orange, respectively. The dashed lines mark the average over the realisations. The black line marks the effective number of parameters inferred from the fit of a $\chi^2$-distribution to a histogram of $\chi^2_{\rm min}$ (cf. Eq.~\ref{eq:alphap}).}
\label{fig:par_eff}
\end{figure*}

Figure~\ref{fig:par_eff} shows the performance of the estimators in relation to $N_{\rm d} - k_{\rm dof,eff}$ as inferred from the $\chi^2_{\rm min}$ sampling distribution (cf. Fig.~\ref{fig:chiSq_goodness}). Neither estimator is a good match per se. Since $N_{\Theta, \rm var}$ avoids a point estimate, it is the least noisy, but also furthest from the target. We saw hints that the accuracy of this estimator is significantly affected by the presence of informative priors. The other two estimators bracket the target in both scenarios that we can assess, and since neither is a priori more preferable, we resort to define our final estimate as their average, $N_{\Theta, {\rm eff}} := (N_{\Theta, {\rm like}}+N_{\Theta, {\rm post}})/2$.

Table~\ref{tab:gof} quantifies the performance of the various parameter-number estimates, listing their central values, standard deviations, and the quality measure
\eq{
\label{eq:alphap}
\alpha_p := \frac{1 - 2 \mu_p}{\bar{\sigma}_p}\;,
}
with $\mu_p$ and $\bar{\sigma}_p$ the mean of the $p$-values and its standard error calculated for a fit of a $\chi^2$-distribution with $k_{\rm dof,eff}$ degrees of freedom to the $\chi^2_{\rm min}$ sampling distribution. Since the $p$-value is expected to follow a uniform distribution in $[0; 1]$, the expectation value of $\mu_p$ is 0.5 and its standard error is $1/\sqrt{12}$, so that the error on the mean is $\bar{\sigma}_p \approx 0.03$ for the 100 realisations that we use. Smaller absolute values of $\alpha_p$ are preferable, with $|\alpha_p|<1$ indicating that the bias of the estimator is subdominant to its statistical error.

Our choice of estimator performs well in this regard with $\alpha_p \lesssim 1$, while the other effective parameter number measures with absolute values of $\alpha_p$ between 2 and $\sim 4$, and the traditional $N_\Theta$ with $\alpha_p \approx 23$ and 33 for cosmic shear and the combined weak lensing signals, respectively, are strongly disfavoured. The latter performs even slightly worse than assuming no free parameters at all. However, it should be noted that the $N_{\Theta, {\rm eff}}$ is quite noisy if determined from a single maximum likelihood estimate, so that combining at least a few realisations (from mocks or the posterior predictive distribution) is recommended.

It is interesting to note that the effective number of constrained parameters, defined as the equivalent dimension of a linear, unconstrained parameter space, is 4.5 for cosmic shear and approximately 9 for cosmic shear and GGL combined. The classic approach to goodness of fit, employing $N_\Theta$ to calculate the reduced $\chi^2$, thus yields a conservative test statistic if model under-fitting is the primary concern. This means that previous cosmological analyses with similarly structured model parameter spaces tended to report reduced $\chi^2$ values whose expectation was slightly larger than unity. The change due to using $N_{\Theta, {\rm eff}}$ is mild for our fiducial KiDS-1000 setup (e.g. $7\,\%$ in the reduced $\chi^2$ for cosmic shear only), but could become important for a smaller data vector, for instance if data compression is applied.

\subsection{Reporting parameter constraints}
\label{sec:reporting_constraints}

In cosmology it is widely accepted to adopt the Bayesian paradigm for inference and provide the posterior distribution, usually in the form of samples, as the final deliverable of an experiment. Nonetheless, point estimates of some notion of best-fit parameter value are indispensable to report headline results or compare with other experiments at a high level. Moreover, researchers are often interested in the recovery of a \lq true\rq\ underlying value of model parameters. This is perhaps most obvious in the validation of an analysis pipeline -- we seek to demonstrate that we faithfully recover the input parameters of the mock data.

\begin{figure}
	\includegraphics[width=\columnwidth]{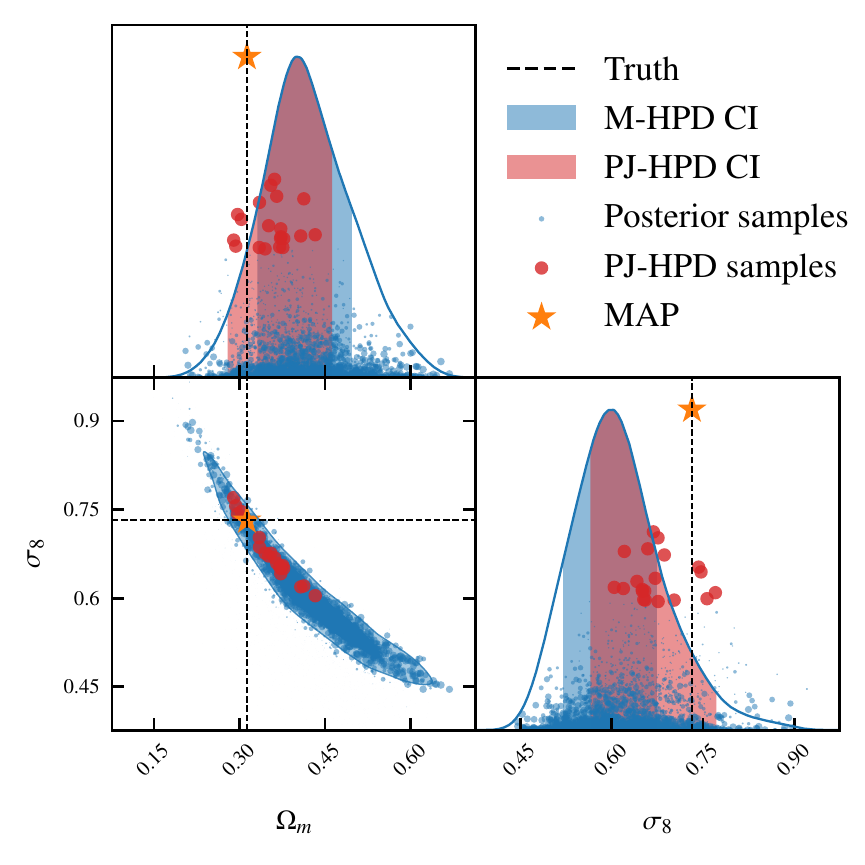}
	\caption{Illustration of $1\sigma$ credible intervals (CI) used to report parameter constraints, for a subset of two parameters for an exemplary mock joint analysis of GGL and cosmic shear with a noiseless data vector. The standard marginal highest posterior density (M-HPD) CIs are shown in blue. The marginal posterior mode is shifted with respect to the true input parameter values (black dashed lines), whereas the maximum a posteriori (MAP) estimate (orange star) tracks the truth well. The CI we associate with the MAP is constructed via the PJ-HPD method (red), using the samples with the highest posterior density (red points). Other posterior samples are shown as blue points, with point size proportional to their posterior mass. In the top and right panels the height of the points and of the MAP indicate their posterior density with arbitrary scaling.}
	\label{fig:reporting_constraints}
\end{figure}

Usually, point estimates are extracted from the one-dimensional marginal distributions of the parameters (as this low-dimensional distribution is well sampled), typically the mode, median, or mean combined with a credible interval that encompasses a defined fraction of the highest marginal posterior density. While these point estimates are unambiguous in their Bayesian interpretation, they do not necessarily peak at, or even cover within a given credible interval, the true parameter value\footnote{Pathological examples with arbitrarily low marginal posterior density at the position of the maximum of the joint posterior are straightforward to construct.}. In our case this is a consequence of a high-dimensional parameter space with multiple, non-linear near-degeneracies, in addition to a wide prior with complex shape. Figure~\ref{fig:reporting_constraints} displays a pertinent example: the strong, banana-shaped degeneracy between $\Omega_{\rm m}$ and $\sigma_8$ present in the posterior of cosmic shear analyses.

We therefore seek to employ a complementary point estimate that more accurately reports the global best fit to the data at hand. The estimate itself is simply the set of parameter values at the maximum (multivariate) posterior, denoted by MAP (maximum a posteriori):
\eq{
	\label{eq:map-def}
	\vec\Theta_{\mathrm{MAP}} = \argmax_{\vec\Theta} {\rm Pr} (\vec{\Theta} | \vec{d}) \;,
}
where $\argmax$ returns the argument of a function at which it attains its maximum. The posterior maximisation advocated in Sect.~\ref{sec:gof} for determining the goodness of fit also provides a precise MAP estimate. In practice, we find good recovery of the global maximum by taking the largest posterior returned by a suite of tens of optimisation runs, each started at the position of a different posterior sample with high probability mass.

The decision-theoretical optimal uncertainty estimate for the MAP is the credible region $\mathcal{H}_{\alpha}$ defined by the highest posterior density (\citealp{robert01}; see also \citealp{price20} for another recent application in cosmology)
\eq{
	\label{eq:hpd-def}
	\mathcal{H}_{\alpha} = \{\vec\Theta : {\rm Pr} (\vec{\Theta} | \vec{d}) > c_\alpha\} \ ,
}
for some threshold $c_{\alpha}$, depending on $\alpha$, the posterior mass enclosed within $\mathcal{H}_{\alpha}$:
\eq{
	\int_{\vec\Theta\in\mathcal{H}_{\alpha} }\!\!\!\! \dd^{N_\Theta} \Theta\; {\rm Pr} (\vec{\Theta} | \vec{d}) = \alpha \ .
}
For our purposes, the definition Eq.~\eqref{eq:hpd-def} is not practical, however. We wish to use a credible interval for the MAP estimate that can be easily compared to established uncertainty estimates, such as the standard deviation of posterior samples or the highest posterior density in the marginal distribution of a given parameter (denoted by M-HPD hereafter). While straightforward in one dimension, this is not the case for the high-dimensional parameter spaces we are considering here. For example, while the notion of $1\sigma$ credible intervals readily translates to $\alpha \approx 0.68$ in one dimension, this is not the case for higher dimensions, where $\alpha$ rapidly decreases with increasing number of dimensions.

We propose a hybrid credible interval estimator that is based on the joint, multi-dimensional highest posterior density region, but projected onto the marginal posterior of the parameter under consideration (PJ-HPD henceforth, for projected joint highest posterior density). For each parameter $i$, define the credible interval as the highest posterior density region that encompasses a fraction $\alpha$ of its marginal posterior mass:
\eq{
\label{eq:marg-hpd-def}
\mathcal{H}^\mathrm{marg.}_{i} = \{\vec\Theta : {\rm Pr} (\vec{\Theta} | \vec{d}) > c_i\}\;,
}
such that
\eq{
\int_{\Theta_{i}\in\mathcal{H}^\mathrm{marg.}_{i} } \hspace*{-0.6cm} \dd\Theta_{i}\; {\rm Pr} (\Theta_{i} | \vec{d})  = \alpha \,,
}
where ${\rm Pr} (\Theta_{i} | \vec{d})$ is the posterior of $\Theta_{i}$ marginalised over all other parameters of the model. In general, this does not correspond to the region of the highest marginal posterior. The PJ-HPD estimate is guaranteed to include the MAP (since by definition MAP lies within the highest posterior density region) and readily allows for the comparison of the credible intervals between analyses with a different dimension of parameter space. It is intuitive in that the PJ-HPD reduces to the standard M-HPD credible interval in one dimension. Moreover, in the case of a multivariate Gaussian posterior, the PJ-HPD credible interval coincides with the marginal $1\sigma$ interval. 

To compute our PJ-HPD credible intervals, we take the posterior samples $\bc{\vec{\hat{\Theta}}_{n}}$ for $n=1,2, \dots$ from nested sampling chains or MCMC and sort them in decreasing order of their associated posterior density, that is ${\rm Pr} (\vec{\hat{\Theta}}_{n+1} | \vec{d}) < {\rm Pr} (\vec{\hat{\Theta}}_n | \vec{d})$. Stepping through this list, and for each parameter $\Theta_i$ under consideration, we record the interval of parameter values $\bb{\hat{\Theta}_{i, a} ;\, \hat{\Theta}_{i,b} }$ that encompasses all parameter values in the posterior samples up to that point in the list, so $\hat{\Theta}_{i, a} \leq \hat{\Theta}_{i, n} \leq \hat{\Theta}_{i, b}$ for all $n$. Then we measure the posterior mass within this interval in the \emph{marginal} distribution of the parameter, $\int_{\hat{\Theta}_{i, a}}^{\hat{\Theta}_{i, b}} \dd\Theta_{i}\, {\rm Pr} (\Theta_{i} | \vec{d})$. The iteration stops when the posterior mass in the marginal distribution reaches the desired level $\alpha$ (i.e. usually $68\,\%$ or $95\,\%$) upon which the corresponding parameter interval is reported as the PJ-HPD credible interval\footnote{We note that the PJ-HPD approach will not necessarily return a simply connected credible interval in the case of a multimodal posterior, nor do we expect it to yield meaningful credible intervals in pathological cases. However, such outcomes should serve as a stark warning not to summarise the inference process via a point estimate in the first place.}. In practice, due to the sparse sampling of the highest-density posterior regions, we linearly interpolate the interval boundaries between the samples just before and after the stopping criterion is reached.

This approach is illustrated in Fig.~\ref{fig:reporting_constraints} where we show the marginal posterior distributions for the parameters $\Omega_{\rm m}$ and $\sigma_8$ out of a total of 18 model parameters sampled in a mock GGL and weak lensing analysis using a noiseless data vector. The one-dimensional marginal posterior modes do not coincide with the input cosmology, whereas the MAP estimate recovers it well (any residual deviation is due to the limited numerical accuracy of the optimisation). While the one-dimensional marginal posteriors have only moderate skewness, the input parameter values and the MAP estimate lie outside the $68\,\%$ M-HPD credible interval for both $\Omega_{\rm m}$ and $\sigma_8$. In contrast, the PJ-HPD credible interval represents a projection of the highest-density region of the full multivariate posterior that encompasses the MAP.

As the PJ-HPD interval extends to regions of the marginal posterior with lower posterior mass, the resulting constraints are generally slightly less tight than the standard M-HPD credible interval, especially if the posterior deviates strongly from a normal distribution. Moreover, as the PJ-HPD interval is determined from typically tens rather than thousands of posterior samples, its boundaries also scatter more. We find a scatter of order $10\,\%$ of the $1\sigma$ limits for a multivariate Gaussian that has the same dimension and is sampled with the same number of points as the posteriors in our analysis.

We adopt the MAP+PJ-HPD approach for reporting constraints in the KiDS-1000 analysis alongside the standard marginal posterior mode and associated M-HPD, and employ it as the main criterion for validating our inference pipeline; see Sect.~\ref{sec:parameter_constraints}. It should be emphasised that any differences between these two credible intervals are \emph{not} an indication of bias in the parameter they report; rather, this occurs if the posterior that they are applied to deviates from multivariate normality.

Finally, we note that posterior sampling techniques have an inherent limit on the accuracy of parameter constraints, dependent on the number of samples drawn. For settings that our computing resources can deliver in reasonable time (always yielding well in excess of $10^4$ posterior samples), we find this limit to be $\sim 0.1\sigma$ in the key parameter $S_8$, corresponding to the scatter of best-fit values between repeat runs of otherwise identical setups. This floor sets a stopping criterion for validation, as well as for the mitigation of systematic effects in our data (as reported in \citealp{giblin20}).

\section{Validation of the likelihood analysis pipeline}
\label{sec:parameter_constraints}

\begin{figure*}
	\centering
	\includegraphics[width=17cm]{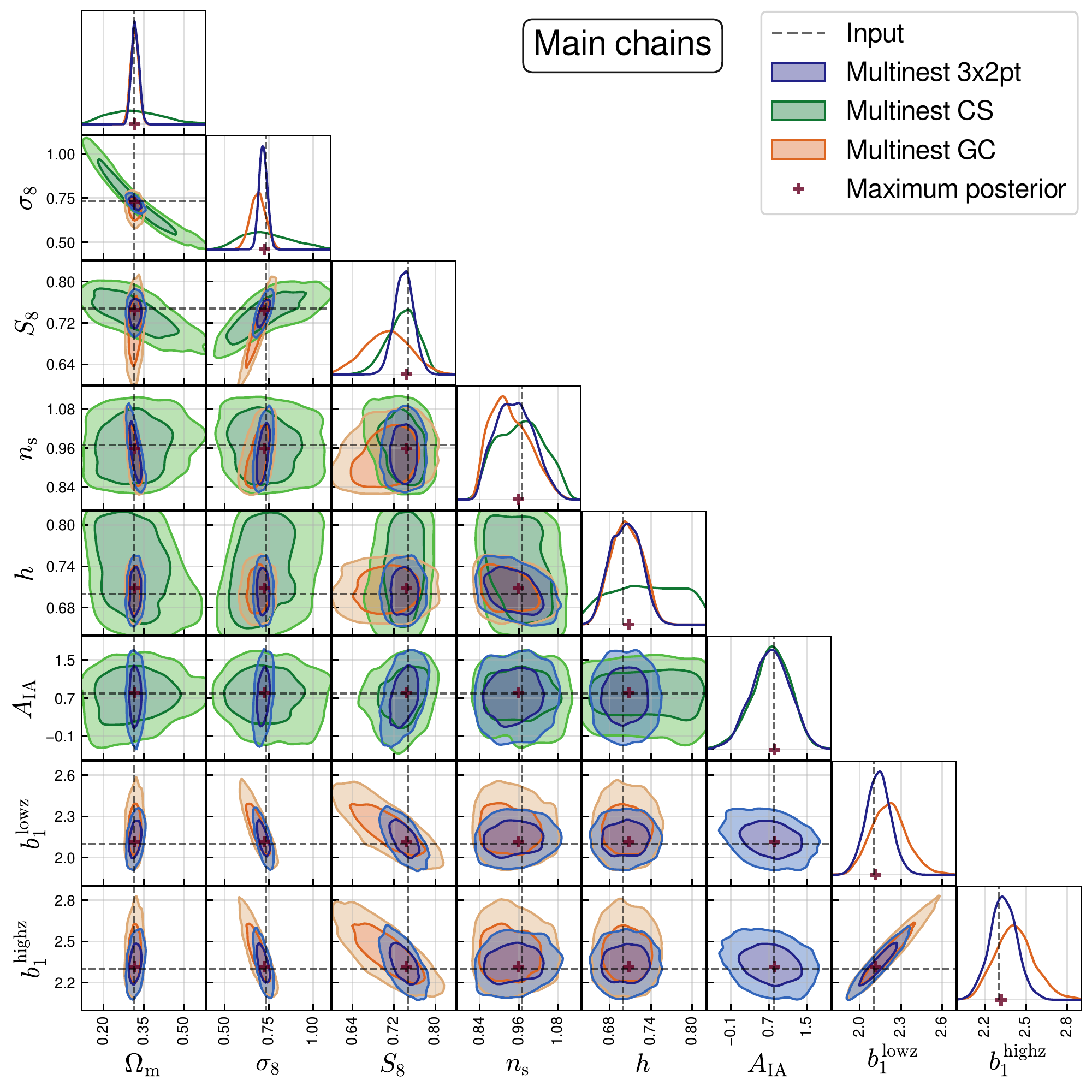}
	\caption{Marginal posterior distributions for a selection of parameters that are significantly constrained by the data, including the derived parameters $\Omega_{\rm m}$ and $\sigma_8$. Shown are one-dimensional marginals along the diagonal and two-dimensional marginal $1\sigma$ and $2\sigma$ credible regions elsewhere. Results are for the reference setup using a noiseless mock data vector of galaxy clustering (GC; red), cosmic shear (CS; green), and the combination of clustering and the weak lensing probes (\lq$3 \!\times\! 2$pt\rq; blue). Input parameter values are indicated by the black dashed lines, which are excellently recovered by the MAP estimate (red crosses).}
	\label{fig:constraints_fiducial}
\end{figure*}

\begin{figure}
	\includegraphics[width=\columnwidth]{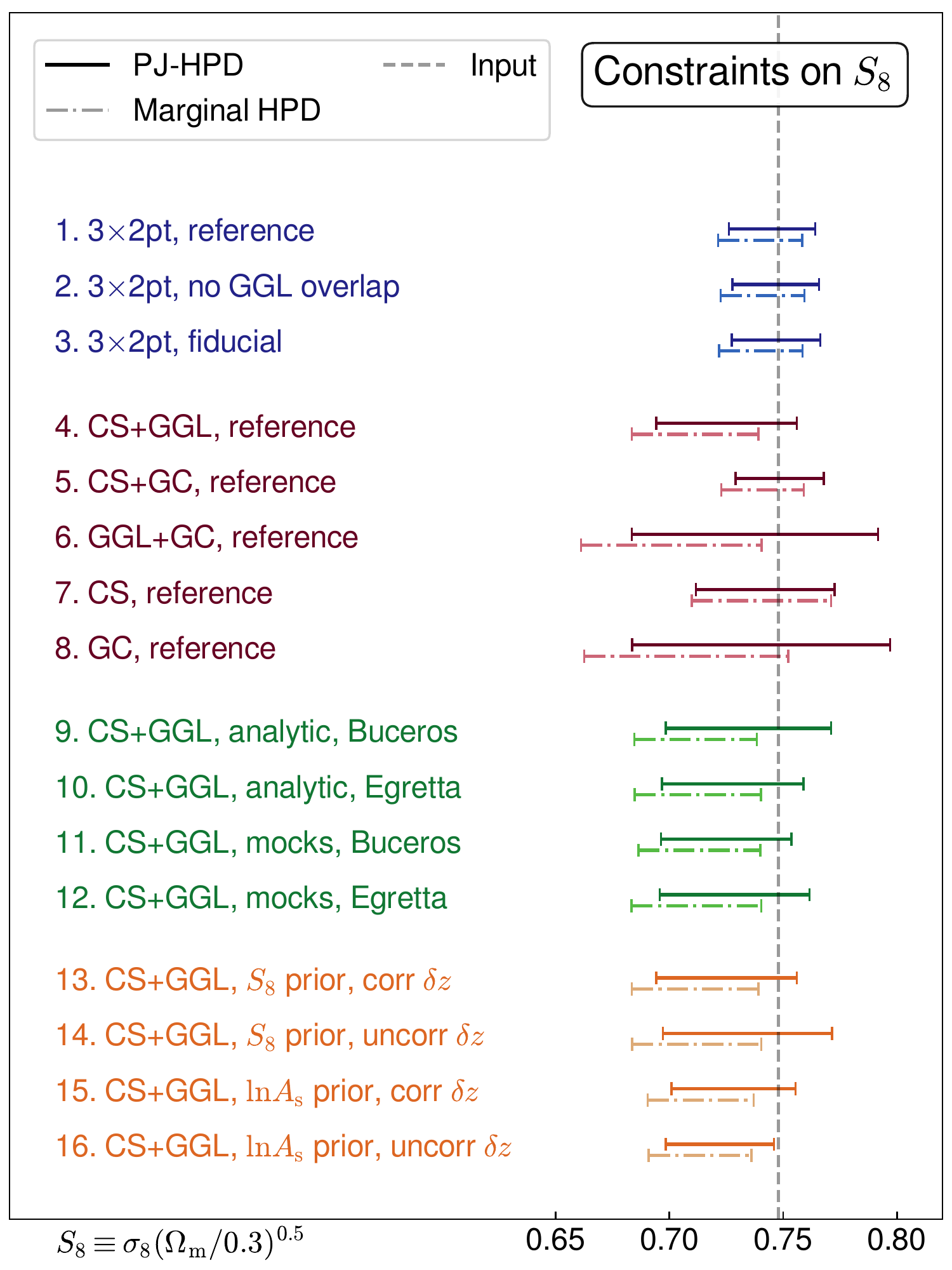}
	\caption{$S_8$ constraints ($1\sigma$) for different compositions of the GGL data vector in the joint analysis (blue), probe combinations (red), different covariance models (green), and different prior assumptions (orange). Standard marginal (M-)HPD credible intervals are shown as dot-dashed lines; PJ-HPD intervals as solid lines. The grey dashed line indicates the input value of $S_8$. The corresponding interval widths, as well as explanations of the case labels, are given in Table~\ref{tab:constraints}. The M-HPD intervals are generally shifted to $S_8$ values lower than the input; see text and Sect.~\ref{sec:reporting_constraints} for a discussion.}
	\label{fig:constraints_validation}
\end{figure}

\begin{figure*}
\includegraphics[width=\columnwidth]{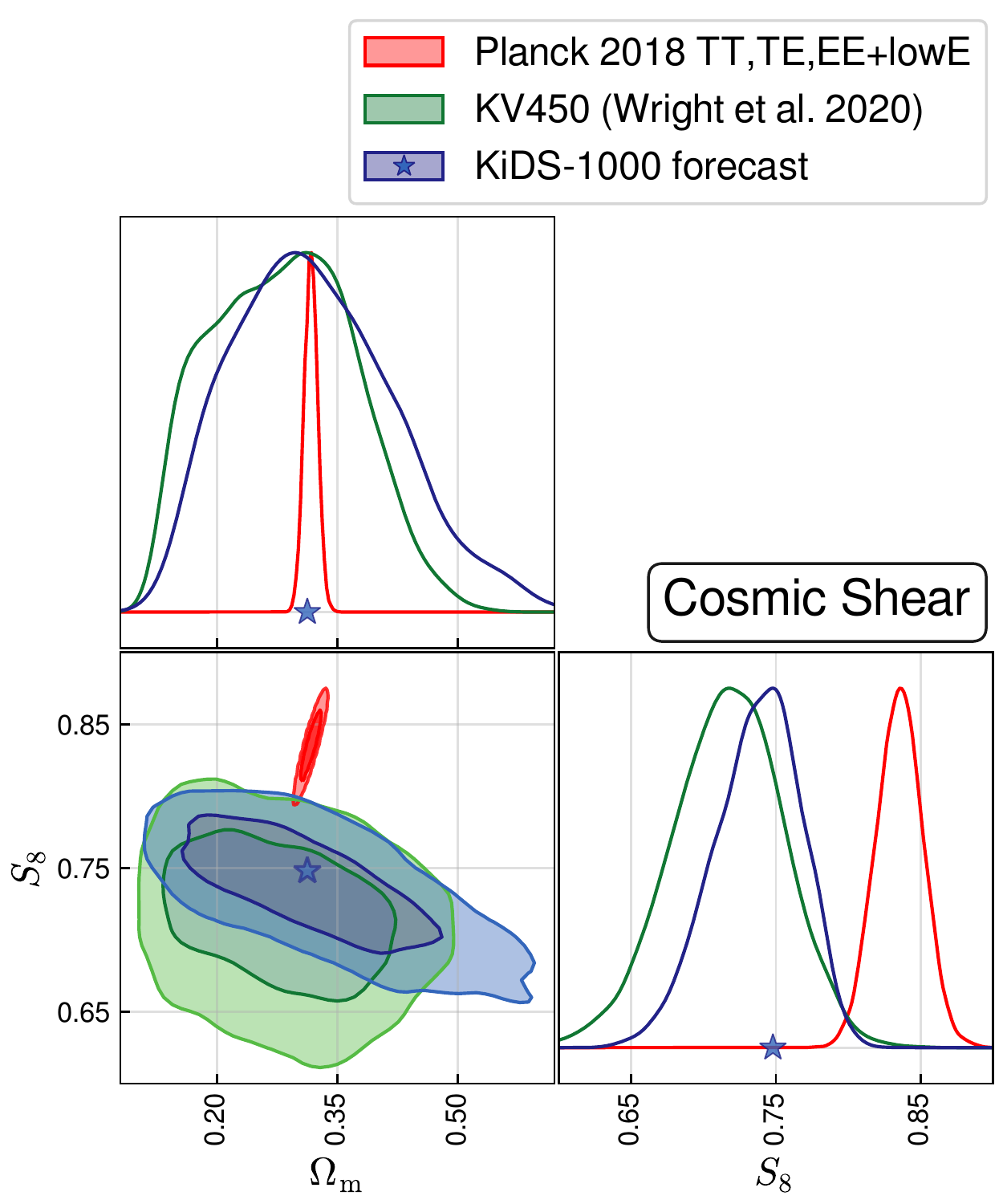}
\includegraphics[width=\columnwidth]{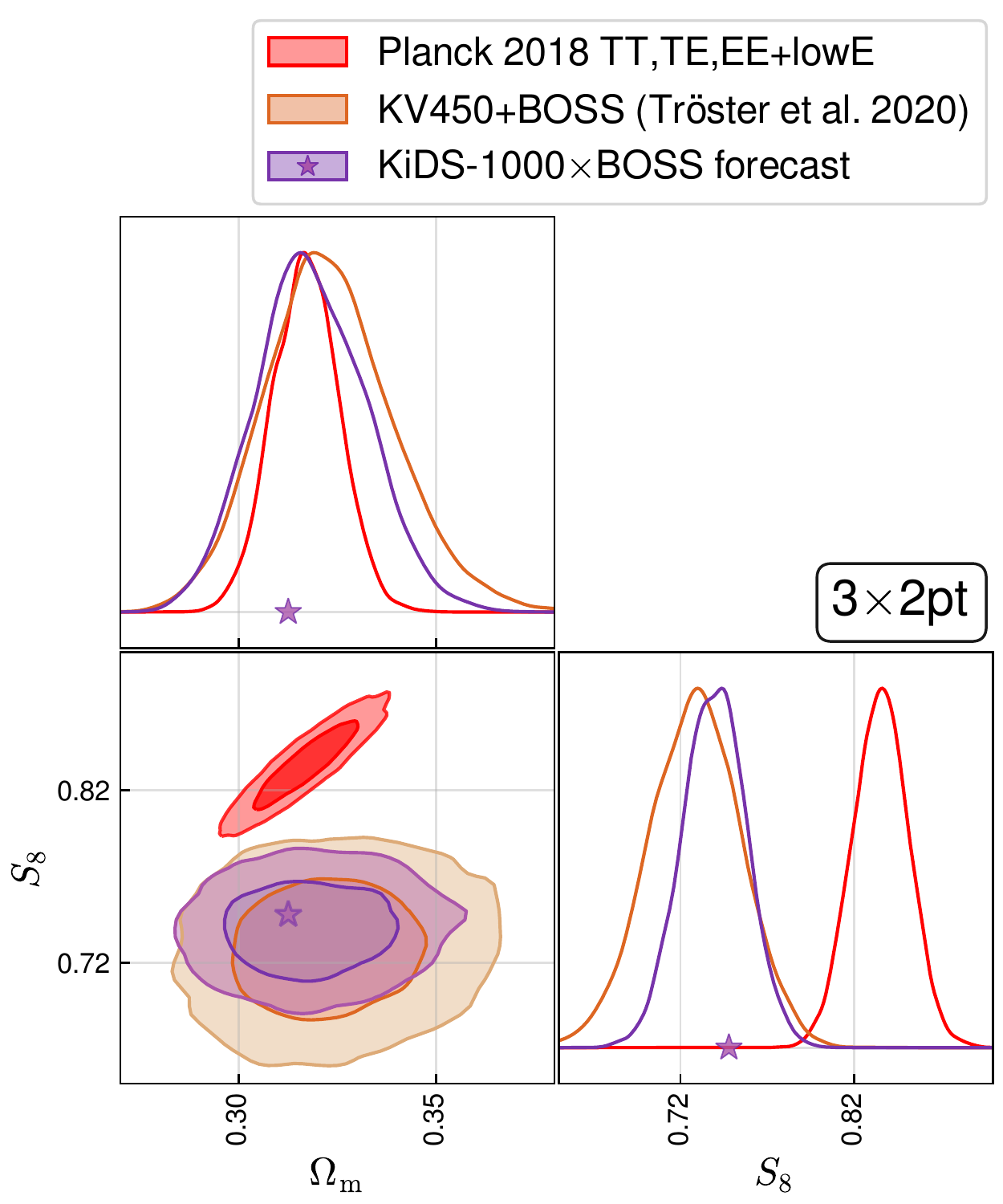}
\caption{Predicted KiDS-1000 constraints (reference setup, IDs 1 and 7) on the key cosmological parameters $\Omega_{\rm m}$ and $S_8$ in relation to previous KiDS results \citep{wright20b,troester19} and the \citet{planck18_parameters} primary CMB constraints, as indicated in the legends. The \textit{left} panel shows results for cosmic shear only; the \textit{right} panel those for the joint analysis of BOSS clustering, galaxy-galaxy lensing between BOSS/2dFLenS and KiDS-1000, and cosmic shear. The star in each panel marks the fiducial cosmology assumed for the forecast. Axis scales differ between the two panels.}
\label{fig:constraints_summary_plot}
\end{figure*}

\begin{table}
\caption{Validation tests and resulting $S_8$ marginal credible intervals.}              
\label{tab:constraints}      
\centering                                      
\begin{tabular}{llcc}          
\hline\hline                        
ID & Setup  & M-HPD  & PJ-HPD  \\    
\hline                                   
1 & $3 \!\times\! 2$pt, reference & $0.037$ & $0.038$ \\
2 & $3 \!\times\! 2$pt, no GGL overlap & $0.037$ & $0.038$ \\
3 & $3 \!\times\! 2$pt, fiducial & $0.037$ & $0.037$ \\
4 & CS+GGL, reference & $0.056$ & $0.062$ \\
5 & CS+GC, reference & $0.036$ & $0.039$ \\
6 & GGL+GC, reference & $0.079$ & $0.108$ \\
7 & CS, reference & $0.061$ & $0.061$ \\
8 & GC, reference & $0.090$ & $0.111$ \\
9 & CS+GGL, analytic, Buceros & $0.054$ & $0.072$ \\
10 & CS+GGL, analytic, Egretta & $0.056$ & $0.062$ \\
11 & CS+GGL, mocks, Buceros & $0.054$ & $0.057$ \\
12 & CS+GGL, mocks, Egretta & $0.057$ & $0.066$ \\
13 & CS+GGL, $S_8$ prior, corr $\delta z$ & $0.056$ & $0.062$ \\
14 & CS+GGL, $S_8$ prior, uncorr $\delta z$ & $0.057$ & $0.074$ \\
15 & CS+GGL, $\ln A_\mathrm{s}$ prior, corr $\delta z$ & $0.047$ & $0.054$ \\
16 & CS+GGL, $\ln A_\mathrm{s}$ prior, uncorr $\delta z$ & $0.045$ & $0.047$ \\
\hline
\end{tabular}
\tablefoot{Setups correspond to those shown in Fig.~\ref{fig:constraints_validation}. The reference includes a more progressive selection of scales in the GGL data vector (Bands 1-7) than our fiducial analysis. GC corresponds to galaxy clustering, CS to cosmic shear, and \lq $3 \!\times\! 2$pt\rq\ to the combination of all probes. Analytic/mocks and Buceros/Egretta indicates analytic and simulated covariances with simple or more realistic survey and sample properties. The choice of an (un)correlated prior on the source redshift distribution shifts is labelled as (un)corr $\delta z$. GGL signals with strong overlap of the lens and source redshift distributions have been removed in the \lq no GGL overlap\rq\ case. We note that setup 4 is identical to 13. The two rightmost columns give the width of the $1\sigma$ credible intervals (CI) for the standard M-HPD credible interval and the PJ-HPD method, i.e. twice the standard deviation if the posterior was Gaussian.}
\end{table}

We proceed to validate our inference pipeline by running it on simulated, noiseless data vectors generated from the same modelling pipeline as used in the inference. The absence of noise guarantees that the likelihood peaks exactly at the parameter combination chosen to generate the data vector and is therefore used to demonstrate recovery of the true underlying model parameters. Since we centre our fiducial set of model parameters at the peak of any of the informative Gaussian priors, we also expect the posterior to peak at the input parameter values. As discussed in Sect.~\ref{sec:reporting_constraints}, this will in general not hold for marginal posterior distributions, so we consider the multivariate maximum a posteriori (MAP) estimate.

Runs with noisy data vectors, which have a zero-mean multivariate Gaussian noise realisation drawn with the full covariance added, resemble more closely the processing of the real data. We repeat our validation tests with ensembles of noisy data but find our conclusions identical to the noiseless case. We also create data vectors that systematically deviate from the ideal model in order to assess biases in parameter constraints caused by effects not accounted for in the model. The results of these runs are discussed in the sections where the effect under consideration is covered.

In Fig.~\ref{fig:constraints_fiducial} we show one- and two-dimensional marginal posteriors of a subset of the 20 sampling parameters that are significantly constrained, alongside projections of the MAP. We also include $\Omega_{\rm m}$ and $\sigma_8$ as derived parameters. The MAP recovers the input values of all parameters shown in the figure to better than $1.7\,\%$. The combination of clustering and weak lensing signals partially lifts the typical degeneracy between $\Omega_{\rm m}$ and $\sigma_8$ for weak lensing-only constraints, through the small clustering errors on $\Omega_{\rm m}$. This benefits the marginal posterior of the key parameter $S_8$ as well since in our analysis setup it is still significantly correlated with $\Omega_{\rm m}$. The joint analysis also leads to visibly tighter and more symmetric marginal posteriors for the linear bias parameters, whereas improvements on the constraints on intrinsic alignments, baryon feedback, and the calibration of the redshift distributions remain marginal.

Figure~\ref{fig:constraints_validation} and Table~\ref{tab:constraints} provide a more quantitative overview of the $S_8$ constraints for the various probe combinations and analysis setups that we consider (identified by their ID). All credible intervals are extracted from a histogram of the posterior samples smoothed with Gaussian kernel density estimation, with the variance determined by Silverman\rq s rule \citep{silverman86}. Both M-HPD and PJ-HPD credible intervals are provided; we use the former to compare $S_8$ constraints with previous measurements (see below) and recommend the latter to assess the recovery of the input $S_8$.

The $1\sigma$ M-HPD credible interval for cosmic shear alone is $20\,\%$ smaller than the KV450 constraint, while the fiducial joint analysis halves the statistical uncertainty on $S_8$ with respect to KV450 (which corresponds to $29\,\%$ smaller errors than the corresponding BOSS+KV450 constraints by \citealp{troester19}). The marginal $1\sigma$ credible intervals shrink by $10\,\%$ ($18\,\%$) for $\Omega_{\rm m}$ over BOSS+KV450 (BOSS alone), and by $31\,\%$ ($40\,\%$) for  $A_{\rm IA}$ over BOSS+KV450 (KV450 alone). The precision on linear and non-linear galaxy bias parameters increases by roughly $30\,\%$ on most parameters with respect to BOSS only, but most of the added constraining power was already achieved in the \citet{troester19} analysis, and non-linear bias constraints are still too weak for informative conclusions on galaxy physics.

The constraints on the key parameters $S_8$ and $\Omega_{\rm m}$ are highlighted in Fig.~\ref{fig:constraints_summary_plot} and put into context with earlier analyses based on $450\,{\rm deg}^2$ of KiDS data \citep{wright20b,troester19}. Both the cosmic shear-only and joint analysis cases are shown, for the reference setup with IDs 7 and 1, respectively. It is evident that in the latter case the tension with \textit{Planck} results will increase if the best-fit remains close to the \citet{troester19} values, with the $S_8$ marginal constraints now on par with those originating from the CMB.

Naively, one might expect the KiDS-only parameter errors to decrease by a factor$\sqrt{2}$ because the KiDS-1000 survey area has doubled with respect to earlier KiDS analyses. However, this does not take into account the highly non-linear parameter dependencies, updated priors, or covariance contributions that do not scale inversely with area, in particular the multiplicative shear bias uncertainty. Most importantly in the cosmic shear case though, for our choices of two-point statistic and scales included in the analysis $S_8$ is not an optimal summary parameter of the constraining power of the data, as significant correlations with $\Omega_{\rm m}$ remain (see Fig.~\ref{fig:constraints_summary_plot}). While a power-law index of 0.5 in the $S_8$ definition was close to optimal for the $\xi_\pm$ data analysed in KV450, we find that larger values of $\sim \! 0.6$ for the index capture the degeneracy between $\Omega_{\rm m}$ and $\sigma_8$ better for band powers (this is explored further in \citealp{asgari20c}). The gain in constraining power for an adjusted definition of $S_8$ is larger and consistent with the naive $\sqrt{2}$ expectation, indicating that the systematic error treatment and the analysis choices have actually evolved in tandem with the statistical power of KiDS-1000.

It is evident that GGL does not add significant constraints to either clustering or cosmic shear, and its addition to their combination has no effect at all on $S_8$ (cf. Fig.~\ref{fig:constraints_validation} and Table~\ref{tab:constraints}, IDs 1 and 4--8). This somewhat sobering finding is caused by the dilution of constraining power due to almost doubling the number of parameters when adding GGL to cosmic shear (by introducing galaxy bias), and due to the more than an order of magnitude larger sky area available for clustering measurements. Accordingly, no palpable difference is seen when cutting the GGL data vector to our conservative, fiducial setup that removes highly non-linear scales and when removing bin combinations with a large fraction of source galaxies in front of lenses (IDs 2,3; see Sects.~\ref{sec:pgm} and \ref{sec:photo-z}). We use the more progressive GGL setting that includes scales with $k \lesssim 1\,h\,{\rm Mpc}^{-1}$ (Bands 1--7) for other comparisons to increase chances to detect any issues with the GGL model during validation (hence referred to as the reference). We also refrain from showing GGL-only results as we find that \textsc{Multinest} struggles to explore the resulting highly degenerate 18-dimensional posterior, producing unreliable results. 

Since we exclude bin combinations with strong overlap of sources and lenses, there is also no self-calibration of intrinsic alignments so that marginal $A_{\rm IA}$ constraints are the same for the joint analysis and cosmic shear alone; see Fig.~\ref{fig:constraints_fiducial}. On the real data the addition of GGL to cosmic shear and clustering does lead to a tightening of the marginal $A_{\rm IA}$ posterior by $17\,\%$, but the marginal posterior maximum also shifts by $12\,\%$ \citep{heymans20}. Both real and mock analysis agree that GGL does not significant constraining power to the galaxy bias parameters either.

Choosing either an analytic or mock-derived covariance (based on $5000$ realisations), with a simple rectangular footprint and uniformly distributed galaxies (Buceros; see Sect.~\ref{sec:mocks_construction} for details about the mock setup) or else the realistic survey mask and depth variations (Egretta), has no significant impact on any of the marginal parameter constraints (cf. IDs 9--12)\footnote{For these analyses we kept the $\delta z$ parameter fixed at 0 to speed up the runs and did not include the uncertainty due to multiplicative shear bias calibration to highlight the differences in the cosmological covariance contributions.}; see also \citet{des_y1_methods} for analogous conclusions drawn for the DES Year-1 analysis. This highlights that, despite the significant patterns of deviations between the covariance approaches and modelling choices analysed in Sect.~\ref{sec:covcomp}, the weak lensing error modelling is robust at the level of constraining power by KiDS-1000 (or indeed the full KiDS survey, which will add another $35\,\%$ of area). It will be interesting to explore in future work how the limitations identified in the covariance model affect next-generation surveys that cover substantial fractions of the sky, balancing overall tighter accuracy requirements due to the increased constraining power with a reduced sensitivity to survey boundaries as coverage becomes more contiguous.

The $S_8$ constraints are also stable under changes from an uncorrelated prior on the source redshift distribution shifts $\delta_z$, to the correlated prior in Fig.~\ref{fig:nz_shifts}, as well as under the choice of parametrisation of the normalisation of the matter power spectrum, where we compare $\ln A_{\rm s}$ with $S_8$ (cf. IDs 13--16). Our choice of using $S_8$ directly as the sampling parameter with an associated wide top-hat prior leads to more conservative marginal $S_8$ constraints, with the $1\sigma$ interval extended by $19\,\%$ mostly towards lower values. We also consider an additional parameter $\eta_{\rm IA}$ that allows for extra freedom in the redshift scaling of the IA signals (cf. Eq.~\ref{eq:kernel_IA}), imposing a wide top-hat prior in the range $\pm 5$. For a cosmic shear-only analysis this has negligible impact on the cosmological parameter constraints, widening the marginal $S_8$ posterior by $3\,\%$.

Figure~\ref{fig:constraints_validation} illustrates that the standard M-HPD credible intervals are all skewed towards lower values of $S_8$ than the input. The displacement is stronger for probe combinations that have weaker constraints on $S_8$ and leave more pronounced degeneracies in their posterior, which is the case for GGL in combination with either clustering or cosmic shear (IDs 4 and 6). As a consequence, the input $S_8$ values lies outside the $1\sigma$ interval for the cosmic shear+GGL case while the shift is mild in the cosmic shear only (ID 7) and joint clustering and weak lensing cases (e.g. ID 1). The PJ-HPD credible intervals encompass the input value by design (however, it can be right at the interval boundary, or just beyond due to numerical inaccuracy; cf. ID 16) and are generally less displaced but also wider than the M-HPD intervals. This increase in width is minor in the well-constrained cases (negligible for cosmic shear only and $3\,\%$ for the joint analysis), but significant in the probe combinations with strong shifts (e.g. $11\,\%$ for cosmic shear+GGL). Larger fluctuations in the PJ-HPD intervals widths are also discernible, for instance for the runs with different covariance models (IDs 9--12) where the standard approach yields nearly identical constraints.

We note that, for an assessment of consistency with \textit{Planck} constraints, the \textit{Planck} likelihood will need to be sampled in $S_8$ as well. The impact of the choice of sampling parameter and the displacement of one-dimensional marginal constraints away from the underlying best-fit value highlight that discrepancies between probes should be interpreted with care if quantified solely in the marginal distributions, as is widespread practice. Tension is assessed most meaningfully in the shared parameter space of the full posterior distributions, but this can be challenging in practice. Therefore, we still assess tension via the marginal distributions of $S_8$, but only for probe combinations for which we find projection effects to be negligible. A detailed discussion of tension assessment is provided in \citet{heymans20}. Bayesian approaches to performing consistency tests in large parameter spaces exist \citep{koehlinger19} and have been applied to the internal consistency of KiDS-1000 cosmic shear \citep{asgari20c}. We advocate the use of a summary measure of tension that minimises the sensitivity to prior choices \citep[see e.g.][]{handley19b,lemos20}.

We find no measurable difference between M-HPD and PJ-HPD $S_8$ credible intervals for \textit{Planck} (as expected since the cosmological posterior is close to Gaussian), and small shifts for KV450 in line with the results of Fig.~\ref{fig:constraints_validation}, so that the previously reported $S_8$ tension between these probes remains valid. The marginal $S_8$ posterior for KiDS-1000 cosmic shear only and for the joint analysis of clustering and weak lensing peaks close to the input and is well approximated by a Gaussian (cf. Fig.~\ref{fig:constraints_summary_plot}), so direct comparison with \textit{Planck} is possible for this case as well.

\section{Summary and conclusions}
\label{sec:conclusions}

In this work we presented the methodology for a joint analysis of spectroscopic galaxy clustering from BOSS and of weak gravitational lensing from the fourth data release of the Kilo-Degree Survey (KiDS-1000). This includes a detailed investigation of the analysis choices from the galaxy catalogue level onwards to cosmological inference, taking into account the tightening of requirements due to the doubled survey area with respect to earlier KiDS cosmological analyses. We summarise here the major updates since the cosmic shear analysis by \citet[KV450]{kv450} and the BOSS+KV450 analysis by \citet{troester19}.

Galaxy-galaxy lensing (GGL), the cross-correlation between lens galaxy positions and source galaxy ellipticities, is now incorporated into the joint analysis, using lens galaxies from both the BOSS and 2dFLenS surveys, which cover $85\,\%$ of the KiDS-1000 area. However, we found that it adds negligible amounts of cosmological constraining power in KiDS-1000 mainly because of an order of magnitude larger survey area available for clustering. Nonetheless, we identified a number of issues that will come to the fore in analyses where clustering, cosmic shear, and GGL are measured over the same footprint: our hybrid matter-galaxy power spectrum model blends perturbative and non-perturbative approaches customary in clustering and weak lensing, respectively, but a satisfactory solution for the deeply non-linear regime remains to be found (Sect.~\ref{sec:pgm}); magnification bias is strongest in the GGL signals and requires dedicated simulations if lens samples deviate from pure flux-limit selection (Appendix~\ref{sec:magbias}); and widely used idealisations in covariance models related to survey geometry fail most prominently in the GGL part (Sect.~\ref{sec:covcomp}).

Weak lensing signals are consistently modelled using band powers derived from correlation function measurements that make them insensitive to survey geometry (Sects.~\ref{sec:cosmicshear_stats} and \ref{sec:ggl_stats}). As opposed to earlier KiDS studies, we now fully incorporate mode mixing in the modelling of both signals and covariances, but we find that on scales that we can model accurately ($\ell \in \bb{100, 1500}$) the recovery of the underlying angular power spectra is highly accurate. The limiting systematic in the cosmological modelling is the knowledge of the highly non-linear matter power spectrum (Sect.~\ref{sec:matterps}), where current fit formulae and emulators disagree at the few per-cent level \citep[cf. the comparison in][]{knabenhans19}.

The main source of uncertainty in the modelling of weak lensing, however, remains the intrinsic alignment of galaxies (Sect.~\ref{sec:cosmicshear_stats}). Direct observations of the effect in galaxy samples typically used for weak lensing do not yet exist, and our physical understanding of alignment mechanisms is still too poor to create predictive ab-initio simulations. In the absence of clear guidance we are thus required to strike a balance between a simple model that risks not capturing the complexity in the data and a flexible model that risks providing catch-all parameters in the likelihood analysis among which other residual systematics, notably those related to the source redshift distributions, can hide.

Recent advances in selecting source galaxy samples whose redshift distribution can be more reliably calibrated \citep{wright20} have been adopted and the corresponding calibration uncertainties fully propagated into the likelihood analysis (Sect.~\ref{sec:photo-z}). We also clarify how the multiplicative shear bias calibration affects estimators and derived survey characteristics critical for covariance calculation (Appendix~\ref{sec:merrors}), and we take a more nuanced approach in quantifying and propagating the uncertainty in this calibration (Sect.~\ref{sec:shear_calibration}).

We build a dedicated suite of more than $20\,000$ mocks to perform an unprecedentedly detailed assessment of our analytic covariance models (Sects.~\ref{sec:mocks} and \ref{sec:covariances}). KiDS-1000 cosmological constraints are demonstrated to remain unchanged when using an analytic or mock-based covariance (Sect.~\ref{sec:parameter_constraints}). We show that the analytic model is capable of capturing the impact of survey geometry and spatial variations in survey depth well and identify the mixed noise-sample variance contribution as the main culprit behind residual differences. We confirm recent work by \citet{heydenreich20} in showing that variable depth biases two-point weak lensing statistics at the few per-cent level (Sect.~\ref{sec:spatial_variability}), which has negligible impact for KiDS-1000, but will challenge widespread analysis approaches that rely on spatial homogeneity of the galaxy sample in future applications.

We explicitly map the complex prior space of cosmic shear cosmological analyses (Sect.~\ref{sec:parameters_priors}), which serves to explain the significant prior dependence of results  \citep[e.g.][]{joudaki20}. That said, the key parameter $S_8$ is robust to different choices of prior, especially the upper tail of its marginal posterior which directly impacts on tension measurements with \textit{Planck}. In contrast to previous analyses, we directly sample in $S_8$, which allows us to impose a wide top-hat, and thus safely uninformative, prior. We employ our mock suite to demonstrate that the weak lensing band power likelihood is consistent with being Gaussian (whereas there is evidence for non-Gaussianity for correlation functions on large scales) and that the $\chi^2$ goodness of fit sampling distribution is indeed $\chi^2$-distributed to good approximation (Sects.~\ref{sec:gaussian_likelihood} and \ref{sec:gof}). However, significant deviations from the textbook approach are seen for the degrees of freedom, or equivalently the effective number of model parameters, and we discuss and advocate alternative estimators.

It is demonstrated that the KiDS-1000 likelihood analysis pipeline recovers the input parameters of a mock data vector with per-cent level accuracy (Sect.~\ref{sec:parameter_constraints}), using a dedicated maximum a posteriori (MAP) estimate of the multivariate posterior density. Experiments have hitherto typically reported the marginal mean or mode of the $S_8$ posterior, which we show to be significantly shifted towards lower values from the truth, beyond $1\sigma$ for some probe combinations. This is not indicative of any biases in the inference process, but simply a consequence of a large parameter space with multiple non-linear degeneracies and complex prior volume. Tension assessment in $S_8$ between experiments should therefore proceed with caution if based on its marginal posterior. We develop an alternative technique that produces a credible interval from the multivariate highest posterior density (PJ-HPD) which is guaranteed to encompass the MAP estimate (Sect.~\ref{sec:reporting_constraints}).

With the fiducial KiDS-1000 analysis setup, we can expect cosmic shear-only $S_8$ constraints that are $20\,\%$ tighter than KV450, and joint clustering and weak lensing constraints that improve in $S_8$ by $29\,\%$ over the previous KiDS+BOSS analysis (Sect.~\ref{sec:parameter_constraints}). These changes undersell the true gain in statistical power somewhat because, as opposed to the correlation function-based analysis in KV450, $S_8$ does not precisely capture the direction transverse to the typical weak lensing $\Omega_{\rm m}$--$\sigma_8$ degeneracy any more.

The key scientific question in the KiDS-1000 analysis is the tentative discrepancy in the amplitude of the matter density fluctuation power spectrum seen between low-redshift large-scale structure probes and the cosmic microwave background as observed by \textit{Planck}. Barring the caveats with regards to the interpretation of marginal $S_8$ constraints and the remaining uncertainties in the modelling of galaxy astrophysics, we confirm our cosmic shear and GGL analysis pipeline to be robust and find no known systematic effect in the cosmological and astrophysical modelling or the analysis methodology that would bias $S_8$ by more than $0.1\sigma$, which is also the numerical accuracy limit for parameter constraints extracted from our sampled posteriors. Together with the expected boost in constraining power, KiDS-1000 thus has the potential to give us a clear steer in the direction of statistical fluke, unknown systematic, or new physics as the most likely explanation for the tension. It should be noted that all methodology investigations presented here were performed while the real shear catalogues were still blinded.

The on-going efforts in maximising the cosmological information and fidelity for the current generation of large-scale structure surveys will feed directly into the forthcoming array of even more ambitious projects. New surveys that will start within the coming three or so years include the ESA \textit{Euclid} mission\footnote{\texttt{https://sci.esa.int/euclid}} \citep{laureijs11}, the Vera C. Rubin Observatory 
LSST\footnote{Legacy Survey of Space and Time; \texttt{www.lsst.org}} \citep{lsst12}, as well as the DESI\footnote{Dark Energy Spectroscopic Instrument; \texttt{www.desi.lbl.gov}} \citep{desi16} , 4MOST\footnote{\texttt{www.4most.eu}} \citep{richard19}, and PFS\footnote{Subaru Prime Focus Spectrograph; \texttt{https://pfs.ipmu.jp}} \citep{takada14} surveys. The combination of spectroscopic \lq lens\rq\ galaxy samples with \lq source\rq\ samples used for weak lensing in overlapping sky regions, as pursued in this work, will remain an important approach to exploiting these data sets. The tighter accuracy requirements in step with the vastly increased raw constraining power, the greater depth of observations, and the sheer number of galaxies observed will pose a multitude of challenges in the analysis methodology that are yet to be tackled.

\begin{acknowledgements}

We would like to thank M. Jarvis for his continued excellent support and maintenance of \textsc{TreeCorr}. We are also grateful to E. Tittley who went out of his way to preserve data in peril. We thank our referee for constructive comments on the manuscript.\\[1ex]

CAL, MA, TT, CH, and BG acknowledge support from the European Research Council under grant number 647112.
TT is also supported by the European Union's Horizon 2020 research and innovation programme under the Marie Sk{l}odowska-Curie grant agreement No 797794.
CH and FK are supported by the Max Planck Society and the Alexander von Humboldt Foundation in the framework of the Max Planck-Humboldt Research Award endowed by the Federal Ministry of Education and Research.
HHi is supported by a Heisenberg grant of the Deutsche Forschungsgemeinschaft (Hi 1495/5-1). HHi, AHW, JLvdB, and AD are supported by an ERC Consolidator Grant (No. 770935).
AGS acknowledges support by the Excellence Cluster ORIGINS, which is funded by the Deutsche Forschungsgemeinschaft (DFG,  German  Research  Foundation)  under  Germany's Excellence  Strategy  -  EXC-2094  -  390783311.
MB is supported by the Polish Ministry of Science and Higher Education through grant DIR/WK/2018/12, and by the Polish National Science Center through grant no. 2018/30/E/ST9/00698.
MC acknowledges  support  by  the  Spanish  Ministry  of  Science MINECO under grant PGC2018-102021.
HHo and AK acknowledge support from Vici grant 639.043.512, financed by the Netherlands Organisation for Scientific Research (NWO).
KK acknowledges support by the Alexander von Humboldt Foundation.
HYS acknowledges the support from NSFC of China under grant 11973070, the Shanghai Committee of Science and Technology grant No.19ZR1466600 and Key Research Program of Frontier Sciences, CAS, Grant No. ZDBS-LY-7013.
MvWK acknowledges the support by the Science and Technology Facilities Council.
\\[1ex]

We  are  indebted  to  the  staff at ESO-Garching and ESO-Paranal for managing the observations at VST and VISTA that yielded  the  data  presented  here.  Based  on  observations  made with ESO Telescopes at the La Silla Paranal Observatory under programme IDs 177.A-3016, 177.A-3017, 177.A-3018, 179.A-2004, 298.A-5015, and on data products produced by the KiDS consortium.
The 2dFLenS survey is based on data acquired through the Australian Astronomical Observatory, under program A/2014B/008. It would not have been possible without the dedicated work of the staff of the AAO in the development and support of the 2dF-AAOmega system, and the running
of the AAT.\\[1ex]

The BOSS-related results in this paper have been made possible thanks to SDSS-III. Funding for SDSS-III has been provided by the Alfred P. Sloan Foundation, the Participating Institutions, the National Science Foundation, and the U.S. Department of Energy Office of Science. SDSS-III is managed by the Astrophysical Research Consortium for the Participating Institutions of the SDSS-III Collaboration including the University of Arizona, the Brazilian Participation Group, Brookhaven National Laboratory, Carnegie Mellon University, University of Florida, the French Participation Group, the German Participation Group, Harvard University, the Instituto de Astrofisica de Canarias, the Michigan State/Notre Dame/JINA Participation Group, Johns Hopkins University, Lawrence Berkeley National Laboratory, Max Planck Institute for Astrophysics, Max Planck Institute for Extraterrestrial Physics, New Mexico State University, New York University, Ohio State University, Pennsylvania State University, University of Portsmouth, Princeton University, the Spanish Participation Group, University of Tokyo, University of Utah, Vanderbilt University, University of Virginia, University of Washington, and Yale University.\\[1ex]

Author Contributions: All authors contributed to the development and writing of this paper. The authorship list is given in three groups: the lead authors (BJ, CAL, MA, TT, CH), followed by two alphabetical groups. The first alphabetical group includes those who are key contributors to both the scientific analysis and the data products.The second group covers those who have either made a significant contribution to the data products or to the scientific analysis.

\end{acknowledgements}

\bibliographystyle{aa}
\bibliography{bibliography}

\begin{appendix}

\section{Additional tables related to signal modelling}
\label{sec:pgmapprox}

In Table~\ref{tab:fiducialpars} the fiducial parameter values for our analyses are shown. Unless stated otherwise, these are adopted throughout. They generally follow the rounded best-fit values from the \citet{troester19} analysis of BOSS and KV450. For weakly constrained parameters we instead choose the centre values of the flat priors adopted to avoid undue skewness in the prior volume. Magnification parameters are determined as described in Appendix~\ref{sec:magbias}. In addition to the parameters defined in the main body of the paper, we list the curvature density parameter $\Omega_K$, the baryon density parameter $\Omega_{\rm b}$, and the dark energy density parameter $\Omega_\Lambda$. Table~\ref{tab:pgm_fitcoeff} shows the best-fit coefficients for the galaxy-matter power spectrum fit function (Eq.~\ref{eq:pgm_fitterms}) introduced in Sect.~\ref{sec:pgm}.

\begin{table}[h]
\caption{Choice of fiducial model parameters.}              
\label{tab:fiducialpars}      
\centering                                      
\begin{tabular}{lll}          
\hline\hline                        
Parameter & Symbol & Value \\    
\hline                                   
CDM density & $\omega_{\rm c}$ & 0.13 \\
Baryon density & $\omega_{\rm b}$ & 0.0225 \\
Scalar spectral amp. & $\ln (10^{10} A_{\rm s}$) & 2.72 \\
Scalar spectral index & $n_{\rm s}$ & 0.97 \\
Hubble constant (scaled) & $h$ & 0.7 \\
Neutrino mass sum & $\sum m_\nu$ & $0.06\,{\rm eV}$  \\
Curvature density parameter & $\Omega_K$  & 0 \\ 
\hline
Matter density parameter & $\Omega_{\rm m}$  & 0.31 \\
Baryon density parameter & $\Omega_{\rm b}$  & 0.046 \\
Dark energy density parameter & $\Omega_\Lambda$  & 0.69 \\
Density fluctuation amp. & $\sigma_8$ & 0.733 \\
Weak lensing amp. parameter & $S_8$ & 0.746 \\
\hline
Linear galaxy bias & $b_1$ & $\bc{2.1, 2.3}$ \\
Quadratic galaxy bias & $b_2$ & $\bc{0.2, 0.5}$ \\
Non-local galaxy bias & $\gamma_3^-$ & $\bc{0.9, 0.1}$ \\
Virial velocity parameter & $a_{\rm vir}$ & $\bc{3.8, 3.0}$ \\
Luminosity function slope & $\alpha_{\rm mag}$ & $\bc{1.8, 2.6}$ \\
Intrinsic alignment (IA) amp. & $A_{\rm IA}$ & 0.8 \\
IA redshift dependence & $\eta_{\rm IA}$ & 0 \\
Baryon feedback amp. & $A_{\rm bary}$ & 2.6 \\
\hline
Redshift offset & $\delta_z$ & $\bc{0,0,0,0,0}$ \\
\hline
\end{tabular}
\tablefoot{The first section contains the primordial $\Lambda$CDM parameters, the second a selection of derived cosmological parameters. The third section lists astrophysical model parameters, with pairs of fiducial values for the two lens bins L1 and L2 in curly brackets. The fourth section comprises the parameters varied in the inference related to measurement systematics (one per source galaxy sample).}
\end{table}

\begin{table}[h]
\caption{Fit coefficients $g_{\alpha,i}^{mn}$ of the approximate galaxy-matter power spectrum model; cf. Eq.$\,$(\ref{eq:pgm_fitterms}).}              
\label{tab:pgm_fitcoeff}      
\centering                                      
\begin{tabular}{llllllll}          
\hline\hline                        
$\alpha$ & $i$ & \multicolumn{6}{l}{polynomial order $m\,n$} \\
\hline
& & $0\,0$ & $0\,1$ & $1\,0$ & $0\,2$ & $1\,1$ & $2\,0$  \\    
\hline                                   
\hline
$b_2$ & 2 & 0.56 & $-$0.96 & $-$2.55 & 0.34 & 2.08 & 2.05 \\
 & 1 & 5.03 & $-$2.71 & $-$5.52 & 0.89 & 5.78 & 3.90 \\
 & 0 & $-$0.45 & $-$2.21 & $-$5.69 & 0.24 & 1.49 & 4.12 \\
\hline
 $\gamma_2$ & 2 & $-$0.03 & $-$0.01 & 0.88 & 0.02 & 0.09 & $-$1.03 \\
 & 1 & 3.45 & $-$0.34 & 1.69 & 0.05 & $-$0.19 & $-$2.47 \\
 & 0 & 0.26 & $-$1.86 & $-$3.68 & 0.51 & 1.50 & 2.39 \\
\hline
$\gamma_3^-$ & 2 & $-$0.01 & 0.00 & 0.81 & 0.01 & 0.11 & $-$1.01 \\
 & 1 & 3.43 & $-$0.32 & 1.17 & 0.01 & $-$0.06 & $-$2.06 \\
 & 0 & $-$1.81 & $-$1.56 & $-$3.08 & 0.49 & 1.50 & 1.93 \\ 
\hline
\end{tabular}
\end{table}

\section{Magnification bias}
\label{sec:magbias}

Gravitational lensing not only modifies the ellipticities of galaxy images but also their apparent size and, therefore, their measured flux. Since the definition of galaxy samples almost universally involves flux-dependent quantities, the selection function acquires a dependency on the foreground large-scale structure and thus on cosmology. This magnification bias is a second-order effect for gravitational shear measurements in our source samples which may need to be accounted for in future weak lensing surveys but which we can safely neglect in KiDS \citep{schmidt09,krause10,deshpande19}. However, it is not a priori clear that magnification bias is negligible in our lens samples, especially since it directly modulates the observable, that is the number counts of galaxies (see e.g. \citealp{duncan14,unruh19,thiele20} for studies of the impact of magnification bias).

If the samples are purely flux-limited, magnification bias is readily modelled as a balance of the local dilution/focussing of solid angle and the modification of apparent magnitudes above or below the flux limit \citep{bartelmann01}, resulting in expressions of the form given in Eq.~(\ref{eq:ggl_magbias}). The only additional information required is the logarithmic slope of the sample's luminosity function at its faint end, $\alpha_{\rm mag}$. However, like other lens samples used for large-scale structure inference jointly with weak lensing \citep[e.g.][]{rozo16}, the two BOSS samples employed in the KiDS-1000 analysis were derived from a complex multi-band selection function, making closed-form analytic modelling impossible \citep[cf.][]{hildebrandt16}.

In a separate publication (von Wietersheim et al., in prep.) we develop a simulation-assisted method to determine an effective luminosity function slope for our BOSS samples that can be used in the standard expressions. Here, we summarise the salient points (for details see \citealp{vdBusch20}): we recreate the LOWZ and CMASS selections in the MICE2 galaxy mocks \citep{fosalba15a,fosalba15b,crocce15} over an octant of the sky. Two variants of the BOSS DR12 low- and high-redshift samples are then derived, one with the selection applied to observed fluxes that include magnification, and one with the selection based on the hypothetical fluxes if no magnification was present. From this we can directly measure $\alpha_{\rm mag}$ via the slope of the linearised relation between the change in galaxy number counts between the magnified and unmagnified catalogues and the weak lensing convergence of these objects. The result is compared to the slope of the magnitude count of the same mock samples near where the counts begin to drop off, and we identify a magnitude range within which the slope measurements agree within the noise.

We find that for both the low- and high-redshift samples $i$-band counts yield the best match, driven by the $i$-band cut of the original CMASS sample which also substantially contributes to L1. The latter measurement is repeated on the real BOSS catalogues, using the same band and magnitude range as identified in the mock. This results in
\eq{
\label{eq:alpha_bestfit}
\alpha_{\rm mag}^{\rm L1} = 1.80 \pm 0.15 \,; ~~~~~~ \alpha_{\rm mag}^{\rm L2} = 2.62 \pm 0.28\,,
}
where the errors are the standard deviations of the magnitude counts among the bins selected for the slope measurements. Rather than pure statistical noise, these errors therefore have contributions from the limited accuracy of the assumption that the BOSS-selected samples have a simple power-law scaling in the magnitude range that we consider.

\begin{figure}
	\includegraphics[width=\columnwidth]{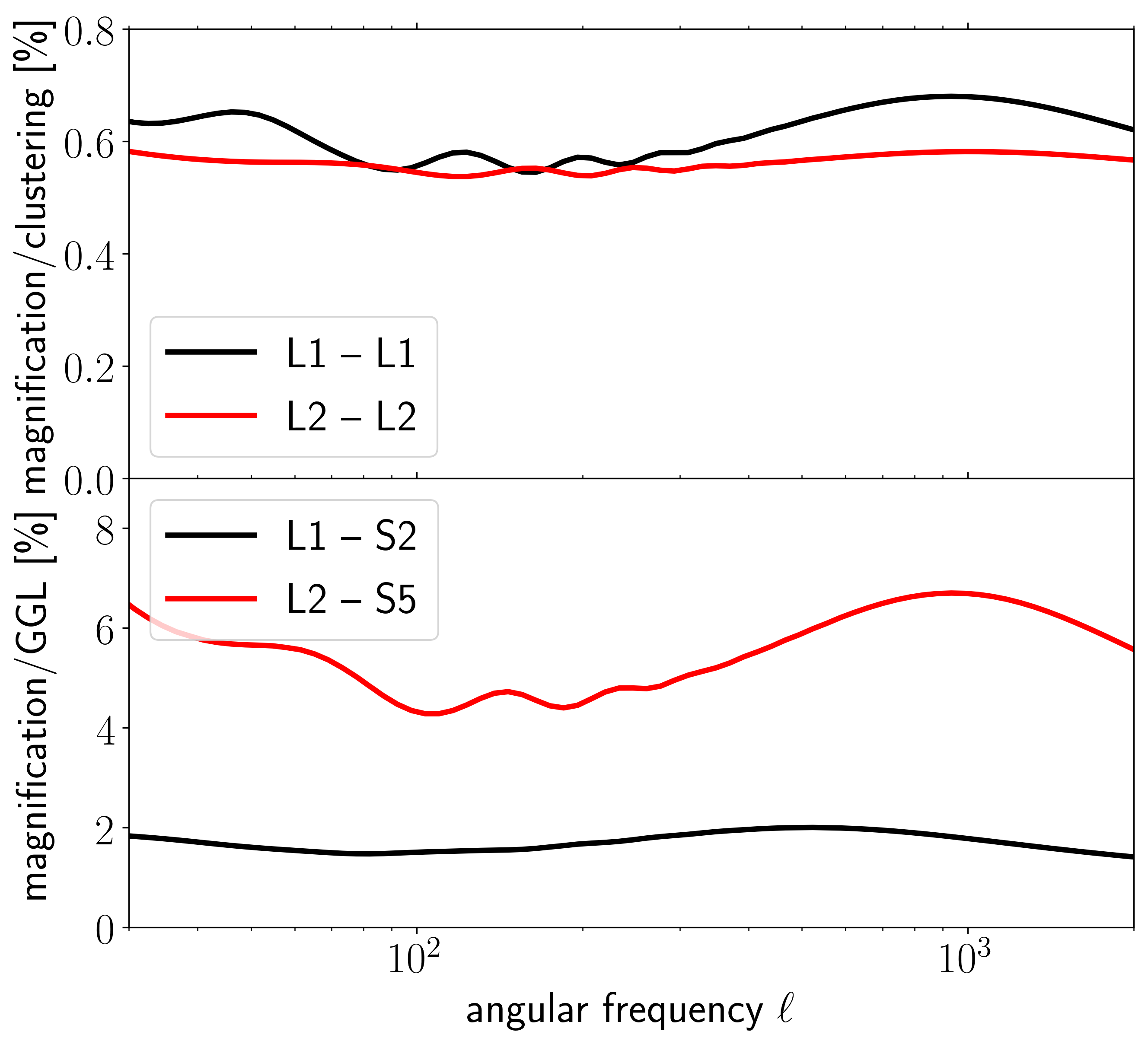}
	\caption{Magnification bias contribution relative to galaxy clustering (top panel) and the galaxy-galaxy lensing signal (bottom panel) for the redshift bin combinations indicated in the legends (cf. Table~\ref{tab:sampleprops}). Shown are angular power spectrum models assuming linear galaxy bias using the values from Table~\ref{tab:fiducialpars}.}
	\label{fig:magbias}
\end{figure}

We propagate these $\alpha_{\rm mag}$ measurements into angular power spectrum predictions for clustering and galaxy-galaxy signals; see Fig.~\ref{fig:magbias}. In both lens bins the magnification contributions (dominated by the magnification-clustering cross-correlation) are at the $0.6\,\%$ level and therefore negligible. The magnification-shear correlation constitutes a few-per cent contribution to the galaxy-galaxy lensing signal and is larger for the high-redshift source bins. This result prompts us to include the latter contributions into our modelling, as further described in Sect.~\ref{sec:ggl_stats}, but they are too small to justify additional freedom in the GGL model, so that we keep the $\alpha_{\rm mag}$ parameters fixed at their best-fit values (Eq.~\ref{eq:alpha_bestfit}) in the analysis.

\section{Propagation of multiplicative shear calibration}
\label{sec:merrors}

In this appendix we show how the multiplicative shear calibration propagates into the correlation function estimators, the sample redshift distributions, and the sample properties that are inputs for the analytic covariance calculation. We work in the weak shear limit where $\gamma \ll 1$, and will model our shear estimator, the observed ellipticity $\epsilon^{\rm obs}$, as
\eq{
\label{eq:eps_shear}
\epsilon^{\rm obs} = (1+m)[\epsilon + \gamma] \,,
}
working in the absence of additive biases such as PSF contamination.   Here the unsheared ellipticity $\epsilon$ is uncorrelated with $\gamma$ (i.e. there are no intrinsic alignments), and is a combination of the true intrinsic ellipticity $\epsilon^{\rm int}$ and random measurement noise. As noise increases in the imaging, galaxy shapes become increasingly round and our ability to recover the underlying shear decreases, independently of the shear estimation method used. We express this inherent fundamental noise bias in terms of a multiplicative shear calibration correction $m$ that can be seen as both a correction that provides an unbiased shear estimator, and also a weight, reflecting that some galaxies carry little shear information. Regarding this latter point, using the term \lq Responsivity\rq\ or \lq Sensitivity\rq\ to describe $(1+m)$ is useful \citep[see e.g.][]{huff17,sheldon17,zuntz18} as this term quantifies how responsive, or sensitive, the observed galaxy is to an induced shear.  Unfortunately we cannot measure $m$ precisely for each galaxy and can only determine an estimate $\hat{m}$, where
\eq{
\hat{m} = m + \eta \, .
}
Here, $\eta$ is a noise term that has zero mean.  It can potentially have a very significant variance such that one should not attempt to de-bias $\epsilon^{\rm obs}$ individually for each galaxy because of the non-linear propagation of the noise in $\hat{m}$.

We note in passing that our multiplicative shear bias calibration takes the impact of variable survey depth on shear measurement into account because $\hat{m}$ is a function of the signal-to-noise and size of the galaxy image, simulated for a range of different seeings (see \citealp{giblin20} for details). While we do not take into account that in a poor-seeing, low-depth region the fraction of blended galaxies will be higher than average, \citet{kannawadi19} showed the calibration to be robust to changes in galaxy background density and thus the levels of blending well within the derived calibration uncertainty. We therefore do not expect the coupling of blending-induced shear bias with variable depth to make a significant impact on the parameter constraints, as artificially changing the blend fraction across the whole simulated survey changes $\hat{m}$ by an insignificant amount.

\subsection{Shear correlation function estimator}

The typical starting point for measuring cosmic shear are the two-point correlation function estimators $\hat{\xi}_\pm$ that we could ideally write as
\eq{
\label{eqn:hatxi}
\hat{\xi}_\pm(\theta) = \frac{\sum_{ij} \,w_i w_j \,(\epsilon^{\rm obs}_{i{\rm t}}\epsilon^{\rm obs}_{j{\rm t}} \pm \epsilon^{\rm obs}_{i\times}\epsilon^{\rm obs}_{j\times})\, \Delta_{ij}(\theta)}{ \sum_{ij} \,w_i w_j \,(1+\hat{m}_i)(1+\hat{m}_j)\, \Delta_{ij}(\theta)} \,,
}
where $w_i$ is a survey-defined weight that has been assigned to galaxy $i$, and the sum is taken over all galaxies $i$ and $j$. We have also introduced the selector function $\Delta_{ij}(\theta)$, which is unity if the angular separation between galaxies $i$ and $j$ lies within a bin centred on $\theta$, and zero otherwise. The tangential and cross components of the ellipticity (and analogously for the shear) are given by $\epsilon_{\rm t} + \ic \epsilon_\times = - \epsilon \expo{-2 \ic \varphi}$, where $\varphi$ is the polar angle of the separation vector between the galaxy pair under consideration.

Ignoring contributions from intrinsic galaxy alignments, we can expand this estimator using Eq.~(\ref{eq:eps_shear}) as
\eqa{
\label{eqn:2pcf}
\hat{\xi}_\pm(\theta) &= \frac{\sum_{ij} \,W_i W_j \,(\epsilon_{i{\rm t}}\epsilon_{j{\rm t}} \pm \epsilon_{i\times}\epsilon_{j\times})\, \Delta_{ij}(\theta)}{ \sum_{ij} \,\widehat{W}_i \widehat{W}_j\, \Delta_{ij}(\theta)} \\ \nn
& \hspace*{1cm}+ \frac{\sum_{ij} \,W_i W_j \,(\gamma_{i{\rm t}}\gamma_{j{\rm t}} \pm \gamma_{i\times}\gamma_{j\times})\, \Delta_{ij}(\theta)}{ \sum_{ij} \,\widehat{W}_i \widehat{W}_j\, \Delta_{ij}(\theta)} \,,
}
where $W_i := w_i(1+m_i)$ and $\widehat{W}_i := w_i(1+\hat{m}_i)$. The first term introduces shape noise into our analysis (which we explore in Sect.~\ref{sec:sigmae}).  The second term is the true weighted cosmic shear signal that we wish to extract, where we can see that our weighted source galaxy distribution now includes the Responsivity, correctly down-weighting galaxies that are unresponsive to shear (as these typically have negative values for $m$).

In practice, Eq.~\ref{eqn:hatxi} is not the actual estimator that we employ, as we assume that the weights and responses are uncorrelated, which is a good approximation, such that the denominator becomes
\eqa{
K(\theta) &:= \sum_{i \in I,j \in J} \,\widehat{W}_i \widehat{W}_j\, \Delta_{ij}(\theta)\\ \nn
 &\approx \big( 1+ \ba{m}_I \big) \br{1+\ba{m}_J} \sum_{i \in I,j \in J} \,w_i \,w_j \, \Delta_{ij}(\theta) \,,
}
where we have made the samples over which the summations run explicit, and where $\ba{m}_I $ denotes the average of $\hat{m}$ over all galaxies in sample $I$.     In taking this sum we reduce the impact of noise in our calibration correction and recover the average noise-bias for the population.   This correction can however lead to the misconception that $K(\theta)$ is only to calibrate the average value of the shear for the galaxy sample.   It is, however, also taking into account the effective down-weighting of unresponsive galaxies that is an inherent part of the shear estimator $\epsilon^{\rm obs}$.

\subsection{Redshift distribution}

From Eq.~\ref{eqn:2pcf} we can see that source galaxies used in the cosmic shear measurement are weighted by $W$, which is a combination of the survey-defined weight $w$ and the shear sensitivity $1+m$.  An estimate of the effective redshift distribution of this source sample is therefore given by
\eq{
\label{eqn:nz}
n_{\rm S}(z) = \frac{\sum_{i} \widehat{W}_i \, n_i(z)}{\sum_{i} \widehat{W}_i} =  \frac{\sum_{i} w_i (1+m_i + \eta_i) \, n_i(z)}{\sum_{i} w_i(1+m_i + \eta_i) } \,,
}
where $n_i(z)$ is the probability distribution of true redshifts for an individual galaxy $i$. We note that in Eq.~(\ref{eqn:nz}) we have made the common approximation that the redshift PDF of a union of galaxy sub-samples is the weighted sum of the sub-sample PDFs. We use $\widehat{W}$ in Eq.~(\ref{eqn:nz}) as only the noisy weight is accessible from observations. However, as the $m$-noise $\eta$ has zero mean, this term does not impact the redshift estimation. If $m$ is uncorrelated with redshift $z$ then this term will cancel in the estimate. As $m$ is correlated with size and magnitude, however, it is likely to correlate with $z$ and therefore it is necessary to include it as part of the weight in the effective redshift distribution.

For the DES Year 1 analysis, Eq.~(\ref{eqn:nz}) corresponds to their weighted stack of individual galaxy PDFs estimated using {\sc BPZ} (\citealp{hoyle18}; see also \citealp{sheldon17}). In KiDS-1000, we determine the true redshift $z_\mu$ per cell $\mu$ in the self-organising map (SOM; see Sect.~\ref{sec:photo-z}) and construct an effective redshift distribution for the full source sample as
\eq{
\label{eqn:nzSOM}
n_{\rm S}(z) = \frac{\sum_{\mu} n_\mu(z) \sum_{i \in \mu} \widehat{W}_i }{\sum_{i} \widehat{W}_i }  \,,
}
where the sum $\sum_{i \in \mu}$ runs over all galaxies $i$ in SOM-cell $\mu$, and $n_\mu(z) = \delta_{\rm D}(z - z_\mu)$, where $\delta_{\rm D}$ denotes the Dirac delta-distribution. We construct a multiplicative shear calibration estimate $m$ per galaxy by applying \lq Method A\rq\ from \citet{fenech17} to the image simulations of \citet{kannawadi19}. This involves fitting $m$ as a function of signal-to-noise and size \citep[see for example Fig.~9 in][]{fenech17}. \citet{fenech17} find the accuracy of these $m$-per galaxy estimates to be lower than the $m$-per sample estimates that we adopt in our fiducial analysis. They are, however, sufficient to determine the impact of including the Responsivity in our redshift estimates. We find the difference to be negligible between the redshift distribution calculated with Eq.~(\ref{eqn:nzSOM}) when incorporating the Responsivity, or when setting $m=0$.  This is because $m$ is typically small for the self-calibrating \textit{lens}fit approach that KiDS takes, but this need not be the case, such as for metacalibration approaches.

\subsection{Shape noise estimates}
\label{sec:sigmae}

Shape noise, quantified via $\sigma_\epsilon$, is often defined in the literature as \lq the intrinsic ellipticity dispersion\rq\ and is a crucial ingredient into the analytical covariance calculation. The definition needs to be reconsidered for the case of a weighted and calibrated ellipticity distribution. \citet{schneider02} derive the analytical covariance for $\xi_{\pm}$. In their Equation~(13), $\sigma_\epsilon$ is defined as 
\eq{
\label{eq:sn_from_xi}
\ba {\epsilon^{\rm obs}_{i{\rm t}}\epsilon^{\rm obs}_{j{\rm t}} + \epsilon^{\rm obs}_{i\times}\epsilon^{\rm obs}_{j\times} }= \sigma_\epsilon^2\, \delta_{ij} + \xi_+(|\vec{\theta}_i - \vec{\theta}_j|)\,,
}
where $\vec{\theta}_i$ denotes the angular position vector of galaxy $i$. This means that the effective $\sigma_\epsilon$ for our shear correlation function estimator is equal to $\hat{\xi}_+(\theta=0)$ in a universe with vanishing gravitational shear (for zero lag, $\delta_{ij}=1$ in Eq.~\ref{eq:sn_from_xi}). Applying this condition to Eq.~(\ref{eqn:hatxi}), the shape noise can therefore be estimated from our weighted and calibrated ellipticity distribution as
\eq{
\label{eqn:sige}
\sigma_\epsilon^2 = K^{-1}(0)\, \sum_i \,w_i^2 \, \bb{ (\epsilon^{\rm obs}_{i1})^2 + (\epsilon^{\rm obs}_{i2})^2 }  \,,
}
where we have approximated the intrinsic ellipticity dispersion by the observed ellipticity dispersion as $|\gamma| \ll 1$, that is $W\epsilon \approx w\epsilon^{\rm obs}$. In evaluating $K(0)$ we use the limit $\Delta_{ij}(0)=\delta_{ij}$. The variances of the ellipticities are calculated after the additive bias correction has been applied (as discussed in \citealp{giblin20}).

\subsection{Effective galaxy pair count and number density}
\label{sec:neff}

Since sample variance contributions are independent of how densely the shear field is sampled, we can restrict ourselves to the shape noise contribution to the correlation function covariance. This term is readily isolated by once again considering the case of vanishing gravitational shear. In this limit the covariances of $\xi_+$ and $\xi_-$ are identical and only have a diagonal contribution. \citet{schneider02} derived a general expression for this covariance term, which we can adapt to our estimator (\ref{eqn:hatxi}) as follows:
\eqa{ \nn
& {\rm Cov}_{\rm G,sn} \bb{\xi_+(\theta);\, \xi_+(\theta)} = {\rm Cov}_{\rm G,sn} \bb{\xi_-(\theta);\, \xi_-(\theta)} \\ \nn
&= K^{-2}(\theta)\, \sum_{ijkl} W_i W_j W_k W_l \, \Delta_{ij}(\theta)\, \Delta_{kl}(\theta)\\ \nn
& \hspace*{0.3cm} \times \ba{ \epsilon_{i{\rm t}}\epsilon_{j{\rm t}} \epsilon_{k{\rm t}}\epsilon_{l{\rm t}} + \epsilon_{i{\rm t}}\epsilon_{j{\rm t}} \epsilon_{k\times}\epsilon_{l\times} + \epsilon_{i\times} \epsilon_{j\times} \epsilon_{k{\rm t}}\epsilon_{l{\rm t}} + \epsilon_{i\times}\epsilon_{j\times} \epsilon_{k\times}\epsilon_{l\times} } \\ 
&= \frac{\sigma_\epsilon^4}{K^2(\theta)}\, \sum_{ij} W_i^2 W_j^2\, \Delta_{ij}(\theta)\;.
}
We note that off-diagonal terms correlating $\xi_\pm$ at different angular separations vanish, and that ${\rm Cov}_{\rm G,sn} \bb{\xi_+(\theta);\, \xi_-(\theta)} =0$. In the absence of weighting and calibration this expression reduces to the intuitive form ${\rm Cov}_{\rm G,sn} \bb{\xi_\pm(\theta);\, \xi_\pm(\theta)}  = \sigma_\epsilon^4/N_{\rm pair}(\theta)$, where $N_{\rm pair}(\theta)$ is the number of pairs among the galaxy samples correlated and within the survey footprint with a separation  that falls into the bin centred on $\theta$. Retaining this expression, we can define an effective number of galaxy pairs in the presence of weights and calibration as
\eq{
\label{eqn:npairs}
N_{\rm pair} (\theta) := \frac{ K^2(\theta) }{\sum_{ij} W_i^2 W_j^2\, \Delta_{ij}(\theta)} \approx \frac{ K^2(\theta) }{\sum_{ij} \widehat{W}_i^2 \widehat{W}_j^2\, \Delta_{ij}(\theta)}\,, 
}
such that the calibration correction carries through correctly into the covariance matrix when we measure this effective number of pairs directly from the data. While the first equality in Eq.~(\ref{eqn:npairs}) is exact, the second one replaces the unobservable noiseless $W$ with the observable quantity $\widehat{W}$ and is in practice used on the data. The effective number density then follows by setting $\Delta_{ij}\,\equiv\,1$, that is by calculating all galaxy pairs in the survey, and using the relation $N_{\rm pair}^{\rm total}= A_{\rm eff}^2 n_{\rm eff,S}^2 $ to obtain:
\eq{
\label{eqn:neff}
n_{\rm eff,S} = \frac{1}{A_{\rm eff}} \frac{(\sum_{i} \widehat{W}_i)^2}{\sum_i \,W_i^2} \approx \frac{1}{A_{\rm eff}} \frac{(\sum_{i} \widehat{W}_i)^2}{\sum_i \,\widehat{W}_i^2} \, ,
}
where $A_{\rm eff}$ is the effective survey area further discussed in Appendix~\ref{sec:anacov}. Again, the second equality is an approximative expression that is applied to the data. The expressions above generalise to the tomographic case in a straightforward manner.

The results for $\sigma_\epsilon$, $N_{\rm pair}$, and $n_{\rm eff}$ derived here reduce to the expressions used in previous analyses \citep{heymans12,kuijken15} for unit Responsivity. The changes in these quantities due to multiplicative shear calibration are small in our analysis, again because $m$ is typically small for the self-calibrating \textit{lens}fit approach that KiDS takes.

\subsection{Tangential shear case}

We repeat the steps above for the case of galaxy-galaxy lensing, using the tangential shear estimator of Eq.~(\ref{eq:ggl_estimator}). Requiring that the noise term in the diagonal elements of the covariance of this estimator is rendered as ${\rm Cov}_{\rm G,sn} \bb{ \ba{\gamma_{\rm t}}(\theta);\, \ba{\gamma_{\rm t}}(\theta)} =  \sigma_\epsilon^2/(2 N_{\rm pair}^{\rm GGL})$, one obtains
\eq{
\label{eqn:npairs_ggl}
N_{\rm pair} ^{\rm GGL}(\theta) := \frac{ \br{ \sum_{i \in R,j \in S} w_i \widehat{W}_j\, \Delta_{ij}(\theta) }^2 }{{\cal N}_{\rm rnd}^2 \br{ \sum_{i \in L,j \in S} w_i^2 W_j^2\, \Delta_{ij}(\theta)} }\,,
}
where $S$, $L$, and $R$ denote the source, lens, and random catalogue, respectively. We again set $\Delta_{ij} \equiv 1$ and identify $n_{\rm eff,S}$ as given by Eq.~(\ref{eqn:neff}) in the resulting expression. Assuming $N_{\rm pair} ^{\rm GGL, total}=A_{\rm eff}^2 n_{\rm eff,S} n_{\rm eff,L}$, we arrive at the following equation for the effective number density of lens galaxies:
\eq{
\label{eqn:neff_ggl}
n_{\rm eff,L} = \frac{1}{A_{\rm eff}} \frac{\br{\sum_{i \in R} w_i}^2}{{\cal N}_{\rm rnd}^2 \br{ \sum_{i \in L} w_i^2} }  = \frac{1}{A_{\rm eff}} \frac{\br{\sum_{i \in L} w_i}^2}{ \br{ \sum_{i \in L} w_i^2} } \, ,
}
so the expression is fully analogous to the one for $ n_{\rm eff,S}$.

\section{Analysis of covariance models}
\label{sec:covcomp_app}

This section covers additional topics in covariance modelling; see Sect.~\ref{sec:covariances} for an overview. The construction of the analytic model employed here is detailed in Appendix~\ref{sec:anacov}.

\subsection{Clustering - weak lensing cross-variance}
\label{sec:bosskidscorr}

\begin{figure}
	\includegraphics[width=\columnwidth]{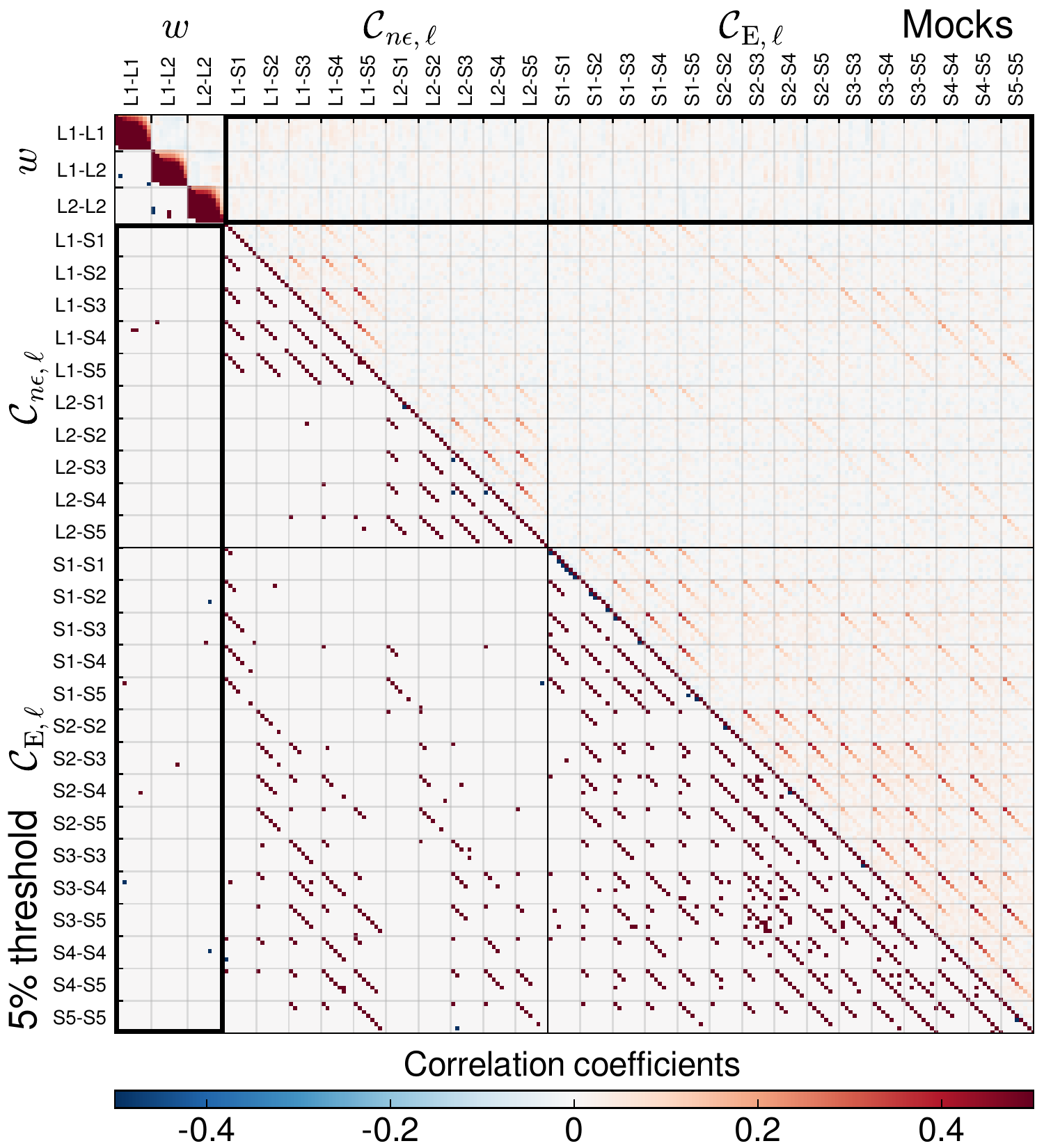}
	\caption{Correlation coefficient matrix for the angular galaxy clustering correlation function $w$, the galaxy-galaxy lensing band power ${\cal C}_{\rm n \epsilon}$, and the cosmic shear band power ${\cal C}_{\rm E}$ calculated over the full BOSS and KiDS-1000 footprints. The upper triangle shows the correlation matrix as calculated from the mocks; the lower triangle indicates correlation coefficients above $\pm 5\,\%$. The black rectangles highlight the cross-correlations between galaxy clustering and weak lensing statistics. These are negligible, with few elements fluctuating above the $5\,\%$ threshold.}
	\label{fig:corr_paper}
\end{figure}

To demonstrate that the clustering and weak lensing signals are uncorrelated and thus statistically independent (in the Gaussian likelihood approximation), we create $4000$ mock realisations with the full BOSS footprint, KiDS, and their overlap. We do not simulate 2dFLenS lens galaxies as the additional GGL measurement outside the area from which clustering was obtained only acts to further reduce correlations. The current \textsc{Flask} implementation does not allow us to incorporate line-of-sight modes so that we cannot model the redshift-space correlation function. However, since weak lensing only depends on transverse modes of the density distribution, it is sufficient to consider the projected angular correlation function,
\eq{
w^{(ij)}(\theta) = \int_0^\infty \frac{\dd \ell\, \ell}{2 \pi}\, {\rm J}_0 (\ell \theta)\, C_{\rm gg}^{(ij)}(\ell)\;, 
}
where $C_{\rm gg}$ is the angular galaxy power spectrum. The correlation function is measured with the standard Landy-Szalay estimator \citep{landy93}, along with our fiducial band power estimates for the weak lensing signals.

Figure~\ref{fig:corr_paper} shows the resulting correlation matrix. Cross-correlations between the clustering correlation functions and any of the weak lensing signals only very rarely exceeds 0.05, and this is largely due to residual noise in the mock covariance. We can therefore safely assume that the clustering and weak lensing parts of the data vector are independent. This trend is driven by the fact that only $3\,\%$ of the BOSS survey area overlaps with KiDS and thus with the weak lensing measurements. Joint clustering and weak lensing measurements over the same sky area do produce significant cross-correlations \citep[e.g.][]{krause17,des_y1_methods} and therefore demand for a more homogeneous approach to summary statistics and their covariance than taken in this work.

\subsection{Correlation function covariance}
\label{sec:corrfctcov}

\begin{figure}
	\includegraphics[width=\columnwidth]{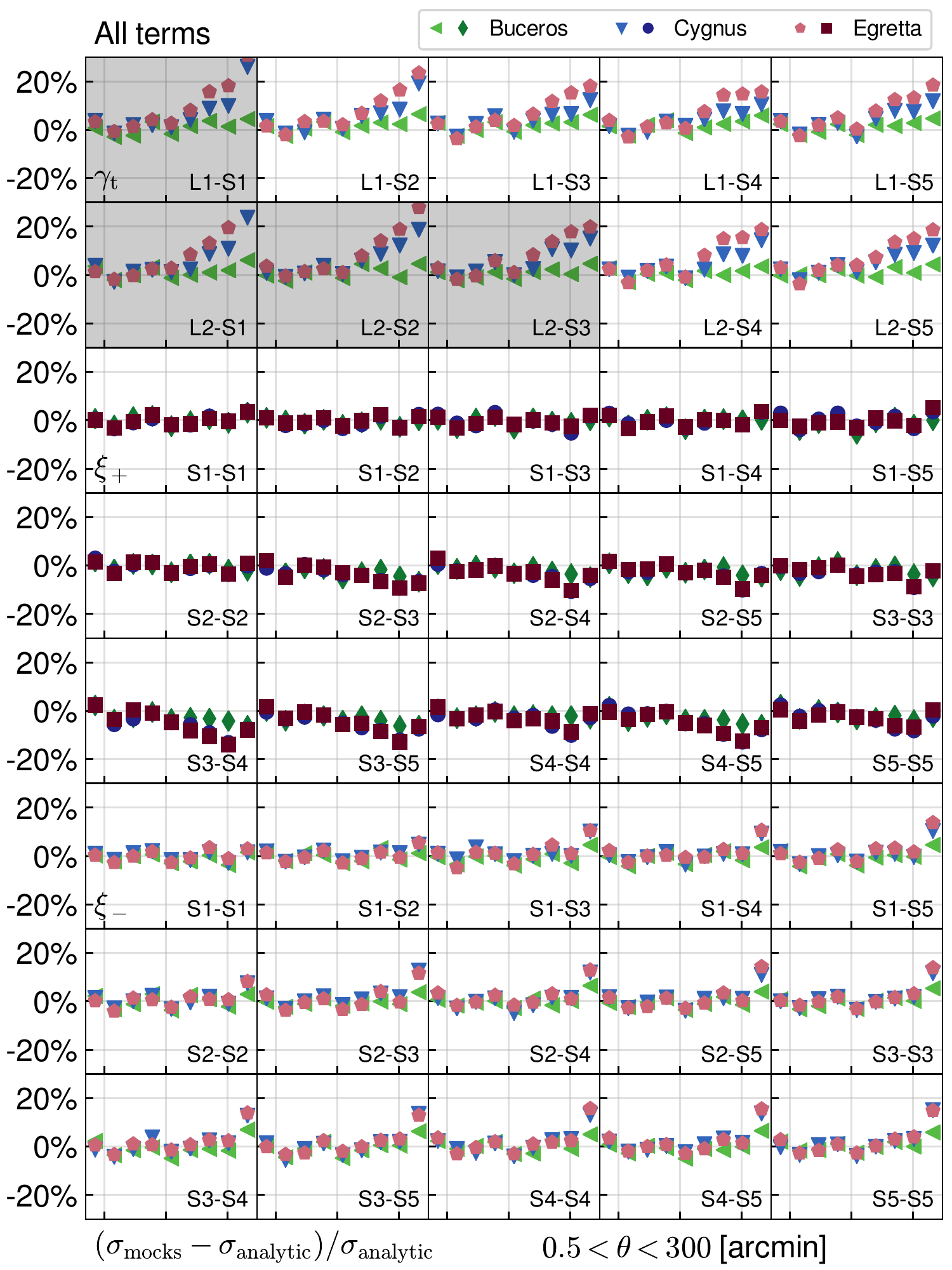}
	\caption{Relative difference between the square root of the diagonals of the mock and analytic covariances of the weak lensing correlation functions, i.e. the real-space analogue of Fig.~\ref{fig:std_diff_egretta}. The top two rows show GGL signals, the centre three rows $\xi_+$, and the bottom three rows $\xi_-$, with bin combinations indicated in the panels. Three cases are shown: spatially uniform galaxy samples in a simple survey footprint (Buceros, green), spatially uniform galaxy samples in the realistic footprints (Cygnus, blue), and spatially varying samples in the realistic footprints (Egretta, red). GGL signals that are not used in the analysis have been greyed out.}
	\label{fig:stdDiff_egretta_direct}
\end{figure}

\begin{figure*}
	\sidecaption
  	\includegraphics[width=12cm]{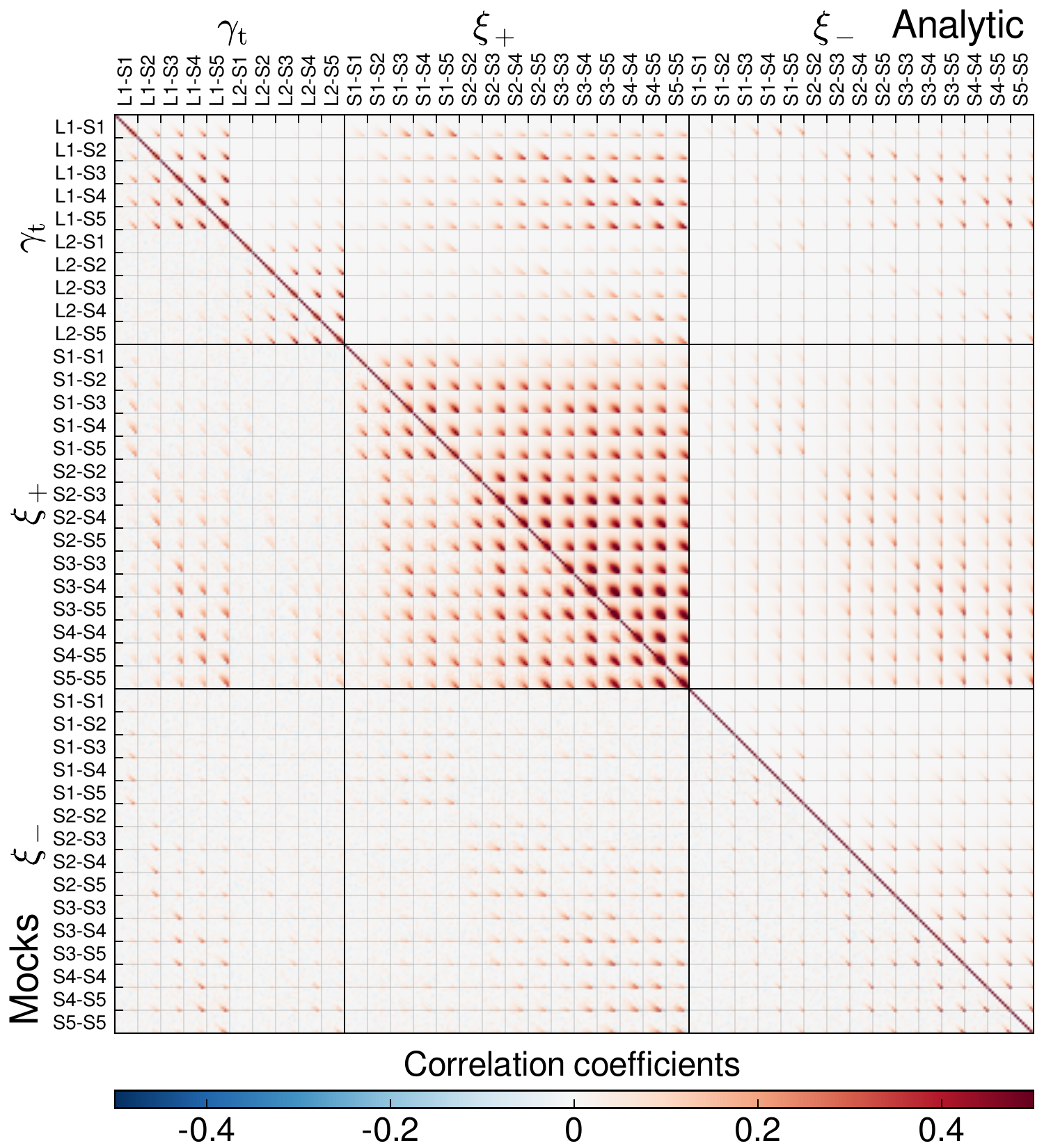}
	\caption{Comparison of the correlation coefficients in the weak lensing correlation function covariance between the mocks (lower left) and analytic approach (upper right) in the Egretta setup. As indicated in the labels, large blocks correspond to the correlation functions $\ba{\gamma_{\rm t}}$, $\xi_+$, and $\xi_-$, while small blocks correspond to the tomographic bin combinations.}
	\label{fig:corrComp_egretta_direct}
\end{figure*}

\begin{figure}
	\includegraphics[width=\columnwidth]{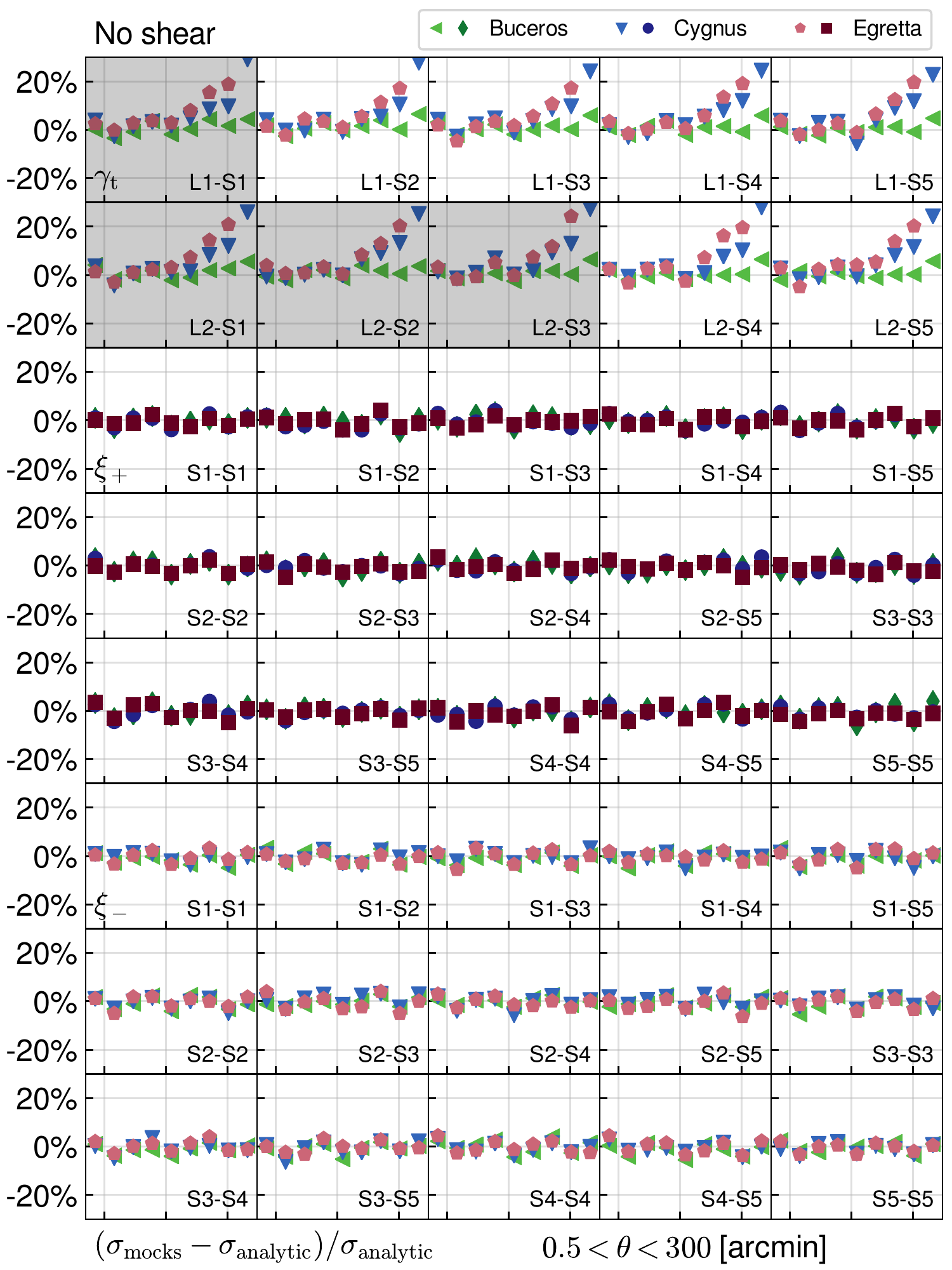}
	\caption{Same as Fig.~\ref{fig:stdDiff_egretta_direct}, but with all weak lensing signals removed from both the analytic and mock covariances, i.e. only noise and sample variance due to clustering contribute.}
	\label{fig:stdDiff_egretta_noshear_direct}
\end{figure}

\begin{figure}
	\includegraphics[width=\columnwidth]{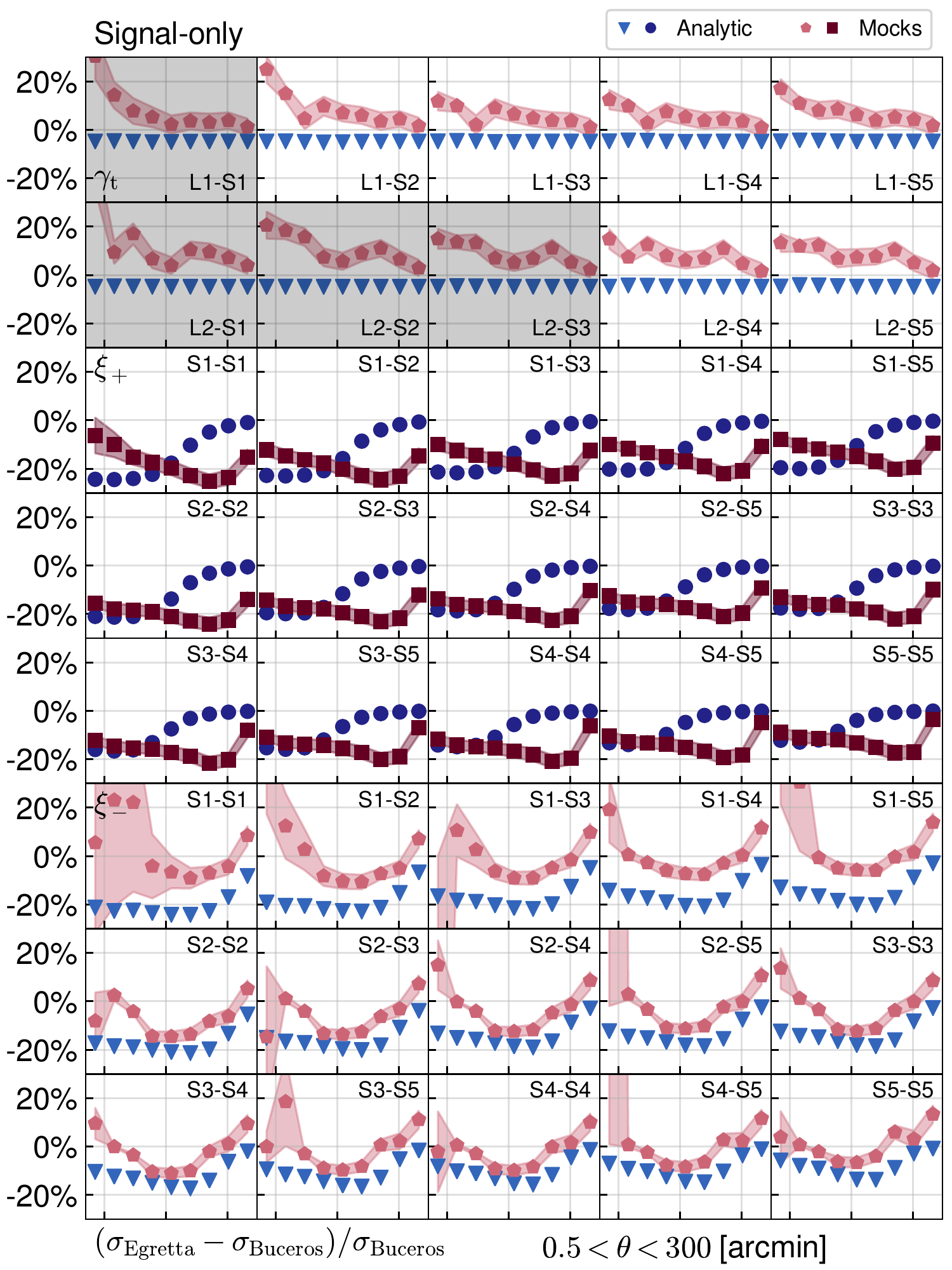}
	\caption{Relative difference between the square root of the diagonals in the Egretta (realistic mask and depth variations) and Buceros (rectangular mask, uniform depth) covariances of the weak lensing correlation functions, with all shape noise contributions removed ($\sigma_\epsilon=0$). Red (blue) symbols show results for the mock (analytic) covariance. Bands around the mock data points indicate the standard error determined from a jackknife estimate of variance. The top two rows show GGL signals, the centre three rows $\xi_+$, and the bottom three rows $\xi_-$, with bin combinations indicated in the panels.}
	\label{fig:stdDiff_egretta_signal_mockcomp}
\end{figure}

\begin{figure}
	\includegraphics[width=\columnwidth]{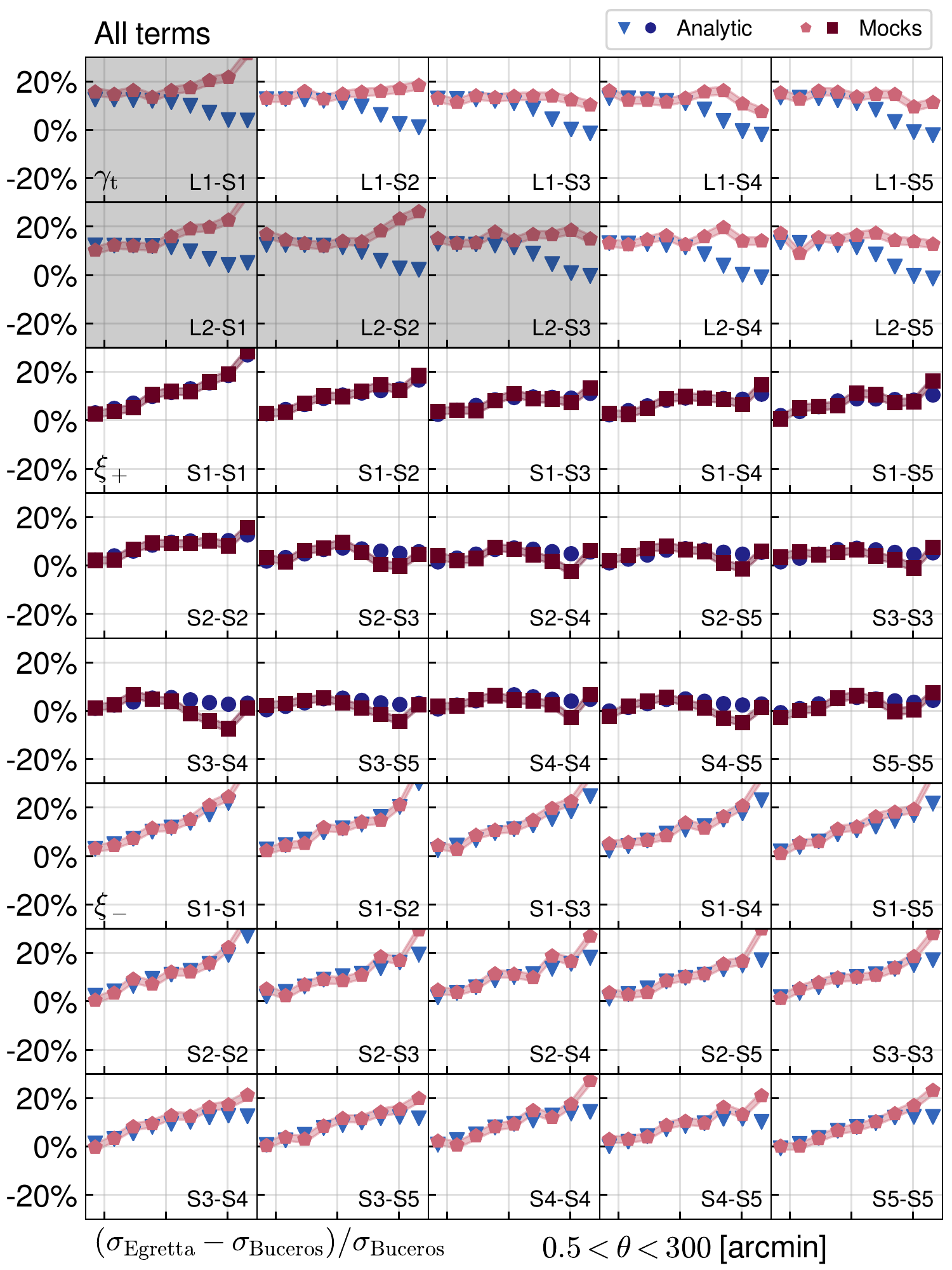}
	\caption{Same as Fig.~\ref{fig:stdDiff_egretta_signal_mockcomp}, but for the full covariance with all cosmological contributions included.}
	\label{fig:stdDiff_footprint_direct}
\end{figure}

It is instructive to study the covariance models of the weak lensing correlation functions as an intermediate step to calculating band power covariance and as a view that isolates any deviations that are localised in configuration space. We compare analytic and mock covariances in the Buceros (simple rectangular survey geometry, uniform galaxy distributions), Cygnus (realistic survey footprints, uniform galaxy distributions), and Egretta (realistic footprint and spatially varying galaxy distributions) cases for tangential shear and the cosmic shear correlation functions $\xi_\pm$ measured in nine angular bins spread equidistantly in the log between $0.5\,{\rm arcmin}$ and $300\,{\rm arcmin}$ (i.e. the same scheme as used in KV450). In these comparisons the GGL estimator was applied to the mock catalogues with 100 times more random points than lens galaxies in order to suppress any residual contributions of terms that the random correction of $\langle \gamma_{\rm t} \rangle$ removes (see Appendix~\ref{sec:anacov} for details).

Figure~\ref{fig:stdDiff_egretta_direct} shows the ratio of the square root of the diagonal elements of the mock and analytic covariances for the three survey configurations. Off-diagonals in the form of correlation coefficients are shown in Fig.~\ref{fig:corrComp_egretta_direct} for the most realistic Egretta case (the Fourier space analogues of these plots are Figs.~\ref{fig:std_diff_egretta} and \ref{fig:corr_comp_egretta}). We generally find very good agreement between the mocks and the analytic approach in all cases. Significant deviations are limited to the largest-scale data point in $\xi_\pm$ and to scales larger than $10\,$arcmin for GGL, with the mock standard deviations up to $20\,\%$ larger. This under-prediction by the analytic model is because its Gaussian sample variance terms ignore survey boundary effects which enhance them by up to a factor two when switching from a simple rectangular footprint to the realistic KiDS survey geometry (Buceros to Cygnus)\footnote{There is a further complication not included in the analytic model in that the GGL correlations include lenses beyond the limits of the source sample footprint. We employ the lens survey area in the analytic calculation of the sample variance terms, but the effective survey area is somewhat reduced because sources with gravitational shear estimates are not available over the full lens survey area. We estimate this to lead to a $2-3\,\%$ under-prediction by the analytic model of the GGL standard deviation on large scales, which can therefore not be the main reason behind the discrepancies seen.}. Variable depth effects (Cygnus to Egretta) have negligible impact on the cosmic shear covariance beyond modifications to the galaxy pair counts, but cause a small additional increase in the GGL covariance.

To gain a better understanding of which covariance terms drive certain discrepancies, we create two special cases that are readily realised in both the analytic and simulation approaches. First, we remove all shear signals, which leaves us with only noise terms and, in the GGL case, with mixed terms that combine clustering with shape noise (referred to as the \lq no-shear\rq\ case). Secondly, we set $\sigma_\epsilon=0$ which removes all terms containing shape noise but keeps all Gaussian and non-Gaussian sample variance contributions, plus GGL mixed terms involving clustering shot noise (referred to as the \lq signal-only\rq case). As discussed in Sect.~\ref{sec:covcomp}, we do not expect the non-Gaussian covariance terms to match quantitatively between mock and analytic model. Hence, instead of directly comparing the two in the signal-only case, we investigate how well the two approaches agree in the changes between the different survey configurations.

The no-shear case in Fig.~\ref{fig:stdDiff_egretta_noshear_direct} demonstrates excellent agreement in the cosmic shear noise terms; using the galaxy pair counts of the measurements in the analytic covariance correctly accounts for survey geometry and variable depth. The large-scale deviations are however still present in the GGL covariance and are caused by the residual mixed noise-sample variance contribution that in our model disregards survey boundary effects. These mixed terms, together with the Gaussian sample variance term which is of similar size (cf. Fig.~\ref{fig:cov_contribution}), also drive the deviations seen in Fig.~\ref{fig:stdDiff_egretta_direct}. We note in passing that using Eq.~(\ref{eqn:npairs_ggl}) for the calculation of the GGL noise is critical for the high accuracy shown here. A naive simple pair count of lens and source galaxy pairs leads to $25\,\%$ difference in the standard deviations on small angular scales when the source and lens bins have substantial overlap.

For the interpretation of the signal-only case in Fig.~\ref{fig:stdDiff_egretta_signal_mockcomp} it is helpful to bear in mind that the only contributing analytic covariance term that is sensitive to survey geometry is the super-sample covariance (SSC). As SSC is suppressed in GGL, the analytic model predicts no geometry dependency, which the mock result suggests is plausible above ca. $10\,$arcmin. Below however, non-Gaussian contributions are seen to cause excess covariance in the Egretta case, but it remains unclear to what degree this effect is influenced by the lognormal and linear galaxy bias assumptions, as well as the resolution limit, in our simulations. As regards the cosmic shear signals, mocks and analytic model generally agree fairly well in that the signal-only covariance is suppressed in the Egretta relative to the Buceros case\footnote{While having the same total area, the Egretta footprint is spread out over a larger fraction of the sky, primarily due the separation between KiDS-N and KiDS-S. Therefore, the Egretta footprint is able to accommodate larger modes of the large-scale matter density distribution, thereby decreasing super-sample covariance.}, while on large scales discrepancies occur, again due to the neglect of survey geometry effects in the Gaussian sample variance. It is interesting to note that even in the signal-only case the differences in the standard deviation between analytic and mock covariances never exceed $30\,\%$.

Patterns in the correlation structure away from the diagonal are well represented in the analytic covariance, with some small deviations discernible that, analogously to the Fourier space case, occur in terms dominated by Gaussian covariance contributions, but interestingly with a reversed sign, that is for correlation functions the analytic correlation coefficients are larger than their mock counterparts (compare Figs.~\ref{fig:corrComp_egretta_direct} and \ref{fig:corr_comp_egretta}). For completeness, we also plot the relative difference between the Egretta and Buceros cases for the full covariance in Fig.~\ref{fig:stdDiff_footprint_direct} (the analogue of Fig.~\ref{fig:std_diff_egretta_footprint}). This figure confirms that the analytic covariance model overall performs very well in recovering the relevant effects of survey geometry and spatial variations. The small residual discrepancies, primarily on large angular scales, mirror those discussed in Fig.~\ref{fig:stdDiff_egretta_direct}.

\subsection{Band power covariance}
\label{sec:bpcov}

\begin{figure}
	\includegraphics[width=\columnwidth]{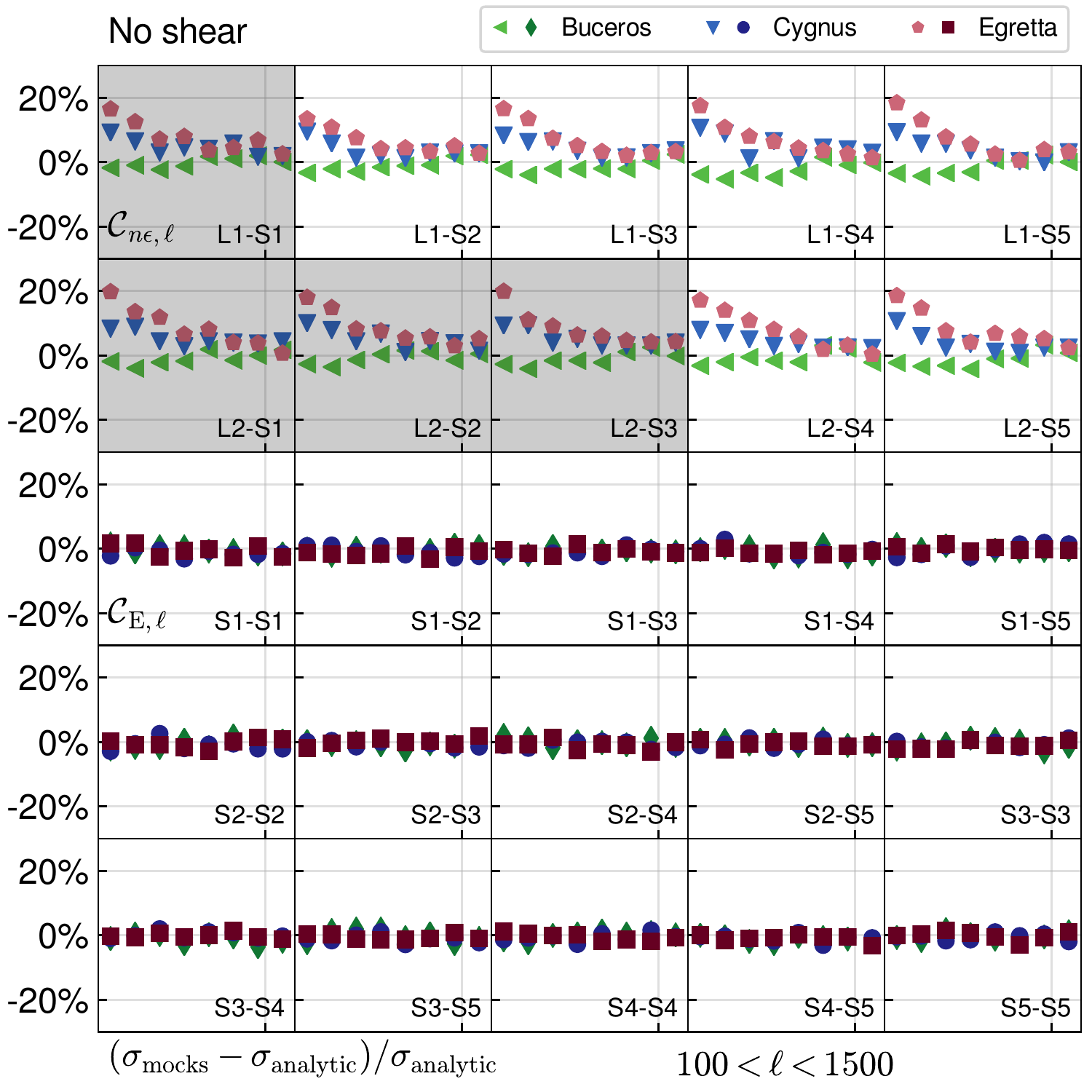}
	\caption{Same as Fig.~\ref{fig:std_diff_egretta}, but with all weak lensing signals removed from both the analytic and mock covariances, i.e. only noise and sample variance due to clustering contribute.}
	\label{fig:stdDiff_egretta_noise}
\end{figure}

\begin{figure}
	\includegraphics[width=\columnwidth]{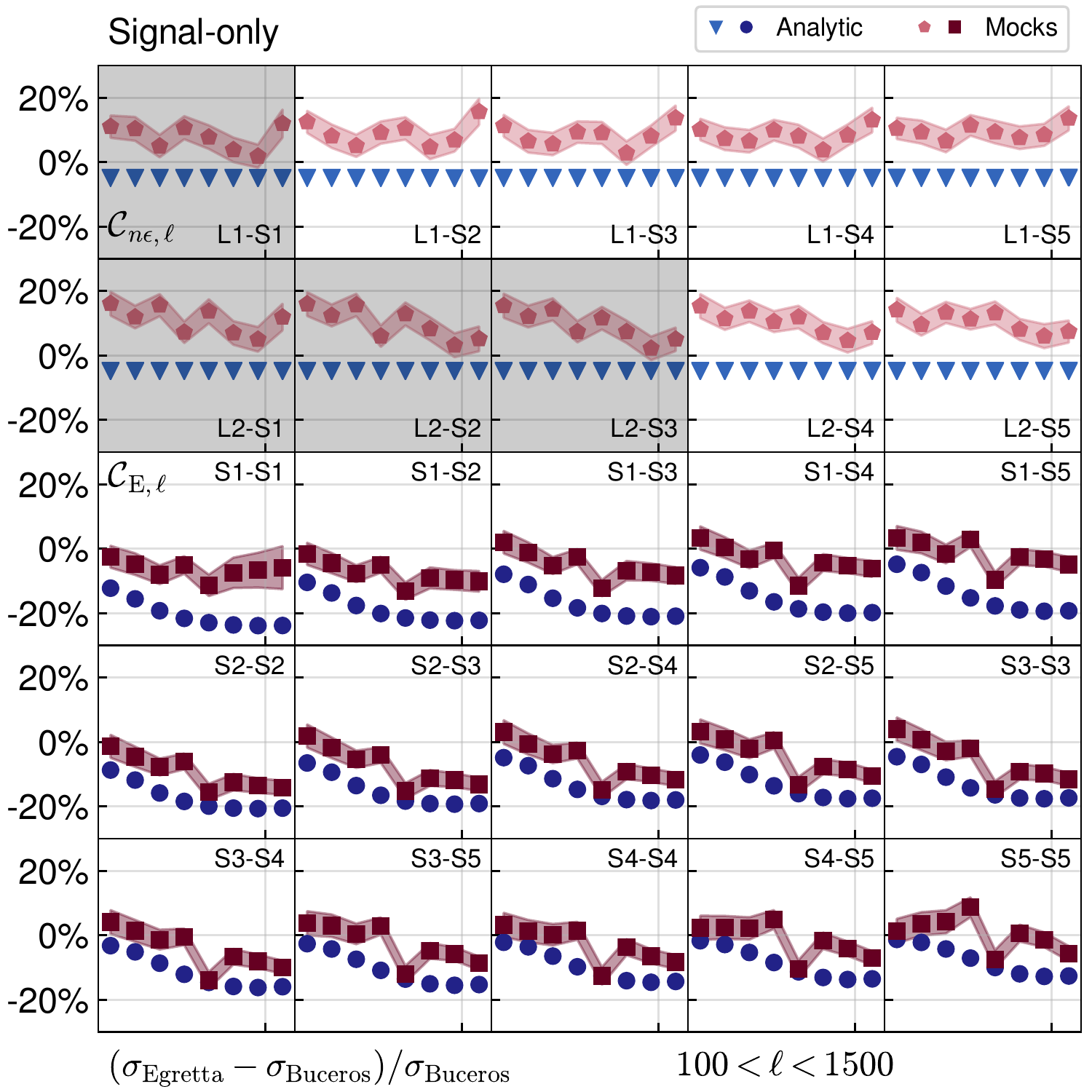}
	\caption{Same as Fig.~\ref{fig:std_diff_egretta_footprint}, but with all shape noise contributions to the analytic and mock covariances removed ($\sigma_\epsilon=0$).}
	\label{fig:stdDiff_egretta_signal_bp_mockcomp}
\end{figure}

\begin{figure}
\includegraphics[width=\columnwidth]{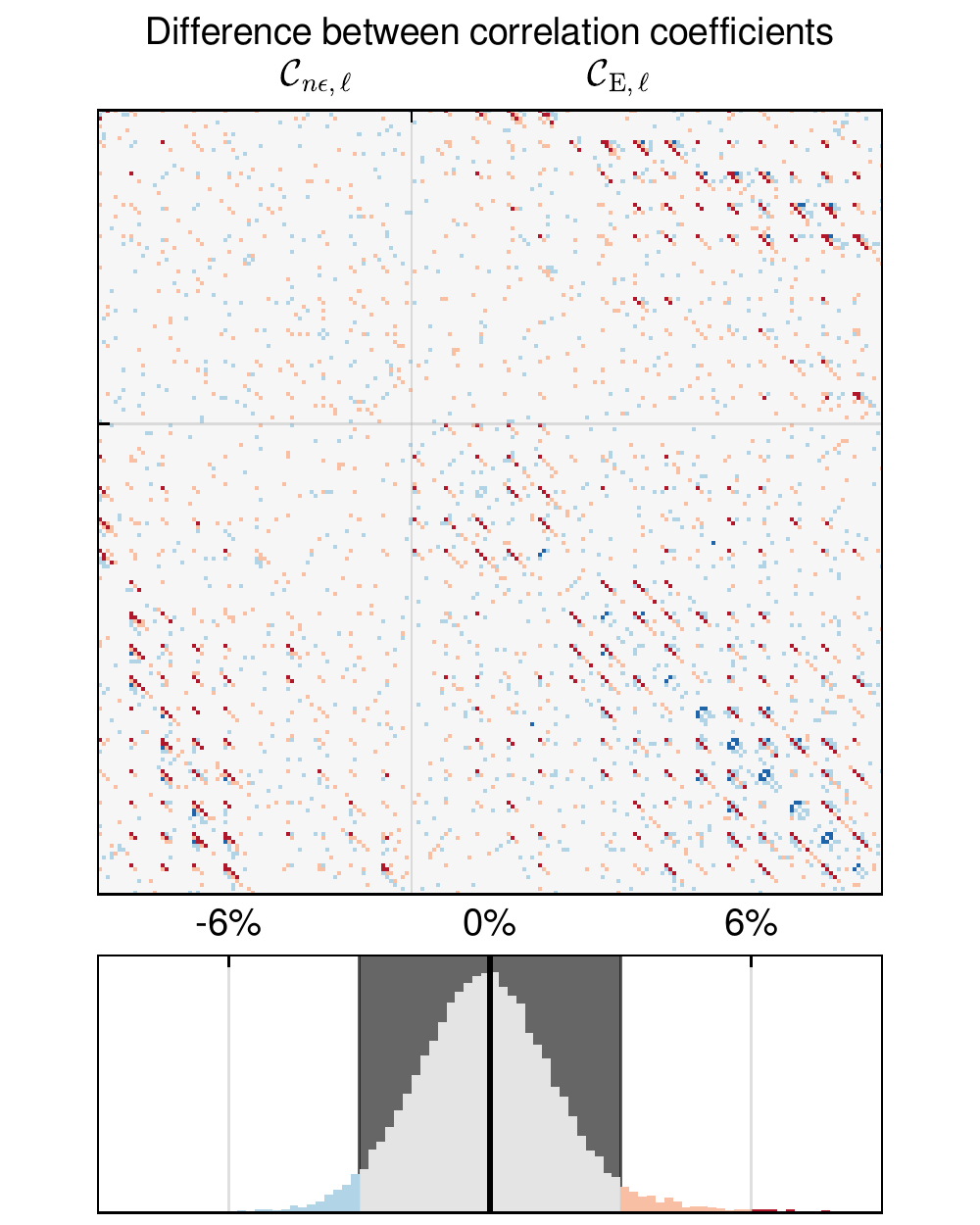}
\caption{Relative difference between the correlation coefficients of the mock and analytic covariances of the weak lensing band power signals. \textit{Bottom}: Histogram of the relative differences marking values beyond $\pm 6\,\% (\pm 3\,\%)$ in dark (light) red and blue. \textit{Top}: Band power correlation matrix with elements coloured according to the value of the difference using the same scheme as in the matrix above.}
\label{fig:corr_diff_hist_egretta}
\end{figure}

Here we provide additional band power covariance comparison plots that further illustrate the fidelity and importance of individual contributions to the covariance model: Fig.~\ref{fig:stdDiff_egretta_noise} compares the mock and analytic approaches in the no-shear case; Fig.~\ref{fig:stdDiff_egretta_signal_bp_mockcomp} shows the relative difference between the Egretta and Buceros setups in the signal-only case. The observed trends are in line with the configuration space covariances. The noise terms agree well, whereas the idealised mixed sample variance-noise terms in the analytic model fail to capture survey geometry effects on large scales in the GGL covariance. While the analytic model suggests minimal impact on sample variance in GGL due to survey geometry and/or variable survey depth, the mocks yield a $10-20\,\%$ excess standard deviation switching from Buceros to Egretta, but it is not clear if this is physical or due to limitations in the mock creation. The trend is reversed for cosmic shear, driven by SSC. Mocks and analytic prediction agree on the scale dependence of this effect, with the latter slightly over-predicting its magnitude.

Figure~\ref{fig:corr_diff_hist_egretta} highlights the differences in correlation coefficients between the mock and analytic covariance in the Egretta case. For the vast majority of coefficients the differences are consistent with scatter due to the finite number of mock realisations. Systematic differences lie in the very thin tails, mostly at the positive end, indicating a larger mock value. These typically occur in the cross-variance between cosmic shear and GGL and are dominated by the mixed term.

\onecolumn  

\section{Analytic covariance model}
\label{sec:anacov}

Here we provide a detailed description of our analytic covariance model. This model is used for the cosmic shear and galaxy-galaxy lensing (GGL) signals only, while the clustering covariance is obtained from the public mocks provided by BOSS. With our own mocks we demonstrate that the cross-variance between the clustering and lensing observables can safely be neglected. For completeness we include the analytic expressions for angular clustering as well. We calculate real-space correlation function covariances first, and then obtain the band power covariance from these expressions analogously to how the signals are derived. While in principle it is possible to go directly from angular power spectrum models to band powers following Eqs.~(\ref{eq:bp_cosmicshear},\ref{eq:bpobsggl}), we choose this approach for two reasons: first, we already have validated correlation function covariance software in hand from previous KiDS cosmology analyses \citep{hildebrandt17,uitert18,kv450}, and secondly, it is easier to incorporate survey effects in real space, such as measured galaxy pair counts.

\subsection{Gaussian real-space covariance}
\label{sec:cov_model_gaussian}

\begin{figure}
\begin{minipage}{0.6\textwidth}
\includegraphics[width=0.95\columnwidth]{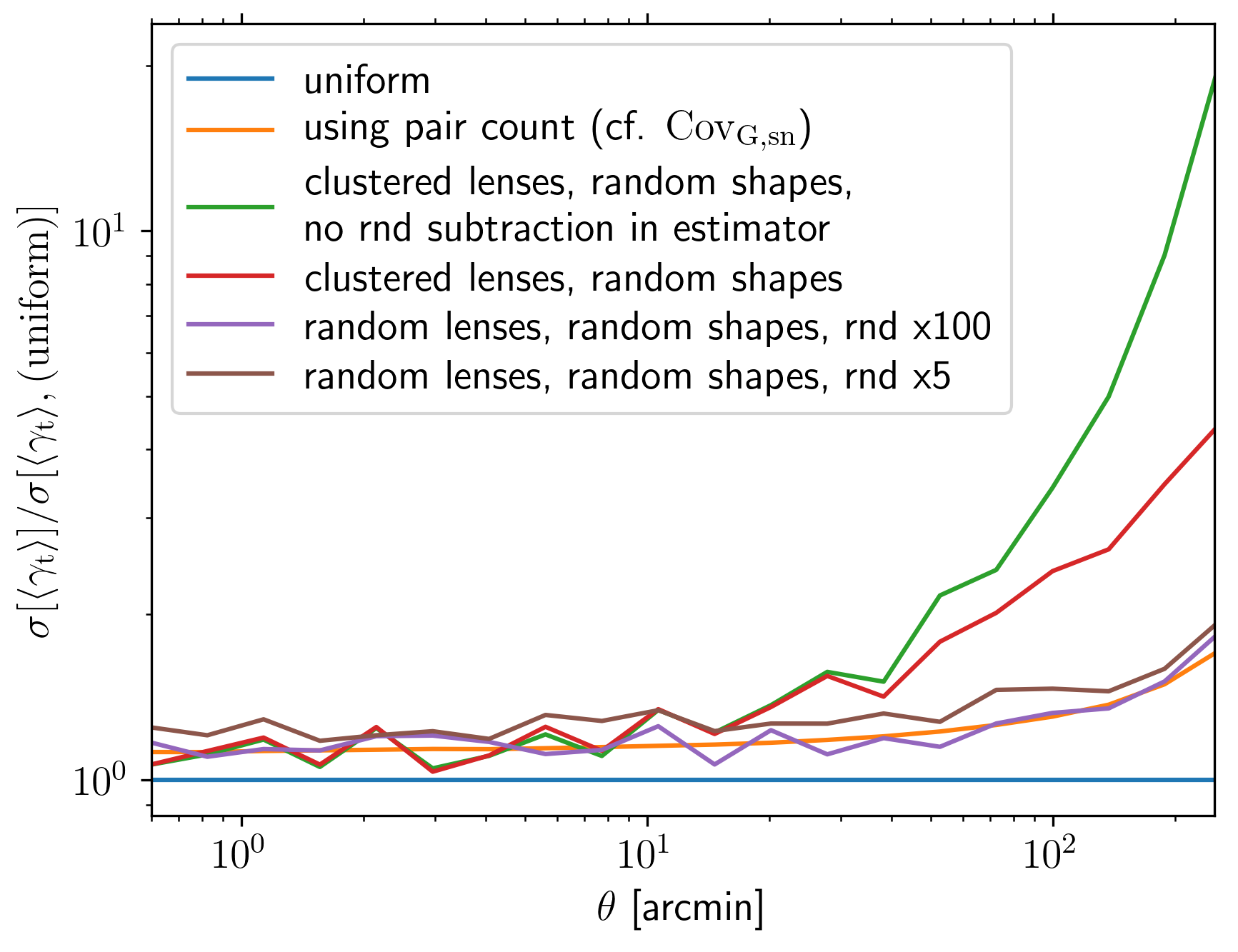}
\end{minipage}%
\begin{minipage}{0.4\textwidth}
\caption{Noise contributions to a GGL tangential shear correlation function measurement for BOSS lenses in the KiDS-1000 overlap (L1) and KiDS high-redshift sources (S5). Shown is the standard deviation as a function of angular separation, normalised by the analytic expectation for uniformly distributed galaxies and neglecting survey boundaries (blue). The noise term as implemented by Eq.~(\ref{eq:g_sn}) is shown in orange. It is in excellent agreement with the sample variance obtained from a measurement with randomised source galaxy shapes and lens galaxy positions, i.e. removing all sample variance contributions (purple). Decreasing the oversampling factor of the random catalogue from our default of ${\cal N}_{\rm rnd}=100$ to only five results in an increased noise level (brown). We also show the signals when only randomising shapes (red) and when additionally not subtracting the GGL signal around random points in the lens sample (green).}
\label{fig:ggl_noiseterms}
\end{minipage}
\end{figure}

We begin with the Gaussian covariance, which corresponds to the full contribution if the underlying gravitational lensing convergence and galaxy number density fields were Gaussian. It consists of a sample variance term (\lq sva\rq), the sampling error due to observing a finite volume of the Universe, shape noise or shot noise (\lq sn\rq), as we observe galaxies as point processes sampling the underlying fields, and a mixed term (\lq mix\rq). We opt to derive the real-space sample variance expression from its Fourier counterpart as proposed by \citet{joachimi08}, leading to
\eq{
\label{eq:cov_g_sva}
{\rm Cov}_{\rm G, sva} \bb{\Xi_\mu^{(ij)}(\bar{\theta}_1);\, \Xi_\nu^{(kl)}(\bar{\theta}_2)} = \frac{1}{2 \pi A_{{\rm max},\mu \nu} } \int_0^\infty \dd \ell\, \ell\; {\cal K}_\mu (\ell \bar{\theta}_1) {\cal K}_\nu (\ell \bar{\theta}_2) \bc{ C^{(ik)}(\ell) C^{(jl)}(\ell) + C^{(il)}(\ell) C^{(jk)}(\ell) } \;, 
}
where we introduced a unified notation for correlation functions with the correspondence $\bc{w, \ba{\gamma_{\rm t}}, \xi_+, \xi_-} \leftrightarrow \bc{\Xi_0, \Xi_2, \Xi_0, \Xi_4}$. The subscripts denote the type of integration kernel that is applicable, with $\xi_+$ and the angular clustering correlation function $w$ sharing the same kernel. The kernels are defined as
\eq{
{\cal K}_\mu (\ell \bar{\theta}_i) := \frac{2}{\theta_{{\rm u},i}^2 - \theta_{{\rm l},i}^2 }\, \int_{\theta_{{\rm l},i}}^{\theta_{{\rm u},i}} \dd \theta'\, \theta' {\rm J}_\mu(\ell \theta') =  \frac{2}{\br{ \theta_{{\rm u},i}^2 - \theta_{{\rm l},i}^2 } \ell^2}\, \times\, \left\{ \begin{matrix} \bb{x {\rm J}_1(x)}_{\ell \theta_{{\rm l},i}}^{\ell \theta_{{\rm u},i}}  & \mu=0 \\ \bb{ -x {\rm J}_1(x) - 2 {\rm J}_0(x) }_{\ell \theta_{{\rm l},i}}^{\ell \theta_{{\rm u},i}}  & \mu=2 \\ \bb{ \br{x -\frac{8}{x}} {\rm J}_1(x) - 8 {\rm J}_2(x) }_{\ell \theta_{{\rm l},i}}^{\ell \theta_{{\rm u},i}}  & \mu=4  \end{matrix} \right.  \,,
}
where the ${\rm J}_\mu$ are cylindrical Bessel functions of the first kind. As opposed to earlier works, we have explicitly averaged over the angular bin centred on $\bar{\theta}_i$ in which $\Xi_\mu$ is measured, delimited by $\bb{\theta_{{\rm l},i} ;\, \theta_{{\rm u},i}}$. The type of angular power spectrum $C^{(ij)}(\ell)$ to be used in the integrand of Eq.~(\ref{eq:cov_g_sva}) is determined by its tomographic bins $i$ and $j$: if both are lens bins, it is a clustering power spectrum; if both are source bins, it is a cosmic shear power spectrum; else one uses a position-shear cross power spectrum. For a more explicit notation in this regard see \citet{joachimi10}. The power spectra are determined under the extended Limber approximation from the full non-linear matter power spectrum evaluated at our fiducial choice of parameters. The power spectra are calculated using the non-linear prescription by \citet{takahashi12} as AGN feedback effects have negligible impact on the covariance in the regime where sample variance contributions are significant \citep[see also][]{schneider20}. Lensing signals generally include intrinsic alignment contributions in the Gaussian terms, but they have been switched off for this study to simplify the comparison with the mocks that do not feature intrinsic alignments. We assume an effective linear galaxy bias and choose the fiducial values of $b_1$ for this purpose as the corrections due to the non-linear bias terms are small.

Following the derivation in \citet{uitert18}, we normalise the covariance term by $A_{{\rm max},\mu \nu} = {\rm max}(A_{{\rm eff},\mu}, A_{{\rm eff},\nu})$, that is the effective survey area applicable to the signal in the case of auto-correlations and the \emph{larger} effective area when cross-correlating signals measured over different parts of the sky. The areas to be used are the full BOSS footprint for clustering, the full KiDS-1000 area for cosmic shear, and the overlap area for GGL; see Table~\ref{tab:surveyareas} for numerical values. The effective survey area is not a quantity defined from first principles and ultimately depends on the chosen resolution at which the survey footprint is considered. We measure the effective area from a binary \textsc{Healpix} mask with $N_{\rm side}=4096$. Hence, mask features of less than arcminute size will not reduce $A_{\rm eff}$. Since we use this area in $n_{\rm eff}$ as well (see Eq.~\ref{eqn:neff}), star masks and other small-scale features are interpreted as diluting the number density of galaxies rather than the survey area. This is in line with the covariance modelling assumptions as long as these small-scale features are below the scales at which the cosmological signal is measured. We have also computed a cosmic shear covariance based on an $N_{\rm side}=2048$ mask, but found no measurable difference on $S_8$ best fits or its errors.

The pure noise term only contributes to the diagonals of auto-correlations and is given by \citep{schneider02}
\eq{
\label{eq:g_sn}
{\rm Cov}_{\rm G, sn} \bb{\Xi_\mu^{(ij)}(\bar{\theta}_1);\, \Xi_\nu^{(kl)}(\bar{\theta}_2)} = \delta_{\bar{\theta}_1 \bar{\theta}_2} \br{\delta_{ik} \delta_{jl} + \delta_{il} \delta_{jk} } \frac{{\cal T}_{\mu \nu}^{\rm sn} }{N_{\rm pair}^{(ij)}(\bar{\theta}_1) } \;, 
}
where $\delta_{ij}$ denotes a Kronecker delta, and where $N_{\rm pair}^{(ij)}$ is given by Eqs.~(\ref{eqn:npairs}) and (\ref{eqn:npairs_ggl}). We defined
\eq{
{\cal T}_{\mu \nu}^{\rm sn} :=  \left\{ \begin{matrix} \sigma_\epsilon^4/2 & \mu=\nu=0 ~\mbox{or}~ \mu=\nu=4 & \mbox{(cosmic shear)}\\ \sigma_\epsilon^2/2 & \mu=\nu=2 & \mbox{(GGL)}\\ 1 & \mu=\nu=0  & \mbox{(clustering)}\\ 0 & \mu \neq \nu & \end{matrix} \right.  \;,
}
where $\sigma_\epsilon$ is the total dispersion of the complex observed galaxy ellipticity (with contributions from the intrinsic ellipticities of galaxies and measurement noise) that is in practice measured via Eq.~(\ref{eqn:sige}). It was furthermore assumed that the noise in galaxy clustering follows a Poisson distribution. The mixed term reads
\eq{
\label{eq:cov_g_mix}
{\rm Cov}_{\rm G, mix} \bb{\Xi_\mu^{(ij)}(\bar{\theta}_1);\, \Xi_\nu^{(kl)}(\bar{\theta}_2)} = \delta_{jl}\, \frac{{\cal T}_j^{\rm mix}}{2 \pi n_{\rm eff}^{(j)}\, A_{{\rm max},\mu \nu} } \int_0^\infty \dd \ell\, \ell\; {\cal K}_\mu (\ell \bar{\theta}_1) {\cal K}_\nu (\ell \bar{\theta}_2)\, C^{(ik)}(\ell) + \mbox{4 perm.}\;, 
}
where we defined
\eq{
{\cal T}_{j}^{\rm mix} :=  \left\{ \begin{matrix}  \sigma_\epsilon^2/2 & j \in {\rm S}1-{\rm S}5 \\ 1 & j \in {\rm L}1-{\rm L}2 \end{matrix} \right.  \;
}
for our lens (L) and source (S) bins, respectively. Figure~\ref{fig:ggl_noiseterms} illustrates the noise contributions to the variance of an exemplary GGL tangential shear signal (L1--S5). Sample variance contributions are switched off selectively by randomising the orientations of source galaxy shapes and/or the positions of lens galaxy positions.

First, the measured sample variance is offset by $13\,\%$ over the expectation for uniformly distributed galaxies in the absence of any survey boundaries for which $N^{(ij)}_{\rm pair}(\theta_i)= \pi (\theta_{{\rm u},i}^2 - \theta_{{\rm l},i}^2) A_{\rm eff} n^{(i)}_{\rm eff} n^{(j)}_{\rm eff}$ is inserted into Eq.~(\ref{eq:g_sn}). This is an example of the aforementioned importance of the choice of effective survey area. We used the BOSS area in the calculation which is larger than the source area due to the more aggressive masking in weak lensing surveys, thereby underestimating the true noise even on small scales where survey boundaries are irrelevant. Our noise term model closely follows the measured GGL standard deviation without any sample variance (i.e. randomised lenses and sources). An oversampling factor of 100 of the random catalogue is sufficient to suppress any additional noise contributions, which we therefore use as our default.

Only randomising the sources preserves a mixed term in the resulting variance, namely a combination of shape noise and angular clustering signal; cf. Eq.~(\ref{eq:cov_g_mix}). Figure~\ref{fig:ggl_noiseterms} shows that this term dominates noise components on large scales, so that idealisations in the modelling of this term will limit the accuracy of the analytic model, as can be seen in Fig.~\ref{fig:stdDiff_egretta_noshear_direct}. Moreover, the figure demonstrates that the subtraction of the GGL signal around random lens points in the estimator (Eq.~\ref{eq:ggl_estimator}) is crucial to suppress additive noise contributions that scale with shape noise and the survey footprint of the lens sample, which would otherwise substantially increase errors on large scales (see the detailed discussion in \citealp{singh17}). Equation~(\ref{eq:g_sn}) was previously shown to be an excellent fit to the noise term of cosmic shear signals; see Fig.~\ref{fig:stdDiff_egretta_noshear_direct} and \citet{troxel18b}.

The ostensibly simplest component, the Gaussian sample variance contribution, currently limits the accuracy of this model (see Appendix~\ref{sec:covcomp_app}). It dominates on the largest scales where the finite extent of the survey footprint affects the sample variance substantially and breaks the assumption of isotropy; see also the discussion in \citet{blake20}. Eqs.~(\ref{eq:cov_g_sva}) and (\ref{eq:cov_g_mix}) acquire their simple form by neglecting survey boundaries altogether. In future, these effects, as well as variations in survey depth, will be easier to account for in a configuration space approach; see \citet{kilbinger04,hikage19}.

\subsection{Non-Gaussian real-space contributions}

Since the galaxy density and weak lensing convergence distributions are highly non-Gaussian on small scales, the covariance picks up additional terms via the connected four-point function of these fields. These are conveniently split into matter trispectrum contributions from modes within the survey footprint and those that link in-survey modes to those with wavelengths larger than the survey which act to rescale the mean of the field inside the survey. The former are given by \citep{takada04}
\eq{
{\rm Cov}_{\rm NG} \bb{\Xi_\mu^{(ij)}(\bar{\theta}_1);\, \Xi_\nu^{(kl)}(\bar{\theta}_2)} = \frac{1}{4 \pi^2 A_{{\rm max},\mu \nu} } \int_0^\infty \dd \ell_1\, \ell_1\;  {\cal K}_\mu (\ell_1 \bar{\theta}_1) \int_0^\infty \dd \ell_2\, \ell_2\;  {\cal K}_\nu (\ell_2 \bar{\theta}_2) \int_0^\pi \frac{\dd \varphi_\ell}{\pi} \; T^{(ijkl)}(\vec{\ell}_1,\vec{\ell}_2,-\vec{\ell}_1,-\vec{\ell}_2)\;, 
}
where $\varphi_\ell$ is the angle between $\vec{\ell}_1$ and $\vec{\ell}_2$. The angular trispectrum is calculated by integrating along the line of sight over the matter trispectrum, $T_{\rm m}$, yielding
\eq{
T^{(ijkl)}(\vec{\ell}_1,\vec{\ell}_2,\vec{\ell}_3,\vec{\ell}_4) = \int_0^{\chi_{\rm hor}} \dd \chi\; \frac{W^{(i)}_{\rm a} (\chi)\, W^{(j)}_{\rm b} (\chi)\, W^{(k)}_{\rm c} (\chi)\, W^{(l)}_{\rm d} (\chi)}{f_{\rm K}^6(\chi)}\; T_{\rm m} \br{\frac{\vec{\ell}_1}{f_{\rm K}(\chi)}, \frac{\vec{\ell}_2}{f_{\rm K}(\chi)}, \frac{\vec{\ell}_3}{f_{\rm K}(\chi)}, \frac{\vec{\ell}_4}{f_{\rm K}(\chi)} }\;
}
under Limber's approximation. The type of kernel $W_{a,b,c,d}$ is chosen according to the probes $\Xi$ under consideration from among Eq.~(\ref{eq:kernel_lensing}) for weak lensing and the comoving distance distribution of lens samples plus an expression for galaxy bias for clustering.

The super-sample covariance (SSC) term reads \citep{takada13}
\eqa{
\label{eq:ssc}
{\rm Cov}_{\rm SSC} \bb{\Xi_\mu^{(ij)}(\bar{\theta}_1);\, \Xi_\nu^{(kl)}(\bar{\theta}_2)} &= \frac{1}{4 \pi^2} \int_0^\infty \dd \ell_1\, \ell_1\;  {\cal K}_\mu (\ell_1 \bar{\theta}_1) \int_0^\infty \dd \ell_2\, \ell_2\;  {\cal K}_\nu (\ell_2 \bar{\theta}_2)\; \\ \nn
&\times\; \int_0^{\chi_{\rm hor}} \dd \chi\; \frac{W^{(i)}_{\rm a} (\chi)\, W^{(j)}_{\rm b} (\chi)\, W^{(k)}_{\rm c} (\chi)\, W^{(l)}_{\rm d} (\chi)}{f_{\rm K}^6(\chi)}\; \frac{\partial P_{\rm m} \bb{\ell_1/f_{\rm K}(\chi)}}{\partial \delta_{\rm b}}\; \frac{\partial P_{\rm m} \bb{\ell_2/f_{\rm K}(\chi)}}{\partial \delta_{\rm b}}\; \sigma_{{\rm bg}, \mu \nu}^2(\chi)\;,
}
where the derivatives denote the response of the matter power spectrum to a change in the density contrast of the background $ \delta_{\rm b}$, which is defined as the average density contrast within the volume of the survey. Here, we defined the variance of background matter fluctuations within the observability masks relevant to the two probes under consideration (indicated by super-/subscripts $\mu$ and $\nu$),
\eq{
\sigma_{{\rm bg}, \mu \nu}^2(\chi) = \frac{1}{A_{{\rm eff},\mu}\, A_{{\rm eff},\nu} } \sum_\ell P_{\rm m,lin} \br{\frac{\ell}{f_{\rm K}(\chi)}} \sum_m a_{\ell m}^\mu\; a_{\ell m}^{\nu *} \;,
}
where the linear matter power spectrum has been used as only linear scales affect the background fluctuations. As surveys now cover substantial fractions of the sky, we drop the flat-sky approximation in this term and express the cross-power of the survey masks via their spherical harmonic coefficients $a_{\ell m}^{\mu,\nu}$. These are determined from the same binary \textsc{Healpix} masks used to calculate the effective survey area. For probes covering the same sky area the summation over $m$ simplifies to the isotropic power spectrum of the mask. We do not include contributions caused by super-survey tidal fields, which are expected to have little impact on our scales of interest but could attain similar levels as the (small) NG contribution \citep{barreira18b}.

To evaluate the matter trispectrum and the matter power spectrum response to the background we opt for a halo model formalism, closely following \citet{takada13,lin14}; see also \citet{krause17} for a similar implementation. We refer the reader to these works for a quantitative description and only summarise the relevant modelling choices here, which are unchanged with respect to earlier KiDS cosmology analyses \citep{hildebrandt17,uitert18,kv450}. Our halo model is based on the halo mass function and halo bias of \citet{tinker10}. It assumes a \citet{navarro96} halo profile with the concentration-mass relation by \citet{duffy08} and employs the analytical form of the profile's Fourier transform by \citet{scoccimarro01}. While implemented, some of the particularly computationally expensive 2-halo terms in the matter trispectrum have been switched off for most practical covariance calculations because they only make negligible contributions to the non-Gaussian covariance, which has little impact on the overall statistical errors to begin with. The logarithmic matter power spectrum response is reduced by two for both clustering \citep{deputter12,takada13} and GGL \citep{singh17} signals as their estimators are normalised to the mean galaxy densities within the survey footprint rather than the global mean density through the use of random catalogues. An effective linear bias is used to translate from polyspectra of matter to those involving galaxy density contrast, and we choose the fiducial values of $b_1$ for this purpose.

\subsection{Covariance of band powers}
\label{sec:cov_model_bp}

Together with the covariance term sourced through uncertainty in the multiplicative bias correction (Eq.~\ref{eq:cov_mbias}), the expressions above constitute the full covariance for angular large-scale structure correlation functions. Various useful two-point statistics can be derived from the correlation functions, which capitalises on the insensitivity of the latter to the often complex survey geometry. Since the relations are linear, simple error propagation allows us to derive corresponding relations between the covariances of these two-point statistics. For the case of band powers, employing Eqs.~(\ref{eq:bp_eb}) and (\ref{eq:bp_ne}) leads to
\eqa{
\label{eq:cov_bp_ee}
{\rm Cov} \bb{ {\cal C}^{(ij)}_{{\rm E/B},m} ;\, {\cal C}^{(kl)}_{{\rm E/B},n} } &=  \frac{\pi^2}{{\cal N}_m\, {\cal N}_n}\; \sum_{a,b} \Delta \theta_a\, \theta_a\; T(\theta_a)\; \Delta \theta_b\, \theta_b\; T(\theta_b)\;
\Bigg\{ g_+^m(\theta_a)\,  g_+^n(\theta_b)\;   {\rm Cov} \bb{\xi_+^{(ij)}(\theta_a);\, \xi_+^{(kl)}(\theta_b)}  \\ \nn
&\hspace*{-2.2cm} + g_-^m(\theta_a)\,  g_-^n(\theta_b)\;  {\rm Cov} \bb{\xi_-^{(ij)}(\theta_a);\, \xi_-^{(kl)}(\theta_b)} \pm g_+^m(\theta_a)\,  g_-^n(\theta_b)\;   {\rm Cov} \bb{\xi_+^{(ij)}(\theta_a);\, \xi_-^{(kl)}(\theta_b)} \pm g_-^m(\theta_a)\,  g_+^n(\theta_b)\;   {\rm Cov} \bb{\xi_-^{(ij)}(\theta_a);\, \xi_+^{(kl)}(\theta_b)}  \Bigg\}  \,
}
for cosmic shear, and 
\eq{
\label{eq:cov_bp_ne}
{\rm Cov} \bb{ {\cal C}^{(ij)}_{{\rm n \epsilon},m} ;\, {\cal C}^{(kl)}_{{\rm n \epsilon},n} } =  \frac{4 \pi^2}{{\cal N}_m\, {\cal N}_n}\; \sum_{a,b} \Delta \theta_a\, \theta_a\; T(\theta_a)\; \Delta \theta_b\, \theta_b\; T(\theta_b)\; h^m(\theta_a)\,  h^n(\theta_b)\;  {\rm Cov} \bb{\ba{\gamma_{\rm t}}^{(ij)}(\theta_a);\, \ba{\gamma_{\rm t}}^{(kl)}(\theta_b)} \,,
}
for GGL. The cross-variances between these signals, and expressions involving clustering (not used in this work), are obtained analogously. The angular binning is the same as that of the correlation functions, which is described in Sect.~\ref{sec:bandpower_measurement}.

\end{appendix}
\end{document}